\documentclass[a4paper,11pt]{article}
\usepackage{epsfig,amsmath,amsfonts,amssymb,mathtools,wasysym,tikz,jheppub,dsfont}
\usepackage[mathscr]{euscript}

\usepackage[format=plain,
            labelfont=it,
            textfont=it]{caption}

\usepackage{tikz}
\usetikzlibrary{arrows, positioning}

\newcommand{\be}{\begin{equation}}
\newcommand{\ee}{\end{equation}}

\newcommand{\doublet}[2]{\left(\begin{array}{c}#1\\#2\end{array}\right)}

\newcommand{\half}{\frac12}

\newcommand{\Urm}{\mathrm{U}}

\newcommand{\SU}{\mathrm{SU}}
\newcommand{\SO}{\mathrm{SO}}

\newcommand{\Acal}{\mathcal{A}}
\newcommand{\Fcal}{\mathcal{F}}

\newcommand{\Rcal}{\mathcal{R}}
\newcommand{\Lcal}{\mathcal{L}}
\newcommand{\Wcal}{\mathcal{W}}

\newcommand{\Zset}{\mathbb{Z}}
\newcommand{\Ncal}{\mathcal{N}}

\newcommand{\Vcal}{\mathcal{V}}
\newcommand{\Vscr}{\mathscr{V}}

\newcommand{\Cset}{{\,\,{{{^{_{\pmb{\mid}}}}\kern-.47em{\mathrm C}}}}}

\newcommand{\ket}[1]{\left|#1\right\rangle}

\newcommand{\diff}{\mathrm{d}}
\newcommand{\p}{\partial}
\newcommand{\ra}{\rightarrow}

\newcommand{\tr}{\mathrm{tr}}

\definecolor{darkgreen}{rgb}{0.0, 0.5, 0.0}

\newcommand{\Xb}{\bar{X}}
\newcommand{\Yb}{\bar{Y}}
\newcommand{\Zb}{\bar{Z}}

\newcommand{\id}{\mathrm{id}}

\newcommand{\comment}[1]{}

\newcommand{\op}{{{}_{\circ}}}

\newcommand{\nocontentsline}[3]{}
\let\origcontentsline\addcontentsline
\newcommand\stoptoc{\let\addcontentsline\nocontentsline}
\newcommand\resumetoc{\let\addcontentsline\origcontentsline}

\newcommand{\R}[2]{\mathcal{R}^{#1}_{\phantom{a}#2}}
\newcommand{\1}{\mathds{1}}
\newcommand{\D}[2]{\Delta(\mathcal{R}^#1_{\phantom{a}#2})}

\newcommand{\Oab}{\Omega^a_{\phantom{a}b}}

\definecolor{BurntOrange}{rgb}{0.8, 0.33, 0.0}

\newcommand{\tallvdots}{%
  \hspace{1pt}\vcenter{%
    \baselineskip=3pt \lineskiplimit=0pt
   \hbox{\textcolor{black}{\textbf{.}}}\hbox{\textcolor{black}{\textbf{.}}}
  }\hspace{1pt}%
}
\makeatother

\renewcommand{\tallvdots}{%
  \hspace{1pt}\vcenter{%
    \baselineskip=2.2pt \lineskiplimit=0pt
  \hbox{\textcolor{black}{.}}\vspace{0.7pt}\hbox{\textcolor{black}{.}}\vspace{0.7pt}\hbox{\textcolor{black}{.}}
  }\hspace{1pt}%
}

\newcommand{\bracketing}[4]{#1 \tallvdots #2 #3 \tallvdots #4}

\newcommand{\bracing}[1]{| #1 \rangle}
\newcommand{\PhiM}{\Phi}

\newcommand*\circled[1]{\tikz[baseline=(char.base)]{
    \node[shape=circle, draw, inner sep=1pt, 
        minimum height=12pt] (char) {\vphantom{1g}#1};}}

\title{Hidden Symmetries of 4D $\Ncal=2$ Gauge Theories}

\author[a]{Hanno Bertle,}
\author[b]{Elli Pomoni,}
\author[c]{Xinyu Zhang,}
\author[d,e]{and Konstantinos Zoubos}

\affiliation[a]{II. Institut f\"ur Theoretische Physik,
Universit\"at Hamburg, \\ 
Luruper Chaussee 149, 22761 Hamburg, Germany}
\affiliation[b]{Deutsches Elektronen-Synchrotron DESY,
Notkestr. 85, 22607 Hamburg, Germany}
\affiliation[c]{Zhejiang Institute of Modern Physics, School of Physics, Zhejiang University, \\ 866 Yuhangtang Rd, Hangzhou, Zhejiang, 310058, China}
\affiliation[d]{Department of Physics,
University of Pretoria, \\
Private Bag X20, Hatfield 0028, South Africa}
\affiliation[e]{National Institute for Theoretical and Computational Sciences (NITheCS),\\ Gauteng, South Africa}

\emailAdd{hanno.bertle@desy.de}
\emailAdd{elli.pomoni@desy.de}
\emailAdd{xinyu.zhang@zju.edu.cn}
\emailAdd{kzoubos@up.ac.za}

\preprint{DESY-24-172}

\abstract{

\bigskip

We study the global symmetries of the $\Zset_2$-orbifold of $\Ncal=4$ Super-Yang-Mills theory and its marginal deformations. The process of orbifolding to obtain an $\Ncal=2$ theory would appear to break the $\SU(4)$ R-symmetry down to $\SU(2)\times\SU(2)\times \Urm(1)$. We show that the broken generators can be recovered by moving beyond the Lie algebraic setting to that of a Lie algebroid. This remains true when marginally deforming away from the orbifold point by allowing the couplings of the $\SU(N)\times\SU(N)$ gauge groups to vary independently. 
The information about the marginal deformation is captured by a Drinfeld-type twist of this SU(4) Lie algebroid.
The twist is read off from the F- and D- terms, and thus directly from the Lagrangian. 
Even though at the orbifold point the algebraic structure is associative, it becomes non-associative
after the marginal deformation. We explicitly check that the planar Lagrangian of the theory is invariant under this twisted version of the $\SU(4)$ algebroid and we discuss implications of this hidden symmetry for the spectrum of the $\Ncal=2$ theory.}

\begin{document}
\maketitle

\section{Introduction}

Symmetries are central to our understanding of the fundamental laws of nature, with Lie groups and their associated Lie algebras serving as the standard framework for describing continuous symmetries. 
However, Lie groups represent a specific case within a broader class of mathematical structures, such as quantum groups (see \cite{Majid,kassel2012quantum} for introductions). These play a crucial role in the study of two-dimensional quantum integrable systems and conformal field theory (see \cite{Gomez96} for an overview). It is also possible to relax some of the group axioms while still capturing the essential notions of symmetry. In particular, abandoning the requirement that all group operations can be composed leads to the concept of a groupoid (see \cite{Brown1987,Weinstein1996} for reviews). 

In four dimensions, there exists a unique maximally supersymmetric gauge theory, $\Ncal=4$ super Yang-Mills (SYM), with gauge group $\SU(N)$ and a marginal coupling constant $g$. The emergence of an integrable structure in the planar limit \cite{Minahan:2002ve} (see \cite{Beisert:2010jr,Bombardelli:2016rwb} for reviews) has enabled the derivation of results that would otherwise be inaccessible. This integrability allows for efficient evaluation of the spectrum and a wide range of observables in $\Ncal=4$ SYM.

The landscape of $\Ncal=2$ superconformal field theories (SCFTs) in four dimensions is vast and not yet fully charted. A complete classification of Lagrangian $\Ncal=2$ SCFTs was achieved in \cite{Bhardwaj:2013qia}, and a significant subset of these theories can be constructed through orbifolding $\Ncal=4$ SYM \cite{Kachru:1998ys,Lawrence:1998ja}. Naturally, one might ask whether the remarkable property of planar integrability in $\Ncal=4$ SYM is inherited by this class of $\Ncal=2$ SCFTs. At the particular submanifold of the conformal manifold where all the marginal couplings are equal, known as the \emph{orbifold point} (which is not really a point but a fixed line parametrised by the remaining coupling), it has been shown that integrability persists \cite{Beisert:2005he}. However, progress towards understanding integrability at generic points of the conformal manifold has remained elusive (see \cite{Pomoni:2019oib} for a discussion). Recent work \cite{Pomoni:2021pbj} revisited these theories from the perspective of dynamical symmetries, though the implications for integrability remain unclear.

A key step in understanding planar integrability is to uncover as fully as possible the set of symmetries governing the planar limit of $\Ncal=2$ SCFTs, which is traditionally thought to consist of the $\Ncal=2$ superconformal symmetry $\SU(2,2|2)$, along with a global flavour symmetry. 
However, in \cite{Pomoni:2021pbj}, it was argued that an enhanced symmetry may be at play --- specifically, a deformation of $\mathrm{PSU}(2,2|4)$, the superconformal group of the parent $\Ncal=4$ SYM. The continuous deformation parameters correspond to the exactly marginal Yang-Mills couplings, which are the coordinates of the $\Ncal=2$ conformal manifold.

In this work, we take further steps towards uncovering the hidden symmetries of the marginally deformed orbifold theories, focusing primarily on the case of the $\Zset_2$ orbifold of $\Ncal=4$ SYM. In Section \ref{OPSymmetries}, starting at the orbifold point, we broaden the mathematical structure with which we describe the symmetry of the theory, from Lie groups to Lie groupoids. This allows us to define generators relating fields in different representations of the $\SU(N)\times \SU(N)$ gauge group of the theory, which would naively be considered broken by the orbifolding process. Through explicit computation, we demonstrate that the action of the theory is invariant under all generators of $\SU(4)$, rather than only those left unbroken by the orbifolding process. It is important to stress that this is only possible after modifying the action of the broken generators in a way that goes beyond the standard Lie-algebraic framework, to that of Lie groupoids.

Next, we focus on the study of marginal deformations of the $\Zset_2$ theory as we move away from the orbifold point by allowing the two gauge couplings to vary independently. This gives us a family of deformations, parametrised by the ratio of the gauge couplings, which still preserves the $\Ncal=2$ superconformal symmetry. 

To uncover the marginally-deformed version of the $\SU(4)$ Lie groupoid which we discovered at the orbifold point, we need to find a way to twist this algebraic structure, analogously to the twists leading to the marginal deformations of the $\Ncal=4$ SYM theory \cite{Beisert:2005if,Ihry:2008gm,Ahn:2010ws,vanTongeren:2015soa,Dlamini:2016aaa,Dlamini:2019zuk}. To achieve this, in Section \ref{sec:quantumplanes}, we turn our attention to the  F- and D-terms of the theory. As already suggested in \cite{Pomoni:2021pbj,Mansson:2008xv}, the F- and D-terms 
define a \emph{quantum plane} structure and thus provide a means to define the twist at the quantum plane (or braid) limit. In our $\Ncal=2$ setting, it is natural to extend these quantum planes using the unbroken $\SU(2)$ R-symmetry to obtain $\SU(2)$ multiplets of quantum planes. This allows us to write the Lagrangian in a form that makes the unbroken R-symmetry explicit.

In Section \ref{TwistsSection}, we construct two-site twists which take the trivial quantum planes at the orbifold point to the marginally deformed ones. In order to define the action of these twists on the Lagrangian, they need to be extended to three and four sites. In Section \ref{Superpotentialinvariance} we find that a simple coassociative extension of the two-site twist correctly twists the superpotential at the orbifold point to that of the marginally deformed theory. This allows us to define a twisted action of the naively broken $\SU(3)$ generators, under which the superpotential is invariant. 

In Section \ref{LagrangianInvariance}
we demonstrate that not only the superpotential but also the full bosonic action can be derived from their counterparts at the orbifold point via a twisting procedure.\footnote{This is important for the following reason. The F- and D-terms allow us to define the two-site twist. On the superpotential, we need to act with a three-site twist, while on the scalar potential of the theory, we need to act with a four-site twist. Having a way to move from two-site, to three-site, and finally to four-site twists is crucial for the precise definition of the algebraic structure we aim to define.} 
Inverting these twists, we show that it is possible to untwist the Lagrangian of the marginally deformed $\Zset_2$ theory, effectively undoing the marginal deformation, and recover the theory at the orbifold point.
Thus, we can define the action of the broken $\SU(4)$ generators also on the marginally deformed quartic terms. 

However, as described in Section \ref{subsec:nonassociativity}, for the construction to be well-defined, associativity must be handled with care. In particular, for the quartic terms in the Lagrangian, there are two inequivalent orderings (a special form of non-associativity), which we denote using different placements of parentheses (bracketings). These orderings are related by a map analogous to what is known as a \emph{coassociator} in the context of quasi-Hopf algebras. Our derivation of the invariance of the quartic terms under all the recovered $\SU(4)$ generators relies on the explicit construction of this coassociator.

Finally, in Section \ref{sec:spectrum},
we examine some aspects of the spectrum of the theory, specifically the 
eigenstates of the one-loop Hamiltonian, and  explore the additional insight that can be gained through the quantum symmetries we have uncovered. In the $\Ncal=4$ SYM theory, the eigenstates of the Hamiltonian are classified by the linear irreducible representations of $\SU(4)$. Our goal is to recover a deformed algebroid version of $\SU(4)$ representation theory which does the same for our $\Ncal=2$ theory. This can be done in full rigour at the orbifold point,  but becomes more challenging in the marginally deformed case. 
Nevertheless, for the special case of BPS multiplets, as well as for all multiplets of length 2, we \emph{are} able to define a suitable action of the broken generators which takes us among the states of the multiplet. This provides a connection between states in the physical spectrum of the theory which would not be visible if one had worked with only the unbroken symmetries. 

Before concluding the introduction, we must acknowledge that our work is inspired by and builds upon \cite{Beisert:2017pnr,Garus:2017bgl,Beisert:2018zxs,Beisert:2018ijg}, where the Yangian symmetry of planar $\Ncal=4$ SYM and related theories is demonstrated at the level of the classical equations of motion, as well as directly at the level of the Lagrangian. As explained in those works, in order to define a consistent action of the Yangian generators on the Lagrangian, the colour trace must be cut open. This is of relevance to our work as well, since the action of the broken generators changes the colour representation of an open state and thus cannot be consistently defined within the trace operation. Given a closed state, an open state is defined through a cyclic opening-up procedure, which we present in Appendix \ref{Ap:OpeningUp}, and this will allow a consistent action of the broken generators.

\section{The $\Zset_2$ orbifold of $\Ncal=4$ SYM}

Following \cite{Gadde:2009dj,Gadde:2010zi}, to define the theory, we start with $\Ncal=4$ SYM with gauge group $\SU(2N)$. In $\Ncal=1$ superspace language, it consists of a vector multiplet $V$ and three chiral multiplets $X,Y,Z$, all in the adjoint representation of the gauge group. The R-symmetry group of the theory is $\SU(4)\simeq\SO(6)$, but only its subgroup $\SU(3)\times \Urm(1)$ is manifest in the superspace formulation. See Appendix \ref{SU4Appendix} for our conventions for the action of R-symmetry generators on the scalar fields. 

\begin{figure}[t]
\begin{center}
\begin{tikzpicture}[scale=0.6]
  \draw[fill=blue] (0,0) circle (2ex);
\draw[fill=red] (6,0) circle (2ex);

\draw[->,blue,thick] (-0.5,-0.8) arc (-60:-310:1);
\draw[->,red,thick] (6.4,0.8) arc (120:-130:1);

\draw[->,purple,thick] (0.2,0.8) arc (110:70:8);
\draw[->,purple,thick] (5.7,0.5) arc (70:110:8);
\draw[->,purple,thick] (5.7,-0.8) arc (-70:-110:8);
\draw[->,purple,thick] (0.2,-0.5) arc (-110:-70:8);

\node at (-0.7,-0) {$1$};\node at (6.8,-0) {$2$};
\node at (3,1.8) {$X_{12}$};\node at (3,-1.7) {$X_{21}$};
\node at (3,0.5) {$Y_{21}$};\node at (3,-0.4) {$Y_{12}$};
\node at (-2.6,0) {$Z_1$};\node at (8.5,0) {$Z_2$};

\end{tikzpicture}
\end{center}
\caption{ The quiver representing the $\Zset_{2}$  $\SU(N)\times\SU(N)$ SCFT. The arrows indicate the order in which fields must be composed, starting from the left, in order to obtain valid gauge index contractions. }
\label{fig:z2}
\end{figure}
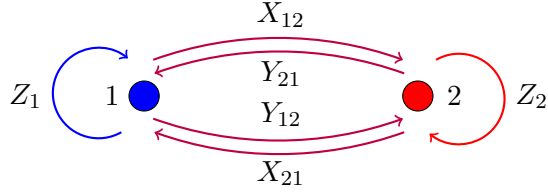

We are interested in reducing the amount of supersymmetry from $\Ncal=4$ to $\Ncal=2$. One way to achieve this is to perform a simultaneous orbifold projection in the R-symmetry space
\be
(V,X,Y,Z)\ra (V,-X,-Y,Z)
\ee
and in the colour space
\be
\varphi\ra \tau^{-1} \varphi \tau ,
\ee
where
\be
\tau =\begin{pmatrix}I_{N\times N} & 0\\
0 & -I_{N\times N}
\end{pmatrix} ,
\ee
and $\varphi$ represents any of the above fields. The orbifolding breaks the gauge group down to $\SU(N)_1\times \SU(N)_2$ and the nonzero components of the chiral fields are:
\be \label{XYZblocks}
X=\begin{pmatrix}0 & X_{12}\\
X_{21} & 0
\end{pmatrix} , \quad 
Y=\begin{pmatrix}0 & Y_{12}\\
Y_{21} & 0
\end{pmatrix} , \quad 
Z=\begin{pmatrix}Z_{1} & 0\\
0 & Z_{2}
\end{pmatrix} ,
\ee
where we have indicated $N\times N$ blocks of the original $2N\times 2N$ matrices. The fields $X_{12}$ and $Y_{12}$ are in the bifundamental representation $\left(\mathbf{N}_1,\overline{\mathbf{N}}_2\right)$ of $\SU(N)_1\times \SU(N)_2$, while $X_{21}$ and $Y_{21}$ are in the conjugate representation $\left(\mathbf{N}_2,\overline{\mathbf{N}}_1\right)$. The fields $Z_1$ and $Z_2$ are in the adjoint representation of each individual gauge group, respectively. These features can be conveniently summarised by a quiver diagram, as in Figure \ref{fig:z2}.

As the $X,Y$ fields now belong to a different representation from the $Z$ fields, the orbifold also breaks the $\SU(4)$ R-symmetry group down to a subgroup, $\SU(2)_L\times\SU(2)_R\times \Urm(1)_r$, with the details of the breaking given in Appendix \ref{SU4Appendix}. Starting in Section \ref{OPSymmetries}, we will explain that the broken generators can be regained by suitably extending our algebraic framework. 

The Lagrangian with generic coupling constants is expressed in the $\Ncal=1$ superspace formulation as
\begin{equation} \label{N2Lagrangian}
\mathcal{L}=\mathcal{L}_{K}+\left(\int\diff^2\theta \ \Wcal+\int\diff^2\bar{\theta} \ \bar{\Wcal}\right)  .
\end{equation}
The K\"ahler part is
\be
\begin{split}
\mathcal{L}_{K} & =  \frac{1}{2}\sum_{i=1}^{2}\left(\int \diff^{2}\theta \ \tr_{i}W_{i}^{\alpha}W_{i\alpha}+\mathrm{h.c.}\right) +\sum_{i=1}^{2}\int \diff^{4}\theta \ \tr_{i}\left(\bar{Z}_{i}e^{g_{i}V_{i}}Z_{i}e^{-g_{i}V_{i}}\right) \\
 &+\int \diff^{4}\theta \ \tr_{1}\left(\bar{X}_{12}e^{g_{2}V_{2}}X_{21}e^{-g_{1}V_{1}}+Y_{12}e^{-g_{2}V_{2}}\bar{Y}_{21}e^{g_{1}V_{1}}\right) \\
 &+ \int \diff^{4}\theta \ \tr_{2}\left(\bar{X}_{21}e^{g_{1}V_{1}}X_{12}e^{-g_{2}V_{2}}+Y_{21}e^{-g_{1}V_{1}}\bar{Y}_{12}e^{g_{2}V_{2}}\right) ,
 \label{eq:LagSP}
\end{split}
\ee
with the kinetic terms canonically normalised. The superpotential is given by
\be \label{superpotential}
\Wcal =  g_{1}\tr_{1}\Big((Y_{12}X_{21}-X_{12}Y_{21})Z_{1}\Big)+ g_{2}\tr_{2}\Big((Y_{21}X_{12}-X_{21}Y_{12})Z_{2}\Big) .
\ee
Note that the traces are with respect to different gauge groups. To obtain the Lagrangian at an \emph{orbifold point}, we simply set $g_1=g_2$. 
The full $\mathcal{N}=2$ superconformal invariance of the orbifold theory is preserved when taking $g_1\neq g_2$.

To write the Lagrangian in components, one expands the superfields and integrates over the Grassmann coordinates $\theta, \bar{\theta}$, see e.g. the lectures \cite{Pomoni:2019oib} for more details.

In this work we will not be interested in the fermionic components of the theory, nor the gauge fields. We will just focus on the scalar fields. Their kinetic terms take the standard forms ($D_\mu X_{12} D^\mu \bar{X}_{21}$, etc.). In order to obtain the quartic terms that contribute to the potential we need to integrate out the auxiliary F and D fields, leading to the F-term and D-term relations:
\be \label{FDg}
\begin{split}
    F_{12}^{Y} &= g_2 X_{12} Z_2 - g_1 Z_1 X_{12} \ ,\quad
    \bar{F}_{12}^{\bar{Y}} =  g_2\bar{X}_{12} \bar{Z}_2 - g_1 \bar{Z}_1 \bar{X}_{12} \ , \\
    F_{12}^{X} &=  g_2 Y_{12} Z_2 - g_1 Z_1 Y_{12} \ , \quad
    \bar{F}_{12}^{\bar{X}} = g_2\bar{Y}_{12} \bar{Z}_2 - g_1 \bar{Z}_1 \bar{Y}_{12} \ , \\
    F_1^{Z} &= g_1( X_{12} Y_{21} - Y_{12} X_{21}) \ , \quad
    \bar{F}_1^{\bar{Z}} = g_1(\bar{X}_{12} \bar{Y}_{21} - \bar{Y}_{12} \bar{X}_{21} )\ ,\\
 D_1 &= g_1\left( \bar{X}_{12} X_{21} + \bar{Y}_{12} Y_{21}- X_{12} \bar{X}_{21}  - Y_{12} \bar{Y}_{21} +Z_1\Zb_1-\Zb_1Z_1\right)\ ,  
\end{split}
\ee
together with their $\Zset_2$-conjugates obtained by the simultaneous exchange of $1\leftrightarrow 2$ indices (including $g_1\leftrightarrow g_2$). These contribute to the quartic terms as 
\be
\begin{split}
\mathcal{V}(g_1,g_2)&=\tr_1\left(F_{12}^X\bar{F}_{21}^{\bar X}+F_{12}^Y\bar{F}_{21}^{\bar Y}+F_{1}^Z\bar{F}_{1}^{\bar Z}+\half D_1^2\right)\\
&+\tr_2\left(F_{21}^X\bar{F}_{12}^{\bar X}+F_{21}^Y\bar{F}_{12}^{\bar Y}+F_{2}^Z\bar{F}_{2}^{\bar Z}+\half D_2^2\right) ,
\end{split}
\ee
which gives in the planar limit\footnote{The full action also contains double-trace terms \cite{Gadde:2009dj}, which we will omit as they are proportional to $\frac{1}{N}$ and should therefore be subleading in the planar limit.}
\begin{align}
    \mathcal{V}(g_1,g_2) = &g_1^2\tr_1\left[ \frac{1}{2} \left[ \bar{Z}_1, Z_1 \right]^2 + \mathcal{M}^{(\mathbf{1})}_1 (Z_1 \bar{Z}_1 + \bar{Z}_1 Z_1 ) + (\mathcal{M}^{(3)}_1)^2 - \frac{1}{2} (\mathcal{M}_1^{(\mathbf{1})})^2 \right] \nonumber \\
    &+g_2^2 \hspace{1pt} \tr_2\left[ \frac{1}{2} \left[ \bar{Z}_2, Z_2 \right]^2 + \mathcal{M}^{(\mathbf{1})}_2 (Z_2 \bar{Z}_2 + \bar{Z}_2 Z_2 ) + (\mathcal{M}^{(3)}_2)^2 - \frac{1}{2} (\mathcal{M}^{(\mathbf{1})}_2)^2 \right] \nonumber \\
    &-2 g_1g_2 \hspace{1pt} \tr_1\left[ Z_1 X_{12} \bar{Z}_2 \bar{X}_{21} + Z_1 Y_{12} \bar{Z}_2 \bar{Y}_{21} + Z_1 \bar{X}_{12} \bar{Z}_2 X_{21}+ Z_1 \bar{Y}_{12} \bar{Z}_2 Y_{21} \nonumber \right] \\
    &-2 g_1g_2 \hspace{1pt} \tr_2\left[ Z_2 X_{21} \bar{Z}_1 \bar{X}_{12} + Z_2 Y_{21} \bar{Z}_1 \bar{Y}_{12} + Z_2 \bar{X}_{21} \bar{Z}_1 X_{12}+ Z_2 \bar{Y}_{21} \bar{Z}_1 Y_{12} \right] ,
    \label{eq:FullQuarticTerms}
\end{align}
where we have defined the $\SU(2)_R$ R-symmetry singlet and triplet mesons with the colour index of the second gauge group $\SU(N)_2$ contracted and the first $\SU(N)_1$ open as
\be
\begin{split}
\mathcal{M}^{(\mathbf{1})}_1 &\coloneq  X_{12} \bar{X}_{21} + Y_{12} \bar{Y}_{21} + \bar{X}_{12} X_{21} + \bar{Y}_{12} Y_{21} \;,\\
    (\mathcal{M}^{(\mathbf{3})}_1)^2 &\coloneq (\bar{Y}_{12}\bar{X}_{21}-\bar{X}_{12}\bar{Y}_{21})(X_{12}Y_{21}-Y_{12}X_{21})+(X_{12}Y_{21}-Y_{12}X_{21})(\bar{Y}_{12}\bar{X}_{21}-\bar{X}_{12}\bar{Y}_{21}) \nonumber \\
    &\quad +(X_{12}\bar{X}_{21}+Y_{12}\bar{Y}_{21})(X_{12}\bar{X}_{21}+Y_{12}\bar{Y}_{21})+(\bar{X}_{12}X_{21}+\bar{Y}_{12}Y_{21})(\bar{X}_{12}X_{21}+\bar{Y}_{12}Y_{21}) \ ,
    \label{eq:DefM}
\end{split}
\ee
and their $\Zset_2$ conjugates, $\mathcal{M}^{(\mathbf{1})}_2$ and $\mathcal{M}^{(\mathbf{3})}_2$, respectively, with
the colour index of the first gauge group $\SU(N)_1$ contracted and the second $\SU(N)_2$ open.

These quartic terms form a singlet under the $\SU(2)_L\times \SU(2)_R\times \Urm(1)_r$ subgroup of $\SU(4)$, defined  by the unbroken generators $\Rcal^a_{\;b}$, with $a,b\in\{1,2\}$ or $a,b\in\{3,4\}$ (see Appendix \ref{SU4Appendix} for more details). In the following we will be interested in showing invariance of the action, in a generalised sense that we will define, for all the $\SU(4)$ generators. In the next section we will start with the orbifold point $g_1=g_2$ and then proceed to the marginally deformed case of $g_1\neq g_2$ in the following sections.

\section{Symmetries at the orbifold point} \label{OPSymmetries}

It has been known since the work of \cite{Ideguchi:2004wm, Beisert:2005he} that planar integrability persists in the $\Zset_k$ orbifolds of $\Ncal=4$ SYM.\footnote{It is noteworthy that this is also the case for extensions to more general orbifolds \cite{Solovyov:2007pw}. See \cite{Zoubos:2010kh,Beccaria:2011qd, deLeeuw:2012hp,Skrzypek:2022cgg} for reviews and further results related to integrability of orbifolds of $\Ncal=4$ SYM.}
Clearly, this implies that the Yangian-type symmetries of $\Ncal=4$ SYM \cite{Beisert:2017pnr} must also be present in the orbifold theories. As the Yangian is an extension of the global symmetry group of the theory, one might expect that there is a way to recover the full $\SU(4)$ symmetry by some kind of ``untwisting'' procedure, similar to \cite{Garus:2017bgl} for the $\beta$-deformation of $\Ncal=4$ SYM.

However, the situation in the orbifold case is not as straightforward. Recall that in $\Ncal=4$ SYM, all the fields are in the adjoint representation of the gauge group. So, for example, the fields $X$ and $Z$ form an $\SU(2)_{XZ}$ doublet. However, after the $\Zset_2$ orbifolding process, $X_{12}$ and $X_{21}$ are in bifundamental representations while $Z_1$ and $Z_2$ are in  adjoint representations. 
Therefore, the descendants of $X$ and $Z$ belong to different vector spaces and do not form a doublet of the standard Lie algebra of $\SU(2)_{XZ}$. 
Thus, when we are restricted at the Lie algebra level, the raising and lowering generators of the $\SU(2)_{XZ}$ group (in our notation  $\sigma^+_{XZ}=\Rcal^3_{\;2}$ and $\sigma^-_{XZ}=\Rcal^2_{\;3}$, see Appendix \ref{SU4Appendix} for our conventions) are broken. In this work we use the notation ``broken" for $\mathfrak{su}(4)$ generators which do not reduce to $\mathfrak{su}(2)_L\times\mathfrak{su}(2)_R\times \mathfrak{u}(1)_r$ generators. Considering the other $\SU(2)$ sectors involving 
$YZ$, $\Xb Z$, and $\Yb Z$, the raising and lowering generators are also broken. In Appendix \ref{SU4Appendix} the reader can find how to embed these $\SU(2)$ sectors in the $\SU(4)$ R-symmetry group of $\Ncal=4$ SYM.

In the following, we will argue that we can consistently recover these broken generators by going beyond the Lie algebraic setting and working instead in the framework of Lie groupoids and their corresponding algebroids \cite{Brown1987,Weinstein1996}. We refer to Appendix \ref{GroupoidAppendix} for the main definitions. The close connection between orbifolds (and quiver theories more generally) and groupoids is well known (see \cite{moerdijk2002orbifolds} for an introduction), but our approach differs from previous treatments in that we are interested in consistently defining the action of the naively broken R-symmetry symmetry generators on products of fields, which are identified with paths on the quiver, as we will now describe. 

\paragraph{The path groupoid}

\mbox{}

Before introducing the Lie algebroid that will replace the R-symmetry Lie algebra, it is crucial to describe the vector space on which it acts. From a practical perspective for physicists, this topic is extensively covered in several papers \cite{Gadde:2009dj,Gadde:2010zi}, including the review \cite{Pomoni:2019oib}.
The construction relies on the planar limit, where spin chain states correspond to single-trace operators in the gauge theory. For $\mathcal{N}=4$ SYM, spin chain states are straightforwardly constructed as a direct product $\mathscr{V} \otimes \mathscr{V} \otimes \cdots \otimes \mathscr{V}$ of the unique singleton representation $\mathscr{V}$ of the $\mathcal{N}=4$ superconformal algebra, defined on a single site and associated with the adjoint representation of the colour group $\SU(N)$.
The $\mathcal{N}=2$ superconformal algebra, however, has multiple ultrashort representations, each associated with different representations of the colour group as dictated by the quiver of the theory. In $\mathcal{N}=2$ quiver theories, the total space of spin chain states is not a simple product because the colour index structure imposes constraints. In the planar limit, the allowed single trace operators are those that follow the arrows in the quiver, as illustrated in Figure \ref{fig:z2}. For instance, $\mathrm{tr}(X_{12}Z_2 X_{21} X_{12} Z_2 X_{21})$ represents a valid single trace operator corresponding to a spin chain state. However, a sequence of fields like $X_{12} Z_2 X_{12} X_{12} Z_2 X_{21}$ is not allowed, as there is no way to contract the colour indices to obtain a single-trace operator and is therefore excluded in the planar limit.
Reference \cite{Pomoni:2021pbj} presents a concise and accessible approach for understanding the total space of spin chain states using the concept of a dynamical spin chain. While this remains the most useful framework for physicists, it can be further abstracted and simplified. The vector space containing the spin chain states is elegantly described by the concept of a \emph{quiver path groupoid}, which can be viewed as a vector space where only a subset of possible element compositions is permitted. These allowed compositions correspond to paths that follow the arrows in the quiver\footnote{For a more formal definition of the path groupoid product $m$, we refer to Appendix \ref{GroupoidAppendix}.}, such as $\mathscr{V}_{(12)} \otimes \mathscr{V}_{(22)} \otimes \mathscr{V}_{(21)} \otimes \mathscr{V}_{(11)} \cdots$.  In our case, the quiver of the $\SU(N)\times \SU(N)$ $\mathcal{N}=2$ SCFT is depicted in Figure \ref{fig:z2}. 

\paragraph{The R-symmetry Lie groupoid}

\mbox{}

Having described the vector space, we now turn to the groupoid structure that will replace the R-symmetry group of $\mathcal{N}=4$ SYM. It is important to emphasise that the quiver path groupoid, which characterises the vector space of spin chains, and the algebraic structure we will introduce as the \emph{R-symmetry groupoid}, are distinct entities. The R-symmetry groupoid, which replaces the R-symmetry group, acts on the quiver path groupoid. 

 Generators in a Lie algebra can be understood as infinitesimal deviations from the Lie group identity element. Similarly, the generators of our R-symmetry algebroid span the infinitesimal version of the R-symmetry groupoid. Hence, in the following we will use both terms depending on the context. 

\begin{figure}
    \centering

\begin{tikzpicture}[
node distance=1.5cm and 2cm,
  box/.style={draw, dashed, minimum size=1cm, text centered},
  arrow/.style={->, >=stealth', thick},
  loop above/.style={->, out=180, in=90, looseness=4, thick}, % Adjusted loop thickness and smaller size
  loop right/.style={->, out=90, in=0, looseness=4, thick}, % Adjusted loop thickness and smaller size
  loop below/.style={->, out=0, in=270, looseness=4, thick}, % Adjusted loop thickness and smaller size
  loop left/.style={->, out=270, in=180, looseness=4, thick}, % Adjusted loop thickness and smaller size
  doublearrow/.style={<->, >=stealth', thick}
]

% Nodes
\node[box] (v11) {$Z_{1}$};
\node[box, right=of v11] (v12) {$X_{12}$};
\node[box, below=of v11] (v22) {$Z_{2}$};
\node[box, right=of v22] (v21) {$X_{21}$};

% Horizontal arrows

\draw[arrow] ([yshift=0.2cm]v11.east) -- node[above] {$\sigma^{(1)}_{+}$} ([yshift=0.2cm]v12.west);
\draw[arrow] ([yshift=-0.2cm]v12.west) -- node[below] {$\sigma^{(1)}_{-}$} ([yshift=-0.2cm]v11.east);
\draw[arrow] ([yshift=-0.2cm]v22.east) -- node[below] {$\sigma^{(2)}_{+}$} ([yshift=-0.2cm]v21.west);
\draw[arrow] ([yshift=0.2cm]v21.west) -- node[above] {$\sigma^{(2)}_{-}$} ([yshift=0.2cm]v22.east);

% Double arrows for gamma
\draw[doublearrow] (v11) -- node[right] {$\gamma$} (v22);
\draw[doublearrow] (v12) -- node[right] {$\gamma$} (v21);

% Loops from middle to orthogonal outermost edges, with thicker arrows and smaller loops
\draw[loop above] (v11) to node[above left] {$\sigma^{(1)}_3$} (v11); % loop from left edge to top edge
\draw[loop right] (v12) to node[above right] {$\sigma^{(1)}_3$} (v12); % loop from top edge to right edge
\draw[loop below] (v21) to node[below right] {$\sigma^{(2)}_3$} (v21); % loop from right edge to bottom edge
\draw[loop left] (v22) to node[below left] {$\sigma^{(2)}_3$} (v22); % loop from bottom edge to left edge

\end{tikzpicture}
 \caption{A graphical depiction of the Lie algebroid that replaces the $\mathfrak{su}(2)_{XZ}$ Lie algebra, acting on single-site letters. The operator $\gamma$ is the odd $\mathbb{Z}_2$ element that flips the quiver diagram in Figure \ref{fig:z2}, exchanging the gauge groups $1\leftrightarrow2$.}
    \label{fig:SU2algebroid}
\end{figure}
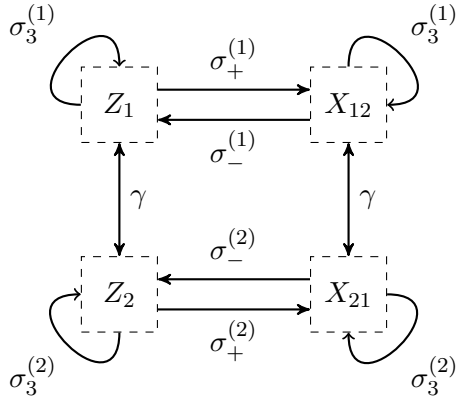

The purpose of the R-symmetry groupoid is to enable a mapping between bifundamental fields, such as $X_{12}$ and $X_{21}$, and the adjoint fields $Z_1$ and $Z_2$. At the level of individual fields (single sites), the algebraic structure that replaces the broken $\mathfrak{su}(2)_{XZ}$ symmetry is illustrated in Figure \ref{fig:SU2algebroid}.
At the level of the algebroid, the naively broken raising and lowering operators act as
\be
\label{eq:brokenSU2}
\sigma_-^{(1)} (X_{12})=Z_{1} \;\;,\quad \;\;
\sigma_+^{(1)} (Z_{1})= X_{12}\;,
\ee
where the ``$=$'' symbol should be understood as a mapping between the two fields, which have different index structures. Here we have adopted the convention that the action of the broken generator flips the second index of the field while preserving the first one. We also note that the planar limit is essential for the bifundamental and adjoint fields to have the same matrix dimension. A way to make sense of expressions like (\ref{eq:brokenSU2})  is to consider that, through the orbifold action, the broken R-symmetry generators have acquired a dependence on the gauge group (specifically, on the labels of the $N\times N$ blocks of the original $\SU(2N)$, as in (\ref{XYZblocks})). This non-direct product form of the R-symmetry and gauge group of the original $\Ncal=4$ SYM theory is what leads to the groupoid structure that we are describing.

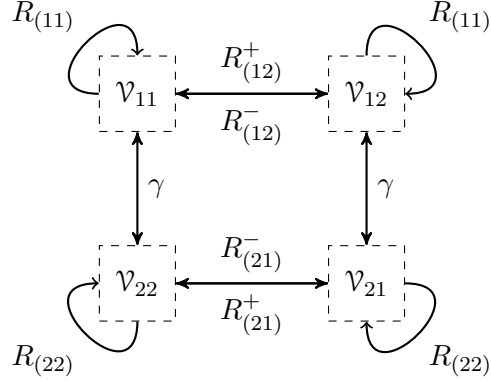
\begin{figure}
    \centering

\begin{tikzpicture}[
node distance=1.5cm and 2cm,
  box/.style={draw, dashed, minimum size=1cm, text centered},
  arrow/.style={->, >=stealth', thick},
  loop above/.style={->, out=180, in=90, looseness=4, thick}, % Adjusted loop thickness and smaller size
  loop right/.style={->, out=90, in=0, looseness=4, thick}, % Adjusted loop thickness and smaller size
  loop below/.style={->, out=0, in=270, looseness=4, thick}, % Adjusted loop thickness and smaller size
  loop left/.style={->, out=270, in=180, looseness=4, thick}, % Adjusted loop thickness and smaller size
  doublearrow/.style={<->, >=stealth', thick}
]

% Nodes
\node[box] (v11) {$\mathscr{V}_{11}$};
\node[box, right=of v11] (v12) {$\mathscr{V}_{12}$};
\node[box, below=of v11] (v22) {$\mathscr{V}_{22}$};
\node[box, right=of v22] (v21) {$\mathscr{V}_{21}$};

% Horizontal arrows
%\draw[arrow] (v11) -- node[above] {$R_{(12)}$} (v12);
%\draw[arrow] (v12) -- node[below] {} (v11);
%\draw[arrow] (v22) -- node[below] {} (v21);
%\draw[arrow] (v21) -- node[above] {$R_{(21)}$} (v22);

%\draw[arrow] ([yshift=0.2cm]v11.east) -- node[above] {$R^{+}_{(12)}$} ([yshift=0.2cm]v12.west);
%\draw[arrow] ([yshift=-0.2cm]v12.west) -- node[below] {$R^{-}_{(12)}$} ([yshift=-0.2cm]v11.east);
%\draw[arrow] ([yshift=-0.2cm]v22.east) -- node[below] {$R^{+}_{(21)}$} ([yshift=-0.2cm]v21.west);
%\draw[arrow] ([yshift=0.2cm]v21.west) -- node[above] {$R^{-}_{(21)}$} ([yshift=0.2cm]v22.east);

\draw[arrow] (v11.east) -- node[above] {$R^{+}_{(12)}$} (v12.west);
\draw[arrow] (v12.west) -- node[below] {$R^{-}_{(12)}$} (v11.east);
\draw[arrow] (v22.east) -- node[below] {$R^{+}_{(21)}$} (v21.west);
\draw[arrow] (v21.west) -- node[above] {$R^{-}_{(21)}$} (v22.east);

% Double arrows for gamma
\draw[doublearrow] (v11) -- node[right] {$\gamma$} (v22);
\draw[doublearrow] (v12) -- node[right] {$\gamma$} (v21);

% Loops from middle to orthogonal outermost edges, with thicker arrows and smaller loops
\draw[loop above] (v11) to node[above left] {$R_{(11)}$} (v11); % loop from left edge to top edge
\draw[loop right] (v12) to node[above right] {$R_{(11)}$} (v12); % loop from top edge to right edge
\draw[loop below] (v21) to node[below right] {$R_{(22)}$} (v21); % loop from right edge to bottom edge
\draw[loop left] (v22) to node[below left] {$R_{(22)}$} (v22); % loop from bottom edge to left edge

\end{tikzpicture}

 \caption{A graphical depiction of the algebraic structure that corresponds to the $\mathfrak{su}(4)$ Lie algebroid.  The vector space on which the algebra acts consists of four distinct sets $\mathscr{V}_{11}$, $\mathscr{V}_{22}$, $\mathscr{V}_{12}$, $\mathscr{V}_{21}$ according to the colour structure of their elements and is itself described by a quiver path groupoid. The algebroid acting on this vector space is made out of unbroken $R_{(11)}$ and $R_{(22)}$, broken  $R^{\pm}_{(12)}$ and $R^{\pm}_{(21)}$ generators as well as $\gamma \in \mathbb{Z}_2$.}
    \label{fig:algebroid}
\end{figure}

The above structure, described for $\mathfrak{su}(2)_{XZ} \subset \mathfrak{su}(4)$, generalises straightforwardly to all generators of $\SU(4)$, allowing us to capture the full $\SO(6)$ scalar sector. The algebraic structure for the entire $\mathfrak{su}(4)$ is depicted in Figure \ref{fig:algebroid}, where the  vector spaces ---adjoint and bifundamental --- are denoted as
\begin{eqnarray}
\label{eq:vectorspace}
&& \Vscr_{11} =
\Big\{
\left\{ Z_1, \bar{Z}_1 \right\}, \left\{ X_{12} X_{21}, Z_1 Z_1,\cdots \right\}
,
\left\{ X_{12} X_{21}Z_1 , Z_1 Z_1 Z_1, \cdots  \right\},
\cdots
\Big\}
\, , \,
\\
&& \Vscr_{12} = \Big\{\left\{ X_{12}, \bar{X}_{12} , Y_{12}, \bar{Y}_{12} \right\}
, \left\{ X_{12}Z_2, Z_1X_{12} \cdots \right\}, \left\{X_{12}X_{21}X_{12}, Z_1X_{12}Z_{2}, \cdots \right\}, \cdots \Big\}  ,
\nonumber
\end{eqnarray}
with $\Vscr_{22}$ and $\Vscr_{21}$ being the $\Zset_2$ conjugates ($1\leftrightarrow 2$) of $\Vscr_{11}$ and $\Vscr_{12}$, respectively.
It is important to stress that the R-symmetry groupoid acts on the entire space of all possible spin chain lengths, i.e. the  entire quiver path groupoid.
This is why in  \eqref{eq:vectorspace}  a set of one-, two-, three- etc. site states appear. Next,
 the sets of unbroken and broken generators acting between these spaces are: 
\begin{eqnarray}
&&
R_{(11)} = R_{(22)}= 
\left\{
\mathcal{R}^1\,_1 , \mathcal{R}^1\,_2  ,
\mathcal{R}^2\,_1  ,
\mathcal{R}^2\,_2 ,
\mathcal{R}^3\,_3 ,
\mathcal{R}^3\,_4 ,
\mathcal{R}^4\,_3  ,
\mathcal{R}^4\,_4
\right\}
\, , \,
\\
&&
 R_{(12)}^{+} =
R_{(21)}^{+} =
\left\{
\mathcal{R}^3\,_1   ,
\mathcal{R}^3\,_2 ,
\mathcal{R}^1\,_4 ,
\mathcal{R}^2\,_4
\right\}
\, , \label{RRaising}
\\
&&
 R_{(12)}^{-} =
R_{(21)}^{-} =
\left\{
\mathcal{R}^1\,_3  ,
\mathcal{R}^2\,_3 ,
\mathcal{R}^4\,_1  ,
\mathcal{R}^4\,_2
\right\}
\, .\label{Rlowering}
\end{eqnarray}
The structure of the diagram in Figure \ref{fig:algebroid} is precisely obtained from the grading that is defined by the orbifold on these generators. 

Note that the unbroken generators  in the set of $R_{(11)} = R_{(22)}$ commute with the orbifolding procedure,
while the  broken generators  in $R^{\pm}_{(12)} = R^{\pm}_{(21)}$ do not. To see this, we can define $\Zset_2$ elements $s_L=(-1)^{i_L}$ and $s_R=(-1)^{i_R}$, where $i_L$ ($i_R$) are the leftmost (rightmost) gauge group indices of an open state. In our conventions, the action of any generator preserves the $s_L$ eigenvalue of any state. Unbroken generators also preserve the $s_R$ eigenvalue, while the action of the broken generators flips the $s_R$ eigenvalue of any state. 

This R-symmetry algebroid should be understood as an extension of the R-symmetry algebra,
 which ensures that only the correct subset of all possible compositions of elements of the basis is allowed. The allowed compositions are those obtained by following the arrows of Figure \ref{fig:algebroid}. 
 Following these arrows it is easy to see that the generators  still obey the $\mathfrak{su}(4)$ algebra
\begin{equation}
    [\R{a}{b},\R{c}{d}] = \delta^c_{\;b} \R{a}{d} - \delta^a_{\;d} \R{c}{b} \ .
    \label{eq:SU(4)algebra}
\end{equation}
As we show in Appendix \ref{Ap:commutators}, the algebroid generators obey the graded structure
\be\begin{split} \label{eq:broken-unbroken}
    [(\text{unbroken}),(\text{unbroken})] &= (\text{unbroken}) ,\\
    [(\text{broken}),(\text{unbroken})] &= (\text{broken}) ,\\ 
    [(\text{broken}),(\text{broken})] &= (\text{unbroken}) \ .
   \end{split}\ee
Finally, we note that the base of this algebroid is a discrete set rather than a continuous space. 

Having written down, in equation \eqref{eq:brokenSU2}, the action of the R-symmetry algebroid on single site elements of the quiver path groupoid, we next turn to the action on spin chain states with more than one site. When a broken generator acts on a product of fields, we need to generalise equation \eqref{eq:brokenSU2} accordingly such that all the indices to the right of the site where the generator acts are changed. This is due to the fact that otherwise we are immediately confronted with the problem that this action will not respect the proper gauge index contraction.
Consider for instance acting with the $\Rcal^2_{\;3}=\sigma^-_{XZ}$ generator on a string of fields:
\be
\sigma^-_{XZ} (\cdots X_{12} \overset{\ell}{X_{21}} X_{12} X_{21}\cdots)\ra \cdots X_{12} \overset{\ell}{Z_{22}}  X_{21} X_{12}\cdots + \dots \ .
  \ee
For concreteness, we have exhibited the action of $\sigma_-$ only on the field at the $\ell$'th site, but of course the full result will be a sum of the actions on all the fields. 
   We notice that all the gauge indices to the right of the action of the generator have flipped from $\SU(N)_1 \leftrightarrow \SU(N)_2$. In the next section, we will define a suitable coproduct which implements this $\Zset_2$ action.

\subsection{Algebroid coproduct}

Recall that the action of Lie algebra generators on products of fields, living in multiple copies of the algebra, is encoded in a coproduct $\Delta$, an operation which tells us how a generator is distributed on two sites. For the unbroken generators, we will of course still have the usual Leibniz rule for the coproduct,
\be
\Delta_{\op}(\Rcal^a_{\;b})=\1\otimes \Rcal^a_{\;b} +\Rcal^a_{\;b}\otimes \1\;, \qquad \mathrm{if~}\R{a}{b}~\mathrm{is~unbroken}\;,
\ee
where the subscript $\op$ indicates that we are working at the orbifold point. 
We also have the usual group-like coproduct for the identity,
\be
\Delta_{\op}(\1)=\1\otimes \1 \ .
\ee
For the broken generators, to enforce the above prescription of the change of indices to the right of where the generators are acting, the coproduct needs to be modified. We define it as
\be \label{opbroken}
\Delta_{\op}(\R{a}{b})=\1\otimes \R{a}{b} +\R{a}{b}\otimes \gamma\;,\qquad \mathrm{if~}\R{a}{b}~\mathrm{is~broken} .
\ee
Here $\gamma$ is a $\Zset_2$ element which exchanges all indices of gauge group 1 with those of gauge group 2,
\be
\gamma(X_{12})=X_{21}, \ \gamma(X_{21})=X_{12}, \ \gamma(Y_{12})=Y_{21}, \ \gamma(Y_{21})=Y_{12},  \ \gamma(Z_{1})=Z_{2}, \ \gamma(Z_{2})=Z_{1} ,
\ee
and similarly for the conjugate fields. We can combine these into a single coproduct as
\begin{equation}
\Delta_{\op}\left(\R{a}{b}\right) \coloneq \1 \otimes \R{a}{b} + \R{a}{b} \otimes \Oab,
\label{eq:CoproductOrbifoldPoint}
\end{equation}
where
\begin{equation}
\Oab=\begin{cases}
\1, & \mathrm{if~}\R{a}{b}~\mathrm{is~unbroken}\\
\gamma, & \mathrm{if~}\R{a}{b}~\mathrm{is~broken}
\end{cases}\, . \label{eq:OmegaCases}
\end{equation}
As a first step towards defining the action of generators on states of arbitrary length, we would like to extend the coproduct to three-sites. As always, there are two ways to do this:
\begin{align}
    \Delta_{\op}^{(3,L)}\left(\R{a}{b}\right) &\coloneq (\Delta_\op \otimes \id)\Delta_{\op}\left(\R{a}{b}\right) = (\Delta_\op \otimes \id) \left(\1 \otimes \R{a}{b} + \R{a}{b} \otimes \Oab \right) \\
    \Delta_{\op}^{(3.R)}\left(\R{a}{b}\right) &\coloneq (\id \otimes \Delta_\op) \Delta_{\op}\left(\R{a}{b}\right) = (\id \otimes \Delta_\op) \left(\1 \otimes \R{a}{b} + \R{a}{b} \otimes \Oab \right) \ ,
\end{align}
We see that for $\Delta_{\op}^{(3,L)}\left(\R{a}{b}\right)$ and $\Delta_{\op}^{(3,R)}\left(\R{a}{b}\right)$ to agree, we need to have
\begin{equation}
    \Delta_\op ( \Oab ) = \Oab \otimes \Oab \ ,
\end{equation}
which  for the broken generators translates to
\begin{equation}
    \Delta_\op ( \gamma ) = \gamma \otimes \gamma \ .
\end{equation}
The fact that this is indeed true
 can be easily understood
via acting with $\gamma \in \Zset_2$ on  spin chains of two sites. $\gamma$ flips the colour indices of both fields. For example,
$\gamma ( X_{12} Z_{22} )= X_{21} Z_{11} $.
Given the above definitions, we have
\be\label{orbifoldpointDelta3}
\begin{split}
\Delta_{\op}^{(3)}\left(\R{a}{b}\right) &= (\Delta_\op \otimes \id)\Delta_{\op}\left(\R{a}{b}\right) = (\id \otimes \Delta_\op) \Delta_{\op}\left(\R{a}{b}\right) \\
&= \1 \otimes \1 \otimes\R{a}{b} + \1 \otimes \R{a}{b} \otimes \Oab + \R{a}{b} \otimes \Oab \otimes \Oab \ .
\end{split}
\ee
So the action of the generators on three sites is coassociative, and  the order in which multiplications are performed is unimportant.  It is straightforward to extend the coproduct to $L$ sites and write down the action of a symmetry generator on a general $L$-site state,
\begin{equation} 
\Delta_{\op}^{(L)}\left(\R{a}{b}\right)=\sum_{\ell=1}^{L} \left(\1\otimes\cdots\otimes \1\otimes{\overset{\ell}{\R{a}{b}}}\otimes \Oab \otimes\cdots\otimes \Oab \right) \ . 
\label{eq:OrbifoldPointDelta}
\end{equation}
Having defined the action of the generators on $L$ sites we can check with an explicit calculation that they also obey the $\mathfrak{su}(4)$ algebra \eqref{eq:SU(4)algebra}, see Appendix \ref{Ap:commutators} for more details.
This relation also obeys the graded structure for the generators in \eqref{eq:broken-unbroken}.

We can finally define the action of any generator of the algebroid on an arbitrary $L$-site state as
\be
\begin{split}
    \Rcal^a_{\;b} \triangleright \ket{\text{state}}_\op&=
\Rcal^a_{\;b}\triangleright \left(c_{i_1i_2\cdots i_L}\varphi^{i_1} \varphi^{i_2}\cdots\varphi^{i_L}\right)\\ &:=c_{i_1i_2\cdots i_L} ~ m(\Delta_\op^{(L)}(\Rcal^a_{\;b}) \triangleright [\varphi^{i_1}\otimes \varphi^{i_2}\otimes \cdots\otimes \varphi^{i_L}])\;,
\end{split}\ee
where $m$ denotes the multiplication in the module, namely the quiver path groupoid (see Appendix \ref{GroupoidAppendix} for the precise definition), and we have collectively denoted the fields by $\varphi^i$. For the unbroken generators, this reduces to the usual product rule for operators, while for broken generators (\ref{eq:OrbifoldPointDelta}) ensures that the direct products one obtains after the action of the coproduct are compatible with $m$. In the following, when we refer to the action of the coproduct of broken generators on states, the definition above will be understood and will not be explicitly indicated.

\subsection{Invariance of the Lagrangian}

Let us now check that, with the coproduct as defined above, the orbifold point Lagrangian (obtained from  (\ref{N2Lagrangian}) by taking $g_1=g_2=g$) is invariant under the action of all the generators of $\SU(4)$. We will look at different terms separately, starting with the K\"ahler part:
\be
\begin{split}
  \mathcal{L}_K & = \tr_1\Big(\bar{X}_{12}e^{gV_{2}}X_{21}e^{-gV_{1}} + \bar{Y}_{12}e^{gV_{2}}Y_{21}e^{-gV_{1}} + \bar{Z}_{1}e^{gV_{1}}Z_{1}e^{-gV_{1}} \Big) \ \\
  &+ \tr_2\Big(\bar{X}_{21}e^{gV_{1}}X_{12}e^{-gV_{2}} + \bar{Y}_{21}e^{gV_{1}}Y_{12}e^{-gV_{2}} + \bar{Z}_{2}e^{gV_{2}}Z_{2}e^{-gV_{2}} \Big) \;. 
\end{split}
\ee
At the one-loop level\footnote{
For the non-expert reader, an explanation of the ``one-loop level'' is in order.
In $\mathcal{N}=4$ SYM which is the paradigmatic example of an integrable gauge theory, the symmetry generators are understood in the expansion $J (\lambda) = J^{(0)} + J^{(1)} \lambda  + J^{(2)} \lambda^2 + \cdots$ with $\lambda$ being the 't Hooft coupling.
$J (\lambda)$ still obey the Lie algebra commutation relations and they commute with the Hamiltonian $\mathbb{H} (\lambda) = \lambda \mathbb{H}^{(1)} + \lambda^2 \mathbb{H}^{(2)} + \cdots$. The nearest-neighbour one-loop Hamiltonian $\mathbb{H}^{(1)}$ commutes with the classical $J^{(0)}$. The higher loop corrections acquire next to nearest, etc. corrections. See \cite{Beisert:2003ys} for a discussion. In this paper we are working at the level of one-loop for the Hamiltonian and classical for the generators.}, the factors $e^{\pm g V_i}$ do not contribute, making this effectively a two-site expression. 

As the invariance under the unbroken $\SU(2)_L\times\SU(2)_R\times\Urm(1)$ generators is obvious (see Appendix \ref{SU4Appendix}), we focus on the broken generators. As reviewed in Appendix \ref{Ap:OpeningUp}, for the action of the broken generators to make sense, we need to first open up the trace of the single trace operator $\mathcal{\Lcal}_K$ in a cyclic way. For the sector with first index being in gauge group 1, we obtain 
\begin{equation}
|\mathcal{L}_{K,1}\rangle_\op = X_{12}\bar{X}_{21}+\bar{X}_{12} X_{21}+Y_{12} \bar{Y}_{21} +\bar{Y}_{12} Y_{21}+Z_1 \bar{Z}_1 +\bar{Z}_1 Z_1 \ ,
\end{equation}
and similarly in the sector with first index in gauge group 2, where we will find its $\Zset_2$ conjugate. A simple computation shows that this open chain state is annihilated by the two-site coproduct (\ref{eq:CoproductOrbifoldPoint}), for all the broken $\Rcal^a_{\;b}$. 

Next we consider the superpotential (\ref{superpotential}), again specialised to the orbifold point. 
Cutting  the traces open cyclically, we obtain 
\begin{equation}
  \frac{1}{g}|\mathcal{W}_1\rangle_\op =   \left(X_{12} Y_{21} - Y_{12} X_{21} \right) Z_1 + Z_1\left(X_{12} Y_{21} - Y_{12} X_{21} \right) + \left( Y_{12} Z_2 X_{21} - X_{12} Z_2 Y_{21} \right), \label{eq:SP1OP}
\end{equation}
as well as the $\Zset_2$ conjugate with $(1\leftrightarrow 2)$. 
These contributions are annihilated by the raising and lowering operators of $\SU(3)_{XYZ}$, acting via the coproduct (\ref{orbifoldpointDelta3}). For concreteness, let us look at a sample calculation for the raising operator of the $XZ$ sector $\R{3}{2}$. We find
\begin{align}  \label{RWactionop}
    \R{3}{2}\triangleright\frac{1}{g}|\mathcal{W}_1\rangle_\op = &\left(X_{12} Y_{21} \!-\! Y_{12} X_{21} \right) X_{12} \!+\! X_{12}\left(X_{21} Y_{12} \!-\! Y_{21} X_{12} \right)
    +\left( Y_{12} X_{21} X_{12} \!-\! X_{12} X_{21} Y_{12} \right) \nonumber \\
    &= \ 0 \ .
\end{align}
Similar computations for the other $\SU(3)_{XYZ}$ generators $\R{2}{3}$, $\R{4}{2}$ and $\R{2}{4}$ also give zero, so we have found that the superpotential is an $\SU(3)$-groupoid invariant expression, generalising its $\SU(3)$ group invariance in the $\Ncal=4$ SYM theory. (Of course, the superpotential is not $\SU(4)$ invariant, as it belongs to the ${\bf 10}$ of $\SU(4)$.) 

Having shown invariance of the K\"ahler and superpotential terms, we are effectively done, as that proves the invariance of the Lagrangian in the superspace formalism. However, since passing to the component formalism changes the number of sites (from three to four), and the opening-up procedure as well as the coproduct (\ref{eq:OrbifoldPointDelta}) depend on the number of sites, it is important to check the invariance of the quartic terms as well. Again, we will need to cut open the traces in (\ref{eq:FullQuarticTerms}), specialised to $g_1=g_2=g$, before acting with the coproduct. 

A slightly tedious, but straightforward calculation confirms that this combination is indeed invariant under the action of all generators of the coproduct at the orbifold point, i.e. 
\begin{equation}
    \R{a}{b}\triangleright |\mathcal{V}(g,g)\rangle = 0 \ \;,\quad \text{for all} \;\;a,b\;,
\end{equation}
where $\Vcal(g,g)$ is the opened-up version of the scalar potential, which we write explicitly in Appendix \ref{Ap:OpeningUp}. So we have confirmed that the gauge theory Lagrangian at the orbifold point is invariant under the full $\SU(4)$ groupoid symmetry. 

As discussed, the additional symmetries due to the revived generators would be expected to lead to an additional understanding of the integrable structure underlying the twisted Bethe ansatz of \cite{Beisert:2005he}, as well as provide an alternative explanation of the planar equivalence of the correlation functions of the $\Ncal=4$ SYM theory and its orbifolds \cite{Bershadsky:1998mb,Bershadsky:1998cb}. We also note that the $\Ncal=2$ supersymmetry was not really essential for our arguments, so one would expect analogous definitions of the broken generators to apply to $\Ncal=1$ orbifolds as well. 

\section{Quantum plane relations}
\label{sec:quantumplanes}

In the previous section, we explained how to recover the broken $\SU(4)$ generators at the orbifold point, by thinking in terms of a groupoid where not all elements can be multiplied with each other. This was enforced by introducing the coproduct (\ref{eq:CoproductOrbifoldPoint}) which, for the naively broken generators, also involves a flip of the gauge indices for all fields to the right of the generator (in our convention). We will now move on to the more challenging case where we marginally deform away from the orbifold point by allowing the gauge couplings to take different values, $g_1\neq g_2$.

\subsection{F- and D- term quantum planes}
  
Our approach to uncovering the quantum algebra is through the quantum plane relations, which we will read off from the F- and D-term relations. Specifically, we will require that the generators of the algebra, both broken and unbroken, preserve the quantum planes coming from the F- and D-terms. As these are quadratic in the fields, the generators will have to act through a coproduct, and for the broken generators we will look for an appropriate deformation of the coproduct \eqref{eq:CoproductOrbifoldPoint}. For the $\Ncal=1$ Leigh-Strassler marginal deformations of the $\Ncal=4$ theory \cite{Leigh:1995ep}, the link between the F-term relations and quantum planes was noticed in \cite{Berenstein:2000ux}, and was further developed in \cite{Mansson:2008xv, Dlamini:2016aaa, Garus:2017bgl, Dlamini:2019zuk}. Our approach to showing the $\SU(4)$ invariance of the marginally deformed orbifold action will be along similar lines to \cite{Garus:2017bgl, Dlamini:2019zuk} in that, by defining appropriate twists, we will seek to untwist the Lagrangian back to the orbifold point. However, there are some major differences in the current $\Ncal=2$ orbifold context:

\begin{itemize}
\item While the Leigh-Strassler deformations are purely superpotential deformations, and therefore only modify the F-term relations, here the marginal deformations are obtained by rescaling the gauge couplings and modify both the F- and D-terms.
\item The twists need to be compatible with the more complicated groupoid structure, where some products of fields are not allowed. In the language of \cite{Pomoni:2021pbj}, we will call such twists dynamical. In particular, this makes it more challenging to define the twists at three- and four-sites, as will be needed in order to untwist the superpotential and quartic terms, respectively. 
\end{itemize}

As in \cite{Pomoni:2021pbj}, in writing the quantum planes it will be convenient to rescale the quartic terms by a factor of $g_1g_2$ (corresponding to a factor of $\sqrt{g_1g_2}$ in the superpotential), and to define
\be
\kappa=\frac{g_2}{g_1} \ .
\ee
So now $\Zset_2$ acts by taking $1\leftrightarrow 2$ and $\kappa\leftrightarrow \kappa^{-1}$.
The F-term and D-term relations (\ref{FDg}) now take the form 
\be \label{FDterms}
\begin{split}
    F_{12}^{Y} &= X_{12} Z_2 - \frac{1}{{\kappa}} Z_1 X_{12} \ ,\quad
    \bar{F}_{12}^{\bar{Y}} =  \bar{X}_{12} \bar{Z}_2 - \frac{1}{{\kappa}} \bar{Z}_1 \bar{X}_{12} \ , \\
    F_{12}^{X} &=  Y_{12} Z_2 - \frac{1}{{\kappa}} Z_1 Y_{12} \ , \quad
    \bar{F}_{12}^{\bar{X}} = \bar{Y}_{12} \bar{Z}_2 - \frac{1}{{\kappa}} \bar{Z}_1 \bar{Y}_{12} \ , \\
    F_1^{Z} &= \frac{1}{\sqrt{\kappa}}( X_{12} Y_{21} - Y_{12} X_{21}) \ , \quad
    \bar{F}_1^{\bar{Z}} = \frac{1}{\sqrt{\kappa}}(\bar{X}_{12} \bar{Y}_{21} - \bar{Y}_{12} \bar{X}_{21} )\ ,\\
 D_1 &= \frac{1}{\sqrt{\kappa}}\left( \bar{X}_{12} X_{21} + \bar{Y}_{12} Y_{21}- X_{12} \bar{X}_{21}  - Y_{12} \bar{Y}_{21} +Z_1\Zb_1-\Zb_1Z_1\right)\; \\ 
\end{split}
\ee
for the first index in gauge group 1, and similarly for their $\Zset_2$ conjugates.

In the following, we will look for twists which relate these quantum planes to those at the orbifold point. Before that, however, we will need to consider the quantum planes related to the ones above via the action of the unbroken symmetries.

\subsection{Extension using the unbroken symmetries} \label{sec:Extension}

The above F-term relations provide us with quantum planes in the holomorphic $XZ$, $YZ$ and $XY$ and antiholomorphic $\Xb\Zb$, $\Yb\Zb$ and $\Xb\Yb$ $\SU(2)$ subsectors of the theory, among which, as discussed, the first two are ``broken'' $\SU(2)$'s while the third enjoys a standard $\SU(2)$ symmetry. To fully understand the effect of the marginal deformation on the $\SU(4)$  groupoid structure, we need to consider the quantum planes in mixed sectors as well, for instance sectors such as $\bar{X}Z$ and $Z\Zb$. To achieve this, we start from the F- term quantum planes and act with the unbroken $\mathrm{SU}(2)_R$-generators.\footnote{Interestingly, the action of $\mathrm{SU}(2)_L$ does not produce additional quantum plane relations, since $\mathcal{W}$ and $\overline{\mathcal{W}}$ are singlets under this unbroken subsector. This will be key to generalisations to more general orbifold theories.
} In this way, the quantum planes will naturally organise themselves in representations of $\SU(2)_R$. The $F^X$ and $F^Y$ relations will give doublets, for instance: 
\begin{align} \label{FYdoublet}
  \vec{F}^Y_{12}=  \begin{pmatrix}
        F_{12}^{Y} \\
        \D{2}{1}F_{12}^{Y}
    \end{pmatrix} &= \begin{pmatrix}
         X_{12} Z_2 - \frac{1}{\kappa} Z_1 X_{12} \\
        \bar{Y}_{12} Z_2 - \frac{1}{\kappa} Z_1 \bar{Y}_{12}
    \end{pmatrix} \ , \\
   \vec{F}^{\bar{Y}}_{12} =  \begin{pmatrix}
        \bar{F}_{12}^{\bar{Y}} \\
        -\D{1}{2}\bar{F}_{12}^{\bar{Y}}
    \end{pmatrix} &= \begin{pmatrix}
         \bar{X}_{12} \bar{Z}_2 - \frac{1}{\kappa} \bar{Z}_1 \bar{X}_{12} \\
        Y_{12} \bar{Z}_2 - \frac{1}{\kappa} \bar{Z}_1 Y_{12}
    \end{pmatrix} \ ,
\end{align}
and $\vec{F}^X_{12}$,  $\vec{F}^{\bar{Y}}_{12}$ as above with $X$'s and $Y$'s exchanged. On the other hand, the $F_i^{Z}$ and $\bar{F}_i^{\bar{Z}}$ relations combine with the $X,Y$-dependent part of the D-term to form an $\mathrm{SU}(2)_R$ triplet:
\begin{align} \label{Gtriplet}
    G_1\coloneq  \begin{pmatrix}
        G^+_1 \\
        G^0_1 \\
        G^-_1
    \end{pmatrix} = \frac{1}{\sqrt{\kappa}}\begin{pmatrix}
        X_{12} Y_{21} - Y_{12} X_{21} \\
        \frac{1}{\sqrt{2}}\left(\bar{X}_{12} X_{21} + \bar{Y}_{12} Y_{21} - X_{12} \bar{X}_{21} - Y_{12} \bar{Y}_{21}\right) \\
        \bar{X}_{12} \bar{Y}_{21} - \bar{Y}_{12} \bar{X}_{21}
    \end{pmatrix} \ ,
\end{align}
as well as its $\Zset_2$ conjugate. The remaining parts of the D-terms,
\be \label{Zsinglet}
E_1\coloneq\frac{1}{\sqrt{2\kappa}}(Z_1\Zb_1-\Zb_1 Z_1)\;\;\text{and}\;\; E_2\coloneq\sqrt{\frac{\kappa}2}(Z_2\Zb_2-\Zb_2 Z_2) 
\ee
are $\SU(2)_R$ singlets and define the quantum planes in the $Z_i,\Zb_i$ sector. 

In the next section, we will define twists which relate the above quantum plane multiplets to those at the orbifold point. Before doing so, it is useful to see how the scalar potential of the theory can be expressed in terms of the $\SU(2)_R$ multiplets. We define the following inner products of the doublet states:
 \be
 \begin{split}
   \vec{F}^Y_{12} \cdot \vec{F}^{\bar{Y}}_{21} = &\left(X_{12} Z_{21} - \frac{1}{\kappa} Z_{1} X_{12} \right)\left(\bar{X}_{21}\bar{Z}_1 - \kappa \bar{Z}_2\bar{X}_{21}\right) \\&+ \left( \bar{Y}_{12} Z_2 - \frac{1}{\kappa} Z_1 \bar{Y}_{12} \right) \left(Y_{21} \bar{Z}_1 - \kappa \bar{Z}_{2} Y_{21} \right) 
 \end{split}
 \ee
and similarly
 \be
 \begin{split}
\vec{F}^X_{12} \cdot \vec{F}^{\bar{X}}_{21} =   &
  \left(Y_{12} Z_2 - \frac{1}{\kappa} Z_1 Y_{12} \right) \left(\bar{Y}_{21} \bar{Z}_1 - \kappa \bar{Z}_2 \bar{Y}_{21} \right) \\&+ \left( \bar{X}_{12} Z_2 - \frac{1}{\kappa} Z_1 \bar{X}_{12}\right)\left( X_{21} \bar{Z}_1 - \kappa \bar{Z}_2 X_{21} \right)\;,
 \end{split}
 \ee
with analogous relations for the $\Zset_2$-conjugate terms. It is easy to check that these inner products are $\SU(2)_R$ singlets. 
For the triplet, we define the $\SU(2)_R$-singlet combination $(G_1)^2 \coloneq G^+_1\cdot G^-_1+G^-_1 \cdot G^+_1 + G_1^0 \cdot \bar{G}_1^0$. Defining also $|E_1|^2=E_1\cdot \bar{E}_1$, which is of course also a singlet, we can write 
\be
\begin{split}
(&G_1)^2 + |E_1|^2=\\ &=\frac{1}{\kappa}\big( \left( X_{12} Y_{21} - Y_{12} X_{21} \right) \left(\bar{X}_{12} \bar{Y}_{21} - \bar{Y}_{12} \bar{X}_{21} \right) + \left(\bar{X}_{12} \bar{Y}_{21} - \bar{Y}_{12} \bar{X}_{21} \right)\left( X_{12} Y_{21} - Y_{12} X_{21} \right) \\
   &\quad\;+ \frac{1}{2} \left[ \left(Z_1 \bar{Z}_1 - \bar{Z}_1 Z_1 \right)  \left(\bar{Z}_1 Z_1 - Z_1 \bar{Z}_1 \right) - \left(\bar{X}_{12} X_{21} + \bar{Y}_{12} Y_{21} - X_{12}\bar{X}_{21} - Y_{12} \bar{Y}_{21}\right)^2 \right]\big) \ .
\label{eq:TripletSinglet}
\end{split}
\ee
Combining the above expressions with their $\Zset_2$ conjugates, we can finally rewrite the scalar potential (\ref{eq:FullQuarticTerms}) in a way which makes the $\SU(2)_R$ structure clearer:

 \be \label{SU2RV}
\begin{split}
   \mathcal{V}(\kappa)&= \tr_1 \left( (G_1)^2 + |E_1|^2 \right) + \hspace{1pt} \tr_2 \left( (G_2)^2 + |E_2|^2 \right) \\
   &\;+ \frac{1}{2} \tr_1 \left( \vec{F}_{12}^{X} \cdot \vec{F}_{21}^{\bar{X}} + \vec{F}_{12}^{\bar{X}} \cdot \vec{F}_{21}^{X}+ \vec{F}_{12}^{Y} \cdot \vec{F}_{21}^{\bar{Y}} + \vec{F}_{12}^{\bar{Y}} \cdot \vec{F}_{21}^{Y} \right) \\
   &\;+ \frac{1}{2} \tr_2 \left( \vec{F}_{21}^{X} \cdot \vec{F}_{12}^{\bar{X}} + \vec{F}_{21}^{\bar{X}} \cdot \vec{F}_{12}^{X}+ \vec{F}_{21}^{Y} \cdot \vec{F}_{12}^{\bar{Y}} + \vec{F}_{21}^{\bar{Y}} \cdot \vec{F}_{12}^{Y} \right) \;.
\end{split}
\ee
In this expression, all terms in the scalar potential are $\SU(2)_R$ singlets. Although this form of $\Vcal(\kappa)$ is equivalent to (\ref{eq:FullQuarticTerms}), it is better aligned to our $\Ncal=2$ theory with its unbroken $\SU(2)_R$ symmetry. In Section \ref{scalarinvariance}, we will show that this form of the scalar potential can be untwisted back to the orbifold-point expression, which will establish its invariance under the broken $\SU(4)$ generators.

\section{Two-site Twists} \label{TwistsSection}

In this section we will find two-site twists that allow us to deform the groupoid structure we introduced in Section \ref{OPSymmetries}, which is applicable only to the orbifold point, such that we can then discuss invariance of the Lagrangian of the marginally deformed theory.
We wish to emphasise that the twists $\mathcal{F}$ that we will write down here are well-educated guesses and are by no means unique. At the moment we have a set of requirements the twists should satisfy: Firstly, that they  give the correct quantum plane relations, i.e. the F- and D-term relations extended by their $\SU(2)_R$ descendants. Secondly, we will require agreement with the BPS spectrum of the theory (up to a global inversion of $\kappa$). Thirdly, the twists we will write down also have the property that their inverses are their $\Zset_2$ conjugates, $\Fcal^{-1}(\kappa)=\Fcal(\kappa^{-1})$.  Our approach is restricted by the fact that we are only able to construct twists acting on two copies of the fundamental representation, i.e. in this work we do not obtain a universal (representation-independent) form of the twists. 

Our rewriting of the quartic terms in terms of the unbroken $\SU(2)_R$ multiplets (\ref{SU2RV}) is a crucial step in finding the correct twists from the orbifold point to the marginally deformed theory. We will need to choose the various twists in a way that preserves this structure, as otherwise we would be breaking the $\SU(2)_R$ symmetry. In the following we will start with the holomorphic $XZ$ sector, proceed to the holomorphic $XYZ$ sector, and finally consider twists in the full $\SU(4)$.

\subsection{The $XZ$-sector twist}

Let us start by considering the holomorphic $XZ$ sector. A twist for this sector was proposed in \cite{Pomoni:2021pbj}, which had some positive features, in particular that it was triangular, and led to an $R$-matrix capturing the quantum plane relations. However, for our current purposes we do not have a good reason to impose these restrictions. Furthermore, that twist left terms of type $XX$ and $ZZ$ untwisted, a condition which we will now relax.  

We will instead opt for a simpler type of $XZ$-sector twist, one that does not have any direct dependence on the $\SU(2)$ structure but only refers to the $\Zset_2$ structure, i.e. whether the indices at each site belong to the first or to the second gauge group. It is
\be \label{diagonaltwist}
\Fcal=\kappa^{-\frac{s}{2}}\otimes \kappa^{-\frac{s}{2}}\;,
\ee
where we have introduced the $\Zset_2$ element $s$, whose definition on site $\ell$ is
\be \label{sdef}
s(\ell)=\left\{\begin{array}{cc} $1$ & \text{if the first gauge index on site $\ell$ is $1$}\\
$-1$ & \text{if the first gauge index on site $\ell$ is $2$}\end{array}\right.\;.
\ee
Since the $\Zset_2$ generator $\gamma$ flips both gauge indices at a given site, $s$ does not commute with $\gamma$ but rather we have
\be \label{sgamma}
s \gamma=-\gamma s \ .
\ee
It is easy to check that this twist correctly leads to the F-term $XZ$ quantum plane:
\be
\Fcal\triangleright (X_{12}Z_2-Z_1X_{12})=X_{12}Z_2-\frac{1}{\kappa}Z_1X_{12} \ ,
\ee
while also giving $\Fcal\triangleright (X_{12}X_{21})=X_{12}X_{21}$ and $\Fcal\triangleright (Z_{1}Z_{1})=\kappa^{-1}~Z_{1}Z_{1}$.

Turning to the coproduct of the broken generators, writing $\sigma^{+}=\Rcal^{3}_{\;2}$ and $\sigma^{-}=\Rcal^{2}_{\;3}$ we have
\be \label{XZcoproduct}
\begin{split}
  \Delta_\kappa(\sigma^\pm)&=\Fcal_{12} \Delta_{\op} (\sigma^\pm) \Fcal^{-1}_{12}
  =\kappa^{-\frac{s}{2}}\otimes \kappa^{-\frac{s}{2}} (\1\otimes \sigma^\pm+\sigma^\pm\otimes \gamma) \kappa^{\frac{s}{2}}\otimes \kappa^{\frac{s}{2}}\\
&=\1\otimes \sigma^\pm+\sigma^\pm \otimes \kappa^{-\frac{s}{2}} \gamma \kappa^{\frac{s}{2}}
=\1\otimes \sigma^\pm+\sigma^\pm \otimes \gamma \kappa^{s}\\&=\1\otimes \sigma^\pm+\sigma^\pm\otimes  K,
\end{split}
\ee
where we defined $K=\gamma \kappa^s$, and we used that the action of $\sigma^\pm$ does not change the first index of the site on which it acts. So the twist has introduced a factor of $\kappa$ compared to the orbifold-point coproduct (\ref{opbroken}). Of course, since the unbroken $\sigma^3$ generator has a trivial undeformed coproduct (see Section \ref{OPSymmetries}), the twist will have no effect, and it will retain this coproduct in the marginally deformed theory:
\be
\Delta_\kappa(\sigma^3)=\1\otimes \sigma^3+\sigma^3\otimes \1\;.
\ee
We also need to reproduce the coproduct for $K$, which for consistency needs to be
\begin{equation}
\Delta_\kappa(K)=K\otimes K \ .
\end{equation}
We can verify that the twist acts as
\begin{align}
\Delta_\kappa(K)=\Fcal\hspace{1pt}\Delta_{\op}(K) \Fcal^{-1}&=\left(\kappa^{-\frac{s}{2}}\otimes \kappa^{-\frac{s}2}\right) \left(\gamma\otimes \gamma\right)  \left(\kappa^\frac{s}{2}\otimes \kappa^{\frac{s}2}\right) \nonumber \\ 
&= (\gamma\kappa^s\otimes \gamma\kappa^s) =K\otimes K
\end{align}
as required. So we have a consistent coproduct acting on states in the $XZ$ sector.

As can be seen from (\ref{FDterms}), the $YZ$, $\Xb\Zb$ and $\Yb\Zb$ sectors are completely equivalent to the $XZ$ sector and have the same twist.

\subsection{The $XY$-sector twist}

The $XY$ sector exhibits unbroken $\SU(2)$ symmetry, suggesting that the introduction of a nontrivial twist is unnecessary. This viewpoint was adopted previously in \cite{Pomoni:2021pbj}. However, the approach we follow in this work is to obtain all the $\kappa$-dependent coefficients in the quantum planes, as they appear in (\ref{FDterms}), via twists. Accordingly, we do not wish to rescale away the overall $1/\sqrt{\kappa}$ factor in the $XY$ quantum plane, but rather the twist should lead to these factors directly without further rescalings.  In addition, we will require that twisting the symmetric state $XY+YX$ does not result in any overall $\kappa$-dependent factors, which is motivated by the fact that the BPS states in this sector (at any length) do not acquire any factors of $\kappa$. This can be achieved by the following twist:
\be \label{XYtwist}
\Fcal^{XY}=\left(\begin{array}{cccc} 1 &0&0&0\\0& \frac{1}{2} \left(\frac{1}{\sqrt{\kappa }}+1\right) & \frac{1}{2} \left(1-\frac{1}{\sqrt{\kappa }}\right) & 0 \\
  0& \frac{1}{2} \left(1-\frac{1}{\sqrt{\kappa }}\right) & \frac{1}{2} \left(\frac{1}{\sqrt{\kappa }}+1\right) &0 \\
  0 &0&0&1
\end{array}\right) \;\; \text{in the basis}\;\; \left(\begin{array}{c} X_{12}X_{21} \\X_{12}Y_{21}\\ Y_{12}X_{21}\\ Y_{12}Y_{21}\end{array}\right)\;,
\ee
for the first gauge index in gauge group 1, and its $\Zset_2$ conjugate version defined accordingly. We note that the inverse of this twist is its $\Zset_2$ conjugate, $\Fcal^{-1}(\kappa)=\Fcal(\kappa^{-1})$. We can confirm that
\be\begin{split}
\Fcal^{XY}\triangleright& ~(X_{12}Y_{21}-Y_{12}X_{21})=\frac{1}{\sqrt{\kappa}}(X_{12}Y_{21}-Y_{12}X_{21}) \;\;, \\
\Fcal^{XY}\triangleright& ~(X_{12}Y_{21}+Y_{12}X_{21})=X_{12}Y_{21}+Y_{12}X_{21}\;,
\end{split}
\ee
as required. Furthermore, the states $X_{12}X_{21}$ and $Y_{12}Y_{21}$ are invariant under the twist. 

The $\Xb\Yb$ sector is equivalent to the $XY$ sector, so we will define the same twist in that sector as well.

\subsection{The D-term twists}

As we have organised our quantum planes according to their $\SU(2)_R$ quantum numbers, the D-term has been split into a part belonging to the triplet (\ref{Gtriplet}) as well as a part which is a $\SU(2)_R$  singlet (\ref{Zsinglet}). For the triplet D-term quantum plane, any twist must be such that the 
symmetrised state  $X\Xb+\Xb X+Y\Yb+\Yb Y$ does not acquire any overall factors of $\kappa$, since it corresponds to (the opened version of) the kinetic terms of the theory, which are unaffected by the marginal deformation. At the same time, the twist should reproduce the quantum planes in (\ref{Gtriplet}). A twist which meets these requirements is
\be
\Fcal^{G_0}=\frac{1}{4}\left(\begin{array}{cccc}
 3+\frac{1}{\sqrt{\kappa }} \;&\;  1-\frac{1}{\sqrt{\kappa }} \;&\;  \frac{1}{\sqrt{\kappa }}-1 \;&\; 1-\frac{1}{\sqrt{\kappa }}   \\
   1-\frac{1}{\sqrt{\kappa }}\; &\;  3+\frac{1}{\sqrt{\kappa }}\; &\; 1-\frac{1}{\sqrt{\kappa }}\; &\; \frac{1}{\sqrt{\kappa }}-1  \\
  \frac{1}{\sqrt{\kappa }}-1 \;&\; 1-\frac{1}{\sqrt{\kappa }} \;&\;  3+\frac{1}{\sqrt{\kappa }} \;&\;  1-\frac{1}{\sqrt{\kappa }}  \\
   1-\frac{1}{\sqrt{\kappa }} \;& \; \frac{1}{\sqrt{\kappa }}-1 \;&\;  1-\frac{1}{\sqrt{\kappa }} \;&\; 3+ \frac{1}{\sqrt{\kappa }} 
\end{array}
\right) \;\; \text{in the basis}\;\; \left(\begin{array}{c} \Xb_{12}X_{21} \\X_{12}\Xb_{21}\\ \Yb_{12}Y_{21}\\ Y_{12}\Yb_{21}\end{array}\right)\;,
\label{eq:F2Triplet}
\end{equation}
as well as its $\Zset_2$ conjugate. As before, this twist satisfies the relation $\Fcal^{-1}(\kappa)=\Fcal(\kappa^{-1})$

For the $\SU(2)_R$ singlet state (\ref{Zsinglet}), we again require that the symmetrised version $Z_1\Zb_1+\Zb_1 Z_1$ does not acquire any $\kappa$-dependent prefactor, as it corresponds to the opened-up kinetic terms. So we choose a similar twist to the $XY$ sector:
\be \label{F2singlet}
\Fcal^E=\left(\begin{array}{cccc} \kappa^{-1} &0&0&0\\0& \frac{1}{2} \left(\frac{1}{\sqrt{\kappa }}+1\right) & \frac{1}{2} \left(1-\frac{1}{\sqrt{\kappa }}\right) & 0 \\
  0& \frac{1}{2} \left(1-\frac{1}{\sqrt{\kappa }}\right) & \frac{1}{2} \left(\frac{1}{\sqrt{\kappa }}+1\right) &0 \\
  0 &0&0&\kappa^{-1}
\end{array}\right) \;\; \text{in the basis}\;\; \left(\begin{array}{c} Z_1Z_1 \\Z_1\Zb_1\\ \Zb_1Z_1\\ \Zb_1\Zb_1\end{array}\right)\;,
\ee
where the twists of $Z_1Z_1$ and $\Zb_1\Zb_1$ follow from the $XZ$ twist (\ref{diagonaltwist}).  Again, we note that $\Fcal^{-1}(\kappa)=\Fcal(\kappa^{-1})$.

\subsection{$\SU(2)_R$ descendant twists}

In the above, we specified the action of the twists for all the quantum planes appearing in (\ref{FDterms}). In order to be able to twist any two-site state, we still need to define twists on the other quantum planes, which do not directly come from the F- and D-terms. For instance, from the $F^Y_{12}$ doublet (\ref{FYdoublet}) we can define the twist in the $\Yb Z$ sector, and similarly for the $Y\Zb, \Xb Z$ and $X \Zb$ sectors. Since we expect the twists to commute with the action of $\SU(2)_R$, the components of the same doublet will have the same twists, given by (\ref{diagonaltwist}).

This leaves the $\mathrm{SU}(2)_R$-singlet monomials $\{X\bar{Y},\bar{X} Y,Y \bar{X}, \bar{Y}X\}$ which are not directly related to any of the quantum planes. We will take the two-site twists to act in the same way as for the $\{XY,\bar{X}\bar{Y}\}$ sectors, i.e. with off-diagonal actions that have overall normalisation 1 for the symmetrised states (corresponding to descendants of $XX$), and $\frac{1}{\sqrt{\kappa}}$ for the anti-symmetrised states.

The reader might wonder why we need to introduce these twists if our goal is to show the invariance of the gauge theory Lagrangian, which is defined by the F- and D-terms, so these additional quantum planes do not appear. As will become clear in the next section, our approach to showing the invariance of the scalar potential will involve rebracketing terms before acting with the inverse twists, such that for instance a closed D-term like $\tr[(\Yb Y)( \Xb X)]$ can, after opening up, leave us with terms including $(Y\Xb)(X\Yb)$, and to proceed we will need to have a definition of the two-site twists on the two factors in the parentheses.

\section{Twisted $\SU(3)$ groupoid invariance of the Superpotential} \label{Superpotentialinvariance}

In the previous section, guided by the F-term and D-term relations, we defined two-site twists for all the different quantum planes within $\SU(4)$. Acting with these twists on the trivial quantum planes at the orbifold point gives us $\kappa$-deformed quantum planes, and in particular takes us from the F- and D-terms at the orbifold point to those in the marginally deformed theory.

The next step is to use these twists to show that the Lagrangian of the gauge theory, in the planar limit, is invariant under the deformed $\mathfrak{su}(4)$ algebroid defined by these twists. Our approach will be to find an appropriate generalisation of the two-site twists to $L$ sites, in order to define twisted coproducts for the broken $\SU(4)$ generators:
\be \label{DeltakL}
\Delta^{(L)}_\kappa(\Rcal^a_{\;b})=\Fcal^{(L)} ~\Delta^{(L)}_{\op}(\Rcal^a_{\;b})~ (\Fcal^{(L)})^{-1}\;\;,\;\quad \text{if} \; \Rcal^a_{\;b} \; \text{is broken}.
\ee
Using these coproducts, we will define an action of these generators on the various terms appearing in the marginally deformed Lagrangian. However, we have already established the invariance of the orbifold-point Lagrangian under the untwisted coproduct (\ref{eq:OrbifoldPointDelta}). So all that needs to be checked to show invariance is that the inverse twist in (\ref{DeltakL}) correctly untwists the terms in the deformed Lagrangian to the corresponding terms at the orbifold point.

In the superspace formalism, the relevant terms in the Lagrangian are of length two (the kinetic terms for the scalar superfields) and of length three (the superpotential), since the vector multiplets are neutral under the $\SU(4)$ generators. In this section, we will focus on the twisted superpotential, which is of course only expected to be invariant under an $\SU(3)$ subgroup of the $\SU(4)$ groupoid.  
  
The (opened-up) superpotential is a state composed of the holomorphic $X,Y$ and $Z$ fields. Therefore, the relevant two-site twists from Section \ref{TwistsSection} are
\be
\Fcal^{XZ}=\Fcal^{YZ}= \kappa^{-s/2}\otimes \kappa^{-s/2}\;\;\text{and}\;\; \Fcal^{XY}\triangleright (X_{12}Y_{21}-Y_{12}X_{21})=\frac{1}{\sqrt{\kappa}}(X_{12}Y_{21}-Y_{12}X_{21})\;,
\ee
where we only write the action of the $XY$ twist on the antisymmetric combination, which is what appears in the superpotential. Given a two-site twist, the two standard ways to define its action on three sites are
\be \label{FLR}
\Fcal^{(3,L)}=(\Fcal\otimes\1) (\Delta\otimes \id)(\Fcal) \;\;\;\text{or} \;\;\; \Fcal^{(3,R)}=(\1\otimes \Fcal) (\id\otimes \Delta)(\Fcal)\;,
\ee
corresponding to the three-site coproducts $\Delta^{(3,L)}(\Rcal^a_{\;b})=(\Delta\otimes\id)\Delta(\Rcal^a_{\;b})$ and $\Delta^{(3,R)}(\Rcal^a_{\;b})=(\id\otimes \Delta)\Delta(\Rcal^a_{\;b})$, respectively. The $L$ and $R$ subscripts stand for ``left'' and ``right'', as their structures are compatible with the action on the respective module products $(v_1v_2)v_2=m((v_1\otimes v_2)\otimes v_3)$ and $v_1(v_2v_3)=m(v_1\otimes (v_2\otimes v_3))$.

Let us start with a coassociative (Hopf algebra) setting, where $\Delta^{(3,L)}$ and $\Delta^{(3,R)}$ agree. If one wishes to twist these coproducts while preserving coassociativity one needs to impose that $\Fcal^{(3,L)}=\Fcal^{(3,R)}$, known as the cocycle condition \cite{Majid}. Otherwise, one obtains a quasi-Hopf algebra \cite{Drinfeld90}, see Appendix \ref{QuantumPlanesAppendix} for more details. A special quasi-Hopf case arises when the twist depends on an additional, ``dynamical'' parameter, which is shifted by Cartan elements evaluated on the different copies of the vector space that the twist is acting on, and leads to a shifted cocycle condition \cite{Felder:1994be,Babelon:1995rz,fronsdal1996quasi,Jimbo:1999zz}. The associativity structure is captured by a dynamical Yang-Baxter equation, which in this context was investigated in \cite{Pomoni:2021pbj}. In that work, the dynamical parameter dependence of the $R$-matrices was chosen such that the shifts implemented the $\Zset_2$ transformation $\kappa\leftrightarrow\kappa^{-1}$.  

It would certainly be very appealing if our twists satisfied a shifted cocycle condition. However, as we do not yet have a universal (representation-independent) form of our twists, we cannot rigorously act on them with the coproduct and evaluate the expressions in (\ref{FLR}) or their shifted versions. In order to make progress, following e.g. \cite{Reshetikhin90}, we will make the assumption that, at least in the holomorphic sector that we are considering, the twists satisfy a quasitriangular-type condition, 
\be
(\Delta\otimes \id)(\Fcal)=\Fcal_{13}\Fcal_{23} \;\;,\quad\text{and}\;\;(\id \otimes \Delta)(\Fcal)=\Fcal_{13}\Fcal_{12} \;,
\ee
with appropriate shifts in line with the dynamical nature of the problem.\footnote{In the non-dynamical case of $\Ncal=1$ integrable deformations of the $\Ncal=4$ SYM theory, such an approach was taken in \cite{Dlamini:2016aaa}.}  We will then write 
\be \label{dynamical3sites}
\Fcal^{(3,L)}=\Fcal_{12}(\lambda) \Fcal_{13}(\lambda^{(2)}) \Fcal_{23}(\lambda) \;\; \text{and}\;\; \Fcal^{(3,R)}=\Fcal_{23}(\lambda^{(1)}) \Fcal_{13}(\lambda) \Fcal_{12}(\lambda^{(3)})\;,
\ee
where $\lambda$ is the dynamical parameter and the notation $\lambda^{(i)}$ indicates the value of $\lambda$ after crossing line $i$. Here $\lambda$ is thought of as taking two possible values, depending on the gauge index of the first site, which flip if one crosses an $X$ or $Y$ field and stay the same when the line being crossed is a $Z$ field. 
The dependence of the twists on $\lambda$ is assumed to be such that if $\lambda$ corresponds to the first gauge group being 1 then $\Fcal(\lambda)=\Fcal(\kappa)$ while if $\lambda$ corresponds to the first gauge group being 2 then $\Fcal(\lambda)=\Fcal(1/\kappa)$. Using a similar graphical notation to \cite{Pomoni:2021pbj}, we can represent the dynamical twist as
\be
\begin{tikzpicture}[scale=0.6,baseline={(-0.2cm,1.4cm)}]

  \node at (-0.5,3) {$\Fcal^{i\;j}_{\;k\;l}(\lambda) \;=$};
  \draw[->,darkgreen,thick] (2,2)--(4,4);
  \draw[->,darkgreen,thick] (4,2)--(2,4);
  \node at (2,1) {$k$};\node at (4,1) {$l$};
  \node at (2,4.5) {$i$};\node at (4,4.5){$j$};
  \node at (1.5,3) {$\lambda$};

  \filldraw[blue,thick] (3,3) circle (3pt); 

  \node at (-3,3) {(a)};\node at (7,3){(b)};
  
  \draw[->,darkgreen,thick] (1+8,3)--(5+8,3);
\node at (0.5+8,3) {$i$};
  \node at (3+8,3.5) {$\lambda$};
  \node at (3+8,2.3) {$\lambda^{(i)}$};

\end{tikzpicture}
\ee
where (a) illustrates the convention that $\lambda$ tracks the gauge group to the left of each twist in the direction provided by the arrows, and (b) the shift in $\lambda$ as one crosses an index line labelled by $i$. Using this notation, we can represent the two three-site twists (\ref{dynamical3sites}) as

\be
\Fcal^{(3,L),T}=\begin{tikzpicture}[scale=0.7,baseline=1cm]
\draw[->,thick,darkgreen] (0,0)--(4,3);\draw[->,thick,darkgreen] (1.8,0) --(0.8,3);\draw[->,thick,darkgreen] (4,0)--(0,3);
\node at (0.3,1.5) {$\lambda$};\node at (2.2,2.4) {\small$\lambda^{(2)}$};
\draw[->,thick] (2,2.1)-- (1.5,1.5);
\node at (0,-0.5) {$1$};\node at (1.8,-0.5) {$2$};\node at (4,-0.5) {$3$};
\filldraw[blue,thick] (1.42,1.1) circle (3pt);
\filldraw[blue,thick] (2,1.5) circle (3pt);
\filldraw[blue,thick] (1.06,2.2) circle (3pt);
\node at (5.5,1.5){$=$};
\node at (8.8,1.6) {$\Fcal_{23}(\lambda)\Fcal_{13}(\lambda^{(2)})\Fcal_{12}(\lambda)$};
\end{tikzpicture}
\ee
and
\be
\Fcal^{(3,R),T}=\begin{tikzpicture}[scale=0.7,baseline=1cm]
\draw[->,thick,darkgreen] (0,0)--(4,3);\draw[->,thick,darkgreen] (2.2,0) --(3.2,3);\draw[->,thick,darkgreen] (4,0)--(0,3);
\node at (0.3,1.5) {$\lambda$};\node at (1.6,0.6) {$\lambda^{(1)}$};\node at (2,2.4) {$\lambda^{(3)}$};
\node at (0,-0.5) {$1$};\node at (2.2,-0.5) {$2$};\node at (4,-0.5) {$3$};
\filldraw[blue,thick] (2.55,1.09) circle (3pt);
\filldraw[blue,thick] (1.95,1.5) circle (3pt);
\filldraw[blue,thick] (2.9,2.2) circle (3pt);
\node at (5.5,1.5){$=$};
\node at (9,1.6) {$\Fcal_{12}(\lambda^{(1)})\Fcal_{13}(\lambda)\Fcal_{23}(\lambda^{(3)})$};
\end{tikzpicture}
\ee
where we note the transposition relative to (\ref{dynamical3sites}), as in this graphical notation the twists are thought of as acting on monomials by matrix multiplication (see Appendix \ref{QuantumPlanesAppendix}). We do not indicate transposition on the individual two-site twists, as they are symmetric. We will also perform an additional transposition in the final state, such that $\Fcal^{(3,L)}$ takes $(\mathscr{V}_1\otimes \mathscr{V}_2)\otimes \mathscr{V}_3\ra (\mathscr{V}_1\otimes \mathscr{V}_2)\otimes \mathscr{V}_3$, and similarly for $\Fcal^{(3,R)}$, which we will not indicate in this graphical notation.

We emphasise that we do not require equality of $\Fcal^{(3,L)}$ and $\Fcal^{(3,R)}$. So, despite the resemblance to \cite{Felder:1994be}, we are in a quasi-Hopf setting, where to twist a three-site state with a given bracketing we need to choose the three-site twist whose structure is compatible with that bracketing. We can also define a coassociator which takes us between the two bracketings (see Appendix \ref{QuantumPlanesAppendix}) but, as we will see, it will not be necessary. 

To apply the dynamical definition of the three-site twist to the orbifold-point superpotential, we need to open it up and look at one of the sectors, which we will take to be that with first gauge group 1. When doing this, there are various choices of bracketing, which are of course all equivalent at the orbifold point. However, we still need to indicate the bracketing, as the different choices lead to different twistings and thus different marginally deformed superpotentials. The situation is similar to passing from classical expressions such as $xp,px,\half(xp+px)$, which are all equal, to the quantum case, where different orderings  will give inequivalent expressions. Here, of course, the issue is not the ordering, but the choice of bracketing. A preferred bracketing is suggested by observing that the $XY$ twist is only diagonal if, when opening up, we preserve the placement of parentheses around the $(XY-YX)$ factor. This also makes manifest the fact that the superpotential, which is in the $\mathbf 10$ of $\SU(4)$, is an $\SU(2)_L$ singlet and belongs to an $\SU(2)_R$ triplet. To preserve this bracketing, writing the orbifold-point superpotential as
\be
\Wcal=\tr_1\left(Z_1(X_{12}Y_{21}-Y_{12}X_{21})\right)+\tr_2\left(Z_2(X_{21}Y_{12}-Y_{21}X_{12})\right) \;,
\ee
we write the corresponding opened states as  
\be \label{Wopen1}
\ket{\Wcal_1}_\op=\left(Z_1(X_{12}Y_{21}-Y_{12}X_{21})\right)+ \left((X_{12}Y_{21}-Y_{12}X_{21})Z_1\right)+\left( Y_{12}{\tallvdots} Z_2 \tallvdots X_{21}-X_{12}\tallvdots Z_2 \tallvdots Y_{21}\right)
\ee
and
\be  \label{Wopen12}
\ket{\Wcal_2}_\op=\left(Z_2(X_{21}Y_{12}-Y_{21}X_{12})\right)+\left((X_{21}Y_{12}-Y_{21}X_{12})Z_2\right)+\left( Y_{21}\tallvdots Z_1 \tallvdots X_{12}-X_{21}\tallvdots Z_1 \tallvdots Y_{12}\right)\;.
\ee
As indicated by the notation, the first two terms will be twisted by $\Fcal^{(3,R)}$ and $\Fcal^{(3,L)}$, respectively. Here we have introduced the new notation $Y_{12}\tallvdots Z_2\tallvdots X_{21}$ which indicates that, as far as the action of the twists is concerned, the first twist to act should be on the products $X_{21}Y_{12}$, i.e. a $\Fcal_{31}$ twist, so that the $XY$ twist acts antisymmetrically on the last terms in (\ref{Wopen1}) and (\ref{Wopen12}). We call the three-site twist, which wraps the state in this way $\Fcal^{(3,W)}$. To relate it to the standard three-site twists, we introduce a cyclic shift operator on the spin chain, which we call $U$.\footnote{For similar reasons, such an operator was also introduced in \cite{Beisert:2018zxs} in the context of showing Yangian invariance of the $\Ncal=4$ SYM action. However, here we enhance it to preserve bracketings as it shifts the sites.} It acts as
\be
U \triangleright \Vscr_1\otimes \Vscr_2\otimes\cdots \otimes \Vscr_L\ra \Vscr_2\otimes \Vscr_3\otimes\cdots \otimes\Vscr_{1}\;,
\ee
and we will require that it preserves the associative structure:
\be
\begin{split}
 Y_{12}\tallvdots Z_2 \tallvdots X_{21}=U^{-1}\triangleright Z_2(X_{21}Y_{12})=U\triangleright (X_{21}Y_{12})Z_2 \ .
\end{split}
\ee
We can now define the $\Fcal^{(3,W)}$ twist on the wrapped bracketings as
\be\begin{split}
\Fcal^{(3,W)}&\triangleright (Y_{12}\tallvdots Z_2\tallvdots X_{21})= \Fcal^{(3,W)}\triangleright U^{-1}\triangleright ( Z_2(X_{21}Y_{12}))\\ 
&\quad=U^{-1}\triangleright (U\Fcal^{(3,W)}U^{-1})\triangleright  Z_2(X_{21}Y_{12})=U^{-1}\triangleright \Fcal^{(3,R)}\triangleright ( Z_2(X_{21}Y_{12}))\;.
\end{split}
\ee
Of course we could also define, similarly,
\be
\Fcal^{(3,W)}\triangleright (Y_{12}\tallvdots Z_2\tallvdots X_{21})=U\triangleright \Fcal^{(3,L)}\triangleright ((X_{21}Y_{12})Z_2)\;,
\ee
which turns out to be equal. Note that, with either definition, the twists on the wrapped bracketing in the sector with first gauge group 1 act on states with first gauge group 2. 

 After these preliminaries, let us consider how each term in the marginally deformed superpotential is related to the corresponding term at the orbifold point:

\be
\begin{tikzpicture}[scale=0.7,baseline=1cm]
  \draw[->,thick,darkgreen] (0,0)--(4,3);\draw[->,thick,darkgreen] (2.2,0) --(3.2,3);\draw[->,thick,darkgreen] (4,0)--(0,3);
  \node at (0,-1) {$Z_{1}$};\node at (2.2,-1) {$[X_{12}$};\node at (3.1,-1.2) {$,$};\node at (4,-1) {$Y_{21}]$};
  \node at (2.1,0.9) {${}^{Y_{12}}$};\node at (3.2,1.5) {${}^{X_{21}}$};\node at (2.2,2)  {${}^{Z_{2}}$};
\filldraw[blue,thick] (2.55,1.09) circle (3pt);
\filldraw[blue,thick] (1.95,1.5) circle (3pt);
\filldraw[blue,thick] (2.9,2.2) circle (3pt);
\end{tikzpicture}\;\;\begin{array}{l}\\ =\Fcal(\kappa^{-1})^{ZX}_{ZX}\Fcal(\kappa)^{ZY}_{ZY}\Fcal(\kappa)^{XY}_{XY}\;-(X\leftrightarrow Y)\\=(\kappa\cdot\frac{1}{\kappa}\cdot\frac{1}{\sqrt{\kappa}}) \cdot Z_{1}[X_{12},Y_{21}]\\=\frac{1}{\sqrt{\kappa}} \cdot Z_{1}[X_{12},Y_{21}]\end{array}
\ee
where the $[X_{12},Y_{21}]$ notation indicates that the twist acts on the antisymmetric combination, $[X_{12},Y_{21}]=(X_{12}Y_{21}-Y_{12}X_{21})$. The second term in (\ref{Wopen1}) twists as
\be
\begin{tikzpicture}[scale=0.7,baseline=1cm]
\draw[->,thick,darkgreen] (0,0)--(4,3);\draw[->,thick,darkgreen] (1.8,0) --(0.8,3);\draw[->,thick,darkgreen] (4,0)--(0,3);
\node at (0,-1) {$[X_{12}$};\node at (0.9,-1.3) {$,$};\node at (1.8,-1) {$Y_{21}]$};\node at (4,-1) {$Z_{1}$};
\node at (0.8,1.3) {${}^{Y_{12}}$};\node at (2.1,0.9) {${}^{X_{21}}$};\node at (1.9,2.1) {${{}^{Z_{2}}}$};
\filldraw[blue,thick] (1.42,1.1) circle (3pt);
\filldraw[blue,thick] (2,1.5) circle (3pt);
\filldraw[blue,thick] (1.06,2.2) circle (3pt);
\end{tikzpicture}\;\;\begin{array}{l}\\ =\Fcal(\kappa)^{YZ}_{YZ}\Fcal(\kappa^{-1})^{XZ}_{XZ}\Fcal(\kappa)^{XY}_{XY}\;-(X\leftrightarrow Y)\\
=(1\cdot1 \cdot \frac{1}{\sqrt{\kappa}}) \cdot [X_{12},Y_{21}]Z_{1}\\=\frac{1}{\sqrt{\kappa}} \cdot [X_{12},Y_{21}]Z_{2}\end{array}
\ee
As discussed, acting on the third term in (\ref{Wopen1}) is the same as acting on $\left(Z_2(X_{21}Y_{12}-Y_{21}X_{12})\right)$ or 
$\left((X_{21}Y_{12}-Y_{21}X_{12})Z_2\right)$ with the appropriate twists, and since those terms are easier to represent graphically we will compute those instead. We find
\be
\begin{tikzpicture}[scale=0.7,baseline=1cm]
  \draw[->,thick,darkgreen] (0,0)--(4,3);\draw[->,thick,darkgreen] (2.2,0) --(3.2,3);\draw[->,thick,darkgreen] (4,0)--(0,3);
  \node at (0,-1) {$Z_{2}$};\node at (2.2,-1) {$[X_{21}$};\node at (3.1,-1.2) {$,$};\node at (4,-1) {$Y_{12}]$};
  \node at (2.1,0.9) {${}^{Y_{21}}$};\node at (3.2,1.5) {${}^{X_{12}}$};\node at (2.2,2)  {${}^{Z_{1}}$};
\filldraw[blue,thick] (2.55,1.09) circle (3pt);
\filldraw[blue,thick] (1.95,1.5) circle (3pt);
\filldraw[blue,thick] (2.9,2.2) circle (3pt);
\end{tikzpicture}\;\;\begin{array}{l}\\ =\Fcal(\kappa)^{ZX}_{ZX}\Fcal(\kappa^{-1})^{ZY}_{ZY}\Fcal(\kappa^{-1})^{XY}_{XY}\;-(X\leftrightarrow Y)\\
=(\frac{1}{\kappa}\cdot{\kappa}\cdot\sqrt{\kappa}) \cdot Z_{2}[X_{21},Y_{12}]\\=\sqrt{\kappa} \cdot Z_{2}[X_{21},Y_{12}]\end{array}
\ee
and 
\be
\begin{tikzpicture}[scale=0.7,baseline=1cm]
\draw[->,thick,darkgreen] (0,0)--(4,3);\draw[->,thick,darkgreen] (1.8,0) --(0.8,3);\draw[->,thick,darkgreen] (4,0)--(0,3);
\node at (0,-1) {$[X_{21}$};\node at (0.9,-1.3) {$,$};\node at (1.8,-1) {$Y_{12}]$};\node at (4,-1) {$Z_{2}$};
\node at (0.8,1.3) {${}^{Y_{21}}$};\node at (2.1,0.9) {${}^{X_{12}}$};\node at (1.9,2.1) {${}^{Z_{1}}$};
\filldraw[blue,thick] (1.42,1.1) circle (3pt);
\filldraw[blue,thick] (2,1.5) circle (3pt);
\filldraw[blue,thick] (1.06,2.2) circle (3pt);
\end{tikzpicture}\;\;\begin{array}{l}\\ =\Fcal(\kappa^{-1})^{YZ}_{YZ}\Fcal(\kappa)^{XZ}_{XZ}\Fcal(\kappa^{-1})^{XY}_{XY}\;-(X\leftrightarrow Y)\\
=(1\cdot1 \cdot \sqrt{\kappa}) \cdot [X_{21},Y_{12}]Z_{2}\\=\sqrt{\kappa}\cdot [X_{21},Y_{12}]Z_{2}\end{array}
\ee
Indeed the two possibilities for shifting the third term are equal, as required by consistency. Note that, as required, all the three-site twists start by twisting the $(XY-YX)$ term to ensure a diagonal action. We see that the above procedure correctly produces the $\kappa$ factors for each term in the marginally deformed open superpotential
\be
\begin{split}
\label{F3onWop}
  &\frac{1}{\sqrt{\kappa}}Z_1(X_{12}Y_{21}-Y_{12}X_{21})+\frac{1}{\sqrt{\kappa}} (X_{12}Y_{21}-Y_{12}X_{21})Z_1
      +\sqrt{\kappa}\left( Y_{12}\tallvdots Z_2 \tallvdots X_{21}-X_{12}\tallvdots Z_2 \tallvdots Y_{21}\right)\\
      &\quad=\Fcal^{(3,R)}\triangleright Z_1(X_{12}Y_{21}\!-\!Y_{12}X_{21})+ \Fcal^{(3,L)}\triangleright(X_{12}Y_{21}\!-\!Y_{12}X_{21})Z_1\\
      &\quad\;+ \Fcal^{(3,W)}\triangleright\left( Y_{12}\tallvdots Z_2 \tallvdots X_{21}\!-\!X_{12}\tallvdots Z_2 \tallvdots Y_{21}\right)\;.\end{split}
\ee
Inverting the twists (which is trivial since all the actions on this state are diagonal) we can therefore take the deformed superpotential back to the orbifold point, where we have already established invariance.
The above approach is similar to introducing a star product which relates each term in the deformed superpotential to that in the undeformed one, as was done in \cite{Dlamini:2019zuk} for the Leigh-Strassler case. However, from the perspective of this work we would like to express the $\kappa$-deformed superpotential as a single twist on the orbifold one, i.e. to write
\be
\ket{\Wcal}_\kappa=\Fcal^{(3)}\triangleright\ket{\Wcal}_\op\;.
\label{eq:dynamicaltwist3}
\ee
To achieve this, one would have to introduce coassociators, which convert all the bracketings to a single, preferred bracketing. This is what is done in the next section for the four-site scalar potential. However, for the superpotential we note instead that, at the computational level, one can summarise all of the above twists in a simpler, dynamical three-site twist as
\be
\Fcal^{(3)}=\kappa^{-\frac{s}{2}}\otimes\kappa^{-\frac{s}{2}}\otimes\kappa^{-\frac{s}{2}} \ ,
\ee
which is a natural extension of the two-site $XZ$ twist (\ref{diagonaltwist}), with the $\Zset_2$ element $s$ defined as in (\ref{sdef}). One should, however, remember that this twist is only valid for states where the $XY$ terms are antisymmetrised and the bracketings respect this antisymmetrisation. With these implicit assumptions, we can drop the parentheses and write the deformed open superpotential as
\be \label{Wopen2}
\ket{\Wcal_1}_\kappa=\frac{1}{\sqrt{\kappa}}\left(Z_1X_{12}Y_{21}-Z_1 Y_{12}X_{21}+X_{12}Y_{21}Z_1-Y_{12}X_{21}Z_1\right)+\sqrt{\kappa}\left( Y_{12}Z_2 X_{21}-X_{12} Z_2 Y_{21}\right) \ .
\ee
We can then define a coproduct for the action of the $\SU(3)$ generators on this state in the standard way,
\be
\begin{split}
\Delta_\kappa^{(3)}(\Rcal^a_{\;b})&=\Fcal^{(3)} \Delta^{(3)}_\op(\Rcal^a_{\;b}) (\Fcal^{(3)})^{-1}\\
&=\1 \otimes \1 \otimes \Rcal^{a}_{\;b}+\1 \otimes \Rcal^{a}_{\;b}\otimes K^{a}_{\;b}+\Rcal^{a}_{\;b}\otimes K^{a}_{\;b}\otimes K^{a}_{\;b}\;,
\end{split}
\ee
where for unbroken generators we simply have $K^a_{\;b}=\1$, while for the broken ones we find
\be
K^a_{\;b}=\gamma \kappa^s \ .
\ee
It is of course not a coincidence that the coproduct for the superpotential is of the same form as that for the individual $\SU(2)$ sectors, as those coproducts were derived from the superpotential via the F-term relations. Acting on (\ref{Wopen2}) with this coproduct, we find that it is indeed annihilated by all the $\SU(3)$ generators. As an example, we act with the $XZ$-sector raising operator to find
\be
\begin{split}
  \Rcal^3_{\;2}\triangleright_\kappa \ket{\Wcal_1}_\kappa&=\frac{1}{\sqrt{\kappa}}\left(X_{12}(X_{21}Y_{12}- Y_{21}X_{12})+(X_{12}Y_{21}-Y_{12}X_{21})X_{12}\right)\\
  &+\sqrt{\kappa}\frac1{\kappa}\left( Y_{12}X_{21} X_{12}-X_{12} X_{21} Y_{12}\right)=0
\end{split}
\ee
where the precise definition of the twisted action $\triangleright_\kappa$ is given in (\ref{twistedXaction}). This is the marginally deformed version of the orbifold-point action (\ref{RWactionop}). 
Note that the additional power of $1/\kappa$ in the last term came from $K^3_{\;2}(X_{21})$ and $K^3_{\;2}(Y_{21})$.

Through the above heuristic, but we believe natural, assumptions about the twist, we have demonstrated that the opened superpotential of the marginally deformed theory is indeed a singlet of the twisted $\SU(3)$ groupoid symmetry.

\section{Twisted $\SU(4)$ groupoid invariance of the Lagrangian} \label{LagrangianInvariance}

In the previous section, we showed that the superpotential of the marginally deformed theory is invariant under the deformed $\SU(3)$ subgroupoid of the deformed $\SU(4)$ groupoid symmetry. At the same time, one of the main constraints on all the twists we defined in section \ref{TwistsSection} is that they preserve the two-site kinetic terms, i.e. that the open combination
\be
X_{12}\Xb_{21}+\Xb_{12}X_{21}+Y_{12}\Yb_{21}+\Yb_{12}Y_{21}+Z_1\Zb_1+\Zb_1 Z_1\;,
\ee
as well as its $\Zset_2$ conjugate, stays unchanged under twisting. Thus, by construction, the kinetic terms transform as an $\SU(4)$ groupoid singlet, and we have therefore shown the invariance of the superspace Lagrangian under all the $\SU(4)$ generators which are explicitly realised in the $\Ncal=1$ superspace formalism.

In principle, checking invariance of the superspace Lagrangian is sufficient to argue for the invariance of the theory in components as well. The cubic interaction terms containing fermions are expected to work out in a similar way to the superpotential. However, as the process of obtaining the component Lagrangian is non-linear since the auxiliary fields are quadratic in the scalars, it is important to  check also the invariance of the (opened-up) scalar potential, which is quartic (length 4) in fields. Thus, we will need to extend our two-site twists to four sites in order to define the corresponding coproducts. 
These twists will be well-defined on closed states. However, our coproducts are such that after a single action of a broken generator one obtains a state which cannot be gauge contracted. Therefore, we will need to cyclically open up the traces (as explained in Appendix \ref{Ap:OpeningUp}) before acting with the broken generators.

\subsection{Four-site twists and nonassociativity}
\label{subsec:nonassociativity}

Let us now consider the full quartic scalar terms in the Lagrangian, given in (\ref{SU2RV}). As discussed, our approach to showing invariance is simply to untwist the quartic terms to the orbifold point, where we act with the groupoid coproduct $\Delta_{\op}$, thus reducing to the proof of invariance at the orbifold point which we established in Section \ref{OPSymmetries}. In other words, our four-site coproduct for general $\kappa$ will be related to that at $\kappa=1$ by 
\be \label{Coproduct4sites}
\Delta^{(4)}_\kappa(\Rcal^{a}_{\;b})=\Fcal^{(4)}  \Delta_{\op}^{(4)}(\Rcal^{a}_{\;b}) (\Fcal^{(4)})^{-1}
\ee
where $\Delta_{\op}^{(4)}$ was defined in (\ref{eq:OrbifoldPointDelta}). So our goal in the following will be to consistently define a twist for four sites, whose inverse untwists the scalar terms to those at the orbifold point. However, this involves extending our two-site $\SU(4)$ twists to act on four sites, which is mathematically not straightforward due to the groupoid/dynamical nature of our setting. Physically, however, it is evident how the two-site twists should extend to four sites, at least for the terms which appear in the scalar potential. Consider an F-term contribution to the scalar potential at the orbifold point:
\be
\tr_1 [F^Y_{12}\bar{F}^{\bar{Y}}_{21}]=\tr_1[(X_{12}Z_2-Z_1 X_{12}) (\Xb_{21}\Zb_1-\Zb_2 \Xb_{21})] .
\ee
In order for this expression to be transformed into its counterpart in the deformed theory, it is sufficient to twist the two quantum planes independently. Thus, informally we would expect the four-site twist to work as follows:
\be \label{F4XZ}
\begin{split}\Fcal^{(4)} & \triangleright\;\tr_{1}[(X_{12}Z_{2}-Z_{1}X_{12})(\Xb_{21}\Zb_{1}-\Zb_{2}\Xb_{21})]\\
 & =\tr_{1}\left[\left(\Fcal^{(2)}\triangleright(X_{12}Z_{2}-Z_{1}X_{12})\right)\left(\Fcal^{(2)}\triangleright(\Xb_{21}\Zb_{1}-\Zb_{2}\Xb_{21})\right)\right]\\
 & =\tr_{1}[(X_{12}Z_{2}-\frac{1}{\kappa}Z_{1}X_{12})(\Xb_{21}\Zb_{1}-\kappa\Zb_{2}\Xb_{21})]
\end{split} 
\ee
which is the correct F-term contribution in the deformed theory. Here the two-site twists are as in (\ref{diagonaltwist}). Similarly, for a D-term contribution such as
\be
D_1^2=\tr_1[(X_{12}\Xb_{21}-\Xb_{12}X_{21}+X_{12}\Xb_{21}-\Xb_{12}X_{21}+[Z_1,\Zb_1])^2] ,
\ee
we would apply the corresponding (triplet and singlet) two-site twists (\ref{eq:F2Triplet},\ref{F2singlet}) to write
\be
\begin{split}
  \Fcal^{(4)}\triangleright \; &\tr_1[(X_{12}\Xb_{21}-\Xb_{12}X_{21}+X_{12}\Xb_{21}-\Xb_{12}X_{21}+[Z_1,\Zb_1])^2]\\
  &=\tr_1\left[\left(\Fcal^{(2)}\triangleright (X_{12}\Xb_{21}-\Xb_{12}X_{21}+X_{12}\Xb_{21}-\Xb_{12}X_{21}+[Z_1,\Zb_1])\right)^2\right]\\
  &=\tr_1\left[\left(\frac{1}{\sqrt{\kappa}}(X_{12}\Xb_{21}-\Xb_{12}X_{21}+X_{12}\Xb_{21}-\Xb_{12}X_{21}+[Z_1,\Zb_1])\right)^2\right]\;,
\end{split}
\ee
which again correctly produces  the D-term contribution to the scalar potential in the deformed theory. Of course, twisting as above requires us to have organised the undeformed scalar potential terms in the very specific way that they arise through the F- and D-term relations. However, for a given monomial this can be ambiguous. For instance, without additional information we cannot determine whether the term
\be
\tr_1[X_{12}Z_2\Zb_{2} \Xb_{21}]=\tr_2[Z_2\Zb_{2} \Xb_{21}X_{12}]
\ee
is an F-term ($\tr_1[(X_{12}Z_2)(\Zb_{2} \Xb_{21})]$) or a D-term ($\tr_2[(Z_2\Zb_{2})(\Xb_{21}X_{12})]$). For this specific monomial, and for similar terms  involving the $Z$ fields, the factors of $\kappa$ end up being the same ($1\cdot \kappa$ for the F-term bracketing and $\sqrt{\kappa}\cdot\sqrt{\kappa}$ for the D-term one), but their contributions to the scalar potential come with a relative factor of $-\half$. Other contributions lead to different $\kappa$ dependence. To sum up, the quartic terms contain the following ambiguous monomials:

\begin{enumerate}
    \item Terms of type $X_{12} \bar{X}_{21} X_{12} \bar{X}_{21}$ and $Y_{12} \bar{Y}_{21} Y_{12} \bar{Y}_{21}$ have a coefficient $\left( \frac{1}{\kappa} + \kappa \right)$  coming from the D-term contributions $(D_1)^2$ and $(D_2)^2$, respectively. This coefficient reduces to the numerical factor $2$ at the orbifold point.
    \item Terms of type $X_{12} \bar{X}_{21} \bar{Y}_{12} Y_{21}$ have a coefficient  $\left( -\frac{1}{\kappa} + 2 \kappa \right)$ coming from $(D_1)^2$ and $F_2^Z\bar{F}_2^{\Zb}$, respectively. This coefficient reduces to an overall $1$ at the orbifold point.
    \item Terms of type $Y_{12} X_{21} \bar{X}_{12} \bar{Y}_{21}$  have a coefficient $\left( \frac{2}{\kappa} - \kappa \right)$ coming from $F_1^Z\bar{F}_1^{\bar{Z}}$ and $(D_2)^2$, respectively. This coefficient also reduces to $1$ at the orbifold point.
\end{enumerate}

To resolve these ambiguities, we are led to the need to retain the placement of parentheses, or bracketings, in the way that they arise in the F- and D-terms. The inequivalence of terms with different bracketings tells us that the $\kappa$-deformed theory will have a quasi-Hopf structure. This was already the case in our study of the superpotential (Section \ref{Superpotentialinvariance}).  However, there it was possible to find a simple three-site twist (\ref{eq:dynamicaltwist3}) that correctly captured all the $\kappa$-dependent factors arising through a more meticulous treatment. For the quartic terms, we do not have such an ``effective'' twist. If we wish to obtain the correct $\kappa$-deformed Lagrangian by twisting we would have to start with the Lagrangian at the orbifold point  with a specific choice of parentheses indicating the F- and D- terms. At the  ``classical'' (orbifold point) level all bracketings are equivalent, but different bracketings will give different answers at the ``quantum'' ($\kappa$-deformed theory) level. So, in effect, it is supersymmetry which tells us how the Lagrangian at the orbifold point  should be bracketed in order for the twists to directly lead to the correct deformed Lagrangian. 

As previously explained, our current  approach does not allow us to act directly with broken generators on closed states. Instead, our procedure requires us to cyclically open up the trace and then act on the opened states. Clearly, the naive opening-up procedure does not respect the parentheses above, which distinguish between F-terms and D-terms. But from the above discussion, it should be clear that it is not possible to construct an unambiguous four-site twist (which can then be inverted in order to demonstrate invariance) unless the bracketings are taken into account and preserved throughout the opening-up procedure. To illustrate how we achieve this, let us cyclically open up the first term of (\ref{F4XZ}):

\be   \label{eq:UpdatedOpeningProcedure}
\begin{split}
    \tr_1 \left[ (X_{12}Z_2)(\bar{X}_{21}\bar{Z}_1)\right] \rightarrow &\;\;
    \frac{1}{4} \left[ ( X_{12} Z_2 )( \bar{X}_{21} \bar{Z}_1 ) +  \bracketing{Z_2}{\bar{X}_{21}}{\bar{Z}_1}{X_{12}}\right.
    \\
    & \quad +\left. ( \bar{X}_{21} \bar{Z}_1 )( X_{12} Z_2 ) +  \bracketing{\bar{Z}_1}{X_{12}}{Z_2}{\bar{X}_{21}} \right] \ .
    \end{split}\ee
As in Section \ref{Superpotentialinvariance}, we adopted the notation $\bracketing{A}{B}{C}{D}$  which indicates that the $BC$ and $DA$ terms were bracketed together in the original closed expression.  This refined opening-up procedure generates an equal number of monomials with each type of bracketing. To distinguish open expressions in the marginally deformed theory from those at the orbifold point, we will write $\ket{(AB)(CD)}_\kappa$ and $\ket{(AB)(CD)}_{\op}$, which we call the standard, or unshifted, bracketing. Similarly, we will write  $\ket{\bracketing{A}{B}{C}{D})}_\kappa$ and $\ket{\bracketing{A}{B}{C}{D}}_{\op}$, which we call the \emph{shifted} bracketing. They are both quartic expressions in the fields $\varphi^i$, with different coefficients. Clearly, at the orbifold point we have
\be \label{ABCDunshiftedshifted}
\ket{(AB)(CD)}_{\op}=c_{ijkl}^{(u)} \ket{(\varphi^i\varphi^j)(\varphi^k\varphi^l)}_\op   \;\;\text{and}\;\;\ket{\bracketing{A}{B}{C}{D}}_{\op}=c_{ijkl}^{(s)} \ket{\bracketing{\varphi^i}{\varphi^j}{\varphi^k}{\varphi^l}}_\op 
\ee
where $c_{ijkl}^{(u)}$ and $c_{ijkl}^{(s)}$ can be read off from (\ref{Vopenunshifted}) and (\ref{Vopenshifted}), respectively, specialised to $\kappa=1$. We emphasise that although the two types of terms are equal in number, their coefficients are different. Of course, at the orbifold point the bracketing of a given monomial is unimportant, i.e. we have
\be
\ket{\bracketing{\varphi^i}{\varphi^j}{\varphi^k}{\varphi^l}}_\op =
\ket{(\varphi^i\varphi^j)(\varphi^k\varphi^l)}_\op\;.\ee
However, the way in which these two types of bracketing become twisted differs. Extending the discussion in Appendix \ref{QuantumPlanesAppendix} to four sites, given an orbifold-point expression, we have
\be \label{eq:F4SB}
\begin{split}
\mathcal{F}^{(4)}\triangleright \ket{(AB)(CD)}_\op &= \left(\mathcal{F}^{(2)}_{12} \otimes \mathcal{F}^{(2)}_{34} \right) \triangleright \ket{(AB)(CD)}_\op\\
&=c_{ijkl} (\Fcal^T)^{ij}_{\;mn}(\Fcal^T)^{kl}_{\;rs} \ket{(\varphi^m\varphi^n)(\varphi^r\varphi^s)}_\kappa\\
&=c_{mnrs}^{(u)}(\kappa) \ket{(\varphi^m\varphi^n)(\varphi^r\varphi^s)}_\kappa
\end{split}
\ee
and
\be\label{eq:F4NSB}
\begin{split}\mathcal{F}_{\text{shifted}}^{(4)} \triangleright \ket{\bracketing{A}{B}{C}{D}}_\op &= \left( \mathcal{F}^{(2)}_{23} \otimes \mathcal{F}^{(2)}_{41} \right) \triangleright \ket{ \bracketing{A}{B}{C}{D}}_\op \\
&=c_{ijkl}  (\Fcal^T)^{jk}_{\;nr}(\Fcal^T)^{li}_{\;sm} \ket{\bracketing{\varphi^m}{\varphi^n}{\varphi^r}{\varphi^s}}_\kappa\\ &=c_{mnrs}^{(s)}(\kappa) \ket{\bracketing{\varphi^m}{\varphi^n}{\varphi^r}{\varphi^s}}_\kappa
\end{split}
\ee
Both four-site twist actions $\mathcal{F}^{(4)}$ and $\mathcal{F}_{\text{shifted}}^{(4)}$ are composed of a pair of two-site twists $\mathcal{F}^{(2)}_{ab}$ as defined in (\ref{diagonaltwist}), (\ref{XYtwist}), (\ref{eq:F2Triplet}), (\ref{F2singlet}) and their descendants. We have defined the coefficients
\be \label{defcoeffs}
\begin{split}
c_{mnrs}^{(u)}(\kappa)&:=c_{ijkl} (\Fcal^T)^{ij}_{\;mn}(\Fcal^T)^{kl}_{\;rs}=c_{ijkl} (\Fcal^{(4),T})^{ijkl}_{\;mnrs} \;\; ,\;\;\\ c_{mnrs}^{(s)}(\kappa)&:= c_{ijkl}  (\Fcal^T)^{jk}_{\;nr}(\Fcal^T)^{li}_{\;sm}
=c_{ijkl} (\Fcal^{(4),T}_{\text{shifted}})^{ijkl}_{\;mnrs}\;.
\end{split}
\ee
The definitions above apply to any orbifold-point coefficients $c_{ijkl}$. If we twist the state defined by the specific $c_{ijkl}^{(u)}$ in the orbifold-point unshifted terms according to (\ref{eq:F4SB}), the corresponding unshifted coefficients $c_{ijkl}^{(u)}(\kappa)$ will be precisely those in the marginally-deformed theory, and similarly for the shifted $c_{ijkl}^{(s)}$ coefficients. Therefore, we can write the scalar potential as 
\be\begin{split}
\Vcal(\kappa)&=c^{(u)}_{ijkl}(\kappa)\ket{(\varphi^i\varphi^j)(\varphi^k\varphi^l)}_\kappa+c^{(s)}_{ijkl}(\kappa)\ket{\bracketing{\varphi^i}{\varphi^j}{\varphi^k}{\varphi^l}}_\kappa\\
&=\Fcal^{(4)}\triangleright\left(c^{(u)}_{ijkl}\ket{(\varphi^i\varphi^j)(\varphi^k\varphi^l)}_\op\right)+\Fcal^{(4)}_{\text{shifted}}\triangleright\left(c^{(s)}_{ijkl}\ket{\bracketing{\varphi^i}{\varphi^j}{\varphi^k}{\varphi^l}}_\op\right)\;.
\end{split}\ee
Note that the two-site twists within $\mathcal{F}^{(4)}$ and $\mathcal{F}_{\text{shifted}}^{(4)}$ do not overlap in their actions on the individual fields of a monomial, and are invertible. So the inverse of the four-site twist is the tensor product of the inverse of each two-site twist independently. We can accordingly define inverse twists, which act on deformed states with prescribed bracketings and give us orbifold-point expressions:
\be\begin{split}
    \left(\mathcal{F}^{(4)}\right)^{-1} \triangleright \ket{ (AB)(CD)}_\kappa &= \left(\left(\mathcal{F}^{(2)}_{12}\right)^{-1} \otimes \left(\mathcal{F}^{(2)}_{34}\right)^{-1} \right) \triangleright \ket{ (AB)(CD)}_\kappa \\
    = c_{ijkl}^{(u)}& ((\Fcal^T)^{-1})^{ij}_{\;mn}((\Fcal^T)^{-1})^{kl}_{\;rs} \ket{\varphi^m\varphi^n\varphi^r\varphi^s}_\op=c_{mnrs} \ket{\varphi^m\varphi^n\varphi^r\varphi^s}_\op
\end{split}\ee
and
\be\begin{split}
    \left(\mathcal{F}_{\text{shifted}}^{(4)}\right)^{-1} \triangleright \ket{\bracketing{A}{B}{C}{D}}_\kappa &= \left( \left(\mathcal{F}^{(2)}_{23}\right)^{-1} \otimes \left(\mathcal{F}^{(2)}_{41}\right)^{-1} \right) \triangleright \ket{\bracketing{A}{B}{C}{D}}_\kappa  \\
    =c_{ijkl}^{(s)} & ((\Fcal^T)^{-1})^{jk}_{\;nr}((\Fcal^T)^{-1})^{li}_{\;sm} \ket{\varphi^m\varphi^n\varphi^r\varphi^s}_\op=c_{mnrs} \ket{\varphi^m\varphi^n\varphi^r\varphi^s}_\op
\end{split}\ee
Of course,  the actions $ \left(\mathcal{F}^{(4)}\right)^{-1}$ on states of the form  $\ket{\bracketing{A}{B}{C}{D}}_\kappa$ and $ \left(\mathcal{F}_{\text{shifted}}^{(4)}\right)^{-1}$ on states of the form $\ket{ (AB)(CD)}_\kappa$ are not defined at this point, due to the incompatible placement of the parentheses. 
It is important to recall that in order to define the action of the broken $\SU(4)$ generators, we express the $\kappa$-deformed quartic terms as an \emph{overall} four-site twist acting on the undeformed quartic terms. This is required in order to invert that twist when acting with the coproduct (\ref{Coproduct4sites}).\footnote{It is easy to check that the unshifted and shifted quartic terms are not independently $\SU(4)$ invariant at the orbifold point.} It is therefore necessary to express the shifted terms in terms of the unshifted ones, or vice versa. For this purpose, we will define a coassociator in the next subsection.

\subsection{The coassociator}
\label{sec:Coassociator}

In the standard quasi-Hopf setting \cite{Drinfeld90}, the coassociator is an object living in three copies of the algebra, which maps between the two expressions with a priori different choices of bracketing:
\be
A\otimes (B\otimes C)=\Phi\triangleright (A'\otimes B')\otimes C'
\ee
where in general the right-hand side is a linear combination. As also discussed in Section \ref{Superpotentialinvariance}, if we twist away from an associative point (where $\Phi_\op=\1\otimes \1\otimes \1$) with a Drinfeld twist $\Fcal$, the three-site twists are $\Fcal^{(3,L)}=(\Fcal\otimes \1)(\Delta\otimes\id)(\Fcal)$ and $\Fcal^{(3,R)}=(\id\otimes \Fcal)(\id\otimes \Delta)(\Fcal)$. Then the coassociator at the deformed point can be defined as taking the left-bracketed expression to the associative point by acting with the inverse of $\Fcal^{(3,L)}$, switching to the right bracketing using the trivial $\Phi_\op$, and then twisting back with $\Fcal^{(3,R)}$ to obtain the opposite bracketing:
\be
\Phi=(\1\otimes \Fcal)(\id\otimes \Delta_\op)(\Fcal)\Phi_\op (\Delta_\op\otimes \id)(\Fcal^{-1})(\Fcal^{-1}\otimes \1)
\ee
This can be extended to more sites. For instance, for four sites there are five inequivalent choices of bracketings, and one can define four-site coassociators mapping any two of these bracketings to each other by going through the associative point. 

In the following, we will follow a similar procedure to define a coassociator which takes us between the two types of four-site twists that appear in (\ref{eq:F4SB}) and (\ref{eq:F4NSB}). As our four-site twists are built from the products of two-site twists, and at this stage we do not have a way of defining them by going through three sites, we will empirically define a four-site coassociator as the transformation taking us from shifted to unshifted monomials:

\begin{equation}
    \Phi =\mathcal{F}^{(4)}\hspace{1pt} \Phi_\op \hspace{1pt} (\mathcal{F}^{(4)}_{\text{shifted}})^{-1} \ ,
\end{equation}
where $\Phi_\op$ is the coassociator at the orbifold point, which is assumed to be trivial:
\begin{align}
  \ket{\bracketing{A}{B}{C}{D}}_\op &= \Phi_\op \triangleright   \ket{(AB)(CD)}_\op   \\
     \ket{(AB)(CD)}_\op   &= \Phi_\op^{-1} \triangleright \ket{\bracketing{A}{B}{C}{D}}_\op  \ ,
\end{align}
and where to avoid confusion we note that here $\ket{\bracketing{A}{B}{C}{D}}_\op$ and  $\ket{(AB)(CD)}_\op$ are not as in (\ref{ABCDunshiftedshifted}) but denote polynomials with the same coefficients $c_{ijkl}$. 
Consequently, on a shifted-twisted expression we have
\be\begin{split}
   { \Phi}\triangleright \ket{\bracketing{A}{B}{C}{D}}_\kappa&
   =\left( \mathcal{F}^{(2)}_{12} \otimes \mathcal{F}^{(2)}_{34} \right) {\Phi}_\op \left(\left(\mathcal{F}^{(2)}_{23}\right)^{-1} \otimes \left(\mathcal{F}^{(2)}_{41}\right)^{-1} \right) \triangleright \left( \mathcal{F}^{(2)}_{23} \otimes \mathcal{F}^{(2)}_{41} \right)\ket{\bracketing{A}{B}{C}{D}}_\op\\
   &=\left( \mathcal{F}^{(2)}_{12} \otimes \mathcal{F}^{(2)}_{34} \right) {\Phi}_\op\ket{\bracketing{A}{B}{C}{D}}_\op\\
   &=\left( \mathcal{F}^{(2)}_{12} \otimes \mathcal{F}^{(2)}_{34} \right) \ket{(AB)(CD)}_\op=\ket{(AB)(CD)}_\kappa ,
   \end{split}
   \ee
where  both the left- and right-hand sides are in principle linear combinations of monomials. More concisely, we can write
 \be
   \begin{split}
   \Phi\triangleright \ket{\bracketing{A}{B}{C}{D}}_\kappa &=(\Fcal^{(4)}\cdot (\Fcal^{(4)}_{\text{shifted}})^{-1})\triangleright \Fcal^{(4)}_{\text{shifted}}\triangleright \ket{\bracketing{A}{B}{C}{D}}_\op\\
 &=(\Fcal^{(4)})\triangleright \ket{(AB)(CD)}_\op
 = \ket{(AB)(CD)}_{\kappa} ,
   \end{split}
   \ee
   where we used the equality of the bracketings at the orbifold point. We think of this expression as a map from the shifted bracketing to the standard one. In terms of the monomials, we can express this as a rotation by the transpose of $\Phi$:
   \be \label{monomialrotation}
   \ket{\bracketing{\varphi^m}{\varphi^n}{\varphi^r}{\varphi^s}}_\kappa=
(\Phi^{T})^{mnrs}_{m'n'r's'}\ket{(\varphi^{m'}\varphi^{n'})(\varphi^{r'}\varphi^{s'})}_\kappa\;.
   \ee
   This relation allows us to connect shifted and unshifted terms within the twisted Lagrangian: Terms that are twists of the same orbifold-point expression
   \be
c_{ijkl}\ket{\bracketing{\varphi^i}{\varphi^j}{\varphi^k}{\varphi^l}}_\op=c_{ijkl}\ket{(\varphi^{i}\varphi^{j})(\varphi^{k}\varphi^{l})}_\op\;,
   \ee
   can be related as
   \be\begin{split}
   c_{ijkl}&(\Fcal^{(4),T}_{\text{shifted}})^{ijkl}_{mnrs}\ket{\bracketing{\varphi^m}{\varphi^n}{\varphi^r}{\varphi^s}}_\kappa=c_{ijkl}(\Fcal^{(4),T}_{\text{shifted}})^{ijkl}_{mnrs}
(\Phi^{T})^{mnrs}_{m'n'r's'}\ket{(\varphi^{m'}\varphi^{n'})(\varphi^{r'}\varphi^{s'})}_\kappa\\
   &=c_{ijkl}(\Fcal^{(4),T}_{\text{shifted}})^{ijkl}_{mnrs}
((\Fcal^{(4),T}_\text{shifted})^{-1})^{mnrs}_{i'j'k'l'}(\Fcal^{(4),T})^{i'j'k'l'}_{m'n'r's'}\ket{(\varphi^{m'}\varphi^{n'})(\varphi^{r'}\varphi^{s'})}_\kappa\\
   &=c_{ijkl}(\Fcal^{(4),T})^{ijkl}_{mnrs}\ket{(\varphi^{m}\varphi^{n})(\varphi^{r}\varphi^{s})}_\kappa .
\end{split}\ee
By acting on all possible shifted monomials, we can obtain linear combinations of unshifted ones, which in turn allow us to ascertain the tensor coefficient of $\Phi^{T}$. We are of course only interested in $\SU(2)_L\times\SU(2)_R\times\Urm(1)_r$-neutral states, which have equal numbers of fields and their conjugate fields and are the ones appearing in the scalar potential. Since there are 74 neutral four-site states of type $(AB)(CD)$, and the same number of $\bracketing{A}{B}{C}{D}$-type states,\footnote{The actual scalar potential expressions in (\ref{Vopenunshifted}) and (\ref{Vopenshifted}) contain 60 terms each, i.e. do not depend on all possible neutral monomials. However, the remaining 14 terms of each type do appear after rebracketing each expression, so they need to be included in our basis for the coassociator.}  the coassociator can be expressed as a $74\times 74$ matrix. In practice, however, the matrix can be split into smaller blocks (specifically, states with no $Z$'s, two $Z$'s, and four $Z$'s), which are presented in Appendix \ref{CoassociatorAppendix}.

\subsection{Invariance of the scalar potential} \label{scalarinvariance}

Let us recall that our goal is to untwist the opened-up scalar terms in the deformed Lagrangian back to the orbifold point. As we saw, these terms are of two types, which we called $\ket{(AB)(CD)}_\kappa$ and $\ket{\bracketing{A}{B}{C}{D}}_\kappa$, which each can be untwisted by either $(\Fcal^{(4)})^{-1}$ or $(\Fcal^{(4)}_{\text{shifted}})^{-1}$. However, for our purposes we need the action to be untwisted by an \emph{overall} inverse twist, which we will choose to be $(\Fcal^{(4)})^{-1}$. 
For this action to make sense, we will use the coassociator to rebracket all the terms of $\ket{\bracketing{A}{B}{C}{D}}_\kappa$ type to those of $\ket{(AB)(CD)}_\kappa$ type.

As an example of how the procedure works, let us consider the monomial $\ket{\bracketing{Z_1}{Z_1}{\bar{Z}_1}{\bar{Z}_1}}_\kappa$ which is part of the opened-up scalar potential. Clearly this term comes purely from a $(D_1)^2$ contribution, as no F-terms give $Z_1Z_1$ or its conjugate. We find 
\begin{align} \label{PhiZZZZ}
    \Phi &\triangleright \ket{\bracketing{Z_1}{Z_1}{\bar{Z}_1}{\bar{Z}_1}}_\kappa  = \left( \mathcal{F}^{(2)}_{12} \otimes \mathcal{F}^{(2)}_{34} \right) \Phi_\op \left(\left(\mathcal{F}^{(2)}_{23}\right)^{-1} \otimes \left(\mathcal{F}^{(2)}_{41}\right)^{-1} \right) \triangleright \ket{\bracketing{Z_1}{Z_1}{\bar{Z}_1}{\bar{Z}_1}}_\kappa  \nonumber \\
   &= \left( \mathcal{F}^{(2)}_{12} \otimes \mathcal{F}^{(2)}_{34} \right) \Phi_\op \triangleright \frac{1}{4} \left[\left(\sqrt{\kappa }+1\right)^2 \ket{\bracketing{Z_1}{Z_1}{\bar{Z}_1}{\bar{Z}_1}}_\op -(\kappa -1) \ket{\bracketing{Z_1}{\bar{Z}_1}{ Z_1}{\bar{Z}_1}}_\op\right. \nonumber \\
    &\qquad\qquad\qquad\qquad\qquad\left.-(\kappa-1)  \ket{\bracketing{\bar{Z}_1}{Z_1}{\bar{Z}_1}{Z_1}}_\op  + \left(\sqrt{\kappa }-1\right)^2 \ket{\bracketing{\bar{Z}_1}{\bar{Z}_1}{ Z_1}{Z_1}}_\op \right] \nonumber \\
    &=  \left( \mathcal{F}^{(2)}_{12} \otimes \mathcal{F}^{(2)}_{34} \right) \triangleright \left[ \frac{1}{4} \left(\left(\sqrt{\kappa }+1\right)^2  \ket{ (Z_1 Z_1)(\bar{Z}_1 \bar{Z}_1)}_\op  -(\kappa -1) \ket{(Z_1 \bar{Z}_1)(Z_1 \bar{Z}_1}_\op   \right. \right. \nonumber \\
    &\qquad\qquad\qquad\qquad\left. \left.-(\kappa-1)    \ket{(\bar{Z}_1 Z_1)(\bar{Z}_1 Z_1)}_\op    + \left(\sqrt{\kappa }-1\right)^2  \ket{(\bar{Z}_1 \bar{Z}_1)(Z_1 Z_1)}_\op   \right) \right] \nonumber \\
    &= \frac{1}{4\kappa^2} \left[\left(\sqrt{\kappa }-1\right)^2 \ket{(\bar{Z}_1\bar{Z}_1)(Z_1Z_1)}_\kappa+\left(\sqrt{\kappa }+1\right)^2 \ket{(Z_1 Z_1)( \bar{Z}_1 \bar{Z}_1)}_\kappa\right] \nonumber \\
    &\quad-\frac{(\kappa -1)}{8 \kappa } \left[ (\kappa +1) \ket{(Z_1 \bar{Z}_1)( Z_1 \bar{Z}_1)}_\kappa+(\kappa -1) \ket{(Z_1 \bar{Z}_1)( \bar{Z}_1 Z_1)}_\kappa\right.\nonumber \\
&\qquad\qquad\quad\;\left.      +(\kappa-1)  \ket{(\bar{Z}_1 Z_1)( Z_1 \bar{Z}_1)}_\kappa
+(\kappa+1)  \ket{(\bar{Z}_1 Z_1)( \bar{Z}_1 Z_1)}_\kappa\right] \;.
\end{align}
We see that, as expected, a single monomial in the shifted bracketing maps to a linear combination of monomials in the unshifted bracketing. Comparing with (\ref{monomialrotation}), we can read off the corresponding tensor components of the coassociator:
{\small\be
\begin{split}
(\Phi^T)^{ZZ\Zb\Zb}_{ZZ\Zb\Zb}=\frac{(\sqrt{\kappa}+1)^2}{4\kappa^2}\;,\;&(\Phi^T)^{ZZ\Zb\Zb}_{\Zb\Zb ZZ}=\frac{(\sqrt{\kappa}-1)^2}{4\kappa^2}\;,\;
(\Phi^T)^{ZZ\Zb\Zb}_{Z\Zb Z\Zb}=(\Phi^T)^{ZZ\Zb\Zb}_{\Zb Z\Zb Z}=\frac{1-\kappa^2}{8\kappa}\;,\\ &\;(\Phi^T)^{ZZ\Zb\Zb}_{Z\Zb\Zb Z}=(\Phi^T)^{ZZ\Zb\Zb}_{\Zb ZZ\Zb}=-\frac{(\kappa-1)^2}{8\kappa}\;.
\end{split}
\ee
}
It is more insightful to act on the actual linear combination of shifted monomials in this sector, which appears in the quartic terms (\ref{Vopenshifted}). Repeating the steps above, one computes:
\begin{align} \label{PhiZZZbZbaction}
    &\Phi \triangleright \left[\ket{\bracketing{Z_1}{Z_1}{\bar{Z}_1}{\bar{Z}_1}}_\kappa - \ket{\bracketing{Z_1}{\bar{Z}_1}{Z_1}{\bar{Z}_1}}_\kappa - \ket{\bracketing{\bar{Z}_1}{Z_1}{\bar{Z}_1}{Z_1}}_\kappa + \ket{\bracketing{\bar{Z}_1}{\bar{Z}_1}{Z_1}{Z_1}}_\kappa \right]  \nonumber \\
    &\quad= \frac12 \left[\frac{2}{\kappa} \Big(\ket{(Z_1 Z_1)(\bar{Z}_1 \bar{Z}_1)}_\kappa + \ket{(\bar{Z}_1 \bar{Z}_1)(Z_1 Z_1)}_\kappa\Big)- (\kappa+1)\Big(\ket{(Z_1 \bar{Z}_1)(Z_1 \bar{Z}_1)}_\kappa \right. \nonumber \\
    &\qquad\quad \left.+\ket{(\bar{Z}_1 Z_1)(\bar{Z}_1 Z_1)}_\kappa\Big) - (\kappa-1)\Big(\ket{(Z_1 \bar{Z}_1 )(\bar{Z}_1 Z_1)}_\kappa+\ket{(\bar{Z}_1 Z_1)(Z_1 \bar{Z}_1)}_\kappa\Big) \right] \ .
\end{align}
where the notation $\Phi~ \triangleright$ on a shifted state denotes the expansion of that state in the unshifted basis by applying (\ref{monomialrotation}). 
In this sector there is of course also a contribution of unshifted type, coming from $(D_1)^2$ terms which are opened up as $\ket{(AB)(CD)}_\kappa$. Adding those terms as well, we find
\begin{align}
    &\frac{1}{2\kappa}\left[ - \ket{(Z_1 \bar{Z}_1)(Z_1 \bar{Z}_1)}_\kappa + \ket{(Z_1 \bar{Z}_1)(\bar{Z}_1 Z_1)}_\kappa + \ket{(\bar{Z}_1 Z_1)(Z_1 \bar{Z}_1)}_\kappa - \ket{(\bar{Z}_1 Z_1)(\bar{Z}_1 Z_1)}_\kappa \right] \nonumber \\
    &+\Phi \triangleright \frac{1}{2\kappa}\left[ \ket{\bracketing{Z_1}{Z_1}{\bar{Z}_1}{\bar{Z}_1}}_\kappa -\ket{\bracketing{Z_1}{\bar{Z}_1}{Z_1}{\bar{Z}_1}}_\kappa - \ket{\bracketing{\bar{Z}_1}{Z_1}{\bar{Z}_1}{Z_1}}_\kappa + \ket{\bracketing{\bar{Z}_1}{\bar{Z}_1}{Z_1}{Z_1}}_\kappa \right]  \nonumber \\
    &= \frac{1}{4\kappa} \left[\frac{2}{\kappa} \Big(\ket{(Z_1 Z_1)(\bar{Z}_1 \bar{Z}_1)}_\kappa + \ket{(\bar{Z}_1 \bar{Z}_1)(Z_1 Z_1)}_\kappa\Big)- (\kappa+3)\Big(\ket{(Z_1 \bar{Z}_1)(Z_1 \bar{Z}_1)}_\kappa \right. \nonumber \\
    &\qquad\quad\left.+\ket{(\bar{Z}_1 Z_1)(\bar{Z}_1 Z_1)}_\kappa\Big) - (\kappa-3)\Big(\ket{(Z_1 \bar{Z}_1 )(\bar{Z}_1 Z_1)}_\kappa+\ket{(\bar{Z}_1 Z_1)(Z_1 \bar{Z}_1)}_\kappa\Big) \right] \ .
    \label{eq:4sitesZZbStandardBracketing}
\end{align}
This is the $\kappa$-dependent contribution to the scalar potential, now with only one type of bracketing. Note that this expression is quite different to what one obtains by simply forgetting about the bracketing. We have finally found an expression which we can untwist using a single inverse twist. We find
\be
\begin{split}
&\left(\Fcal^{(4)} \right)^{-1}\triangleright \frac{1}{4\kappa} \left[\frac{2}{\kappa} \Big(\ket{(Z_1 Z_1)(\bar{Z}_1 \bar{Z}_1)}_\kappa + \ket{(\bar{Z}_1 \bar{Z}_1)(Z_1 Z_1)}_\kappa\Big)\right.\\
&\qquad\qquad\qquad\quad - (\kappa+3)\Big(\ket{(Z_1 \bar{Z}_1)(Z_1 \bar{Z}_1)}_\kappa 
    +\ket{(\bar{Z}_1 Z_1)(\bar{Z}_1 Z_1)}_\kappa\Big) \\\
    &\qquad\qquad\qquad\quad-\left. (\kappa-3)\Big(\ket{(Z_1 \bar{Z}_1 )(\bar{Z}_1 Z_1)}_\kappa+\ket{(\bar{Z}_1 Z_1)(Z_1 \bar{Z}_1)}_\kappa\Big) \right]  \\
    &\qquad\qquad= \frac12 \left[ -2   \ket{(Z_1 \bar{Z}_1)(Z_1 \bar{Z}_1)}_\op  -2   \ket{(\bar{Z}_1 Z_1)( \bar{Z}_1 Z_1)}_\op  +   \ket{(Z_1 Z_1)(\bar{Z}_1 \bar{Z}_1)}_\op \right. \\
    &\qquad\qquad\qquad \left.+  \ket{(\bar{Z}_1 \bar{Z}_1)(Z_1 Z_1)}_\op +   \ket{(Z_1 \bar{Z}_1)(\bar{Z}_1 Z_1)}_\op  +   \ket{(\bar{Z}_1 Z_1)(Z_1 \bar{Z}_1)}_\op  \right] \ ,
\end{split}
\ee
which precisely matches the result expected at the orbifold point, but now with only one type of bracketing.

A similar, but much more tedious, computation for all the remaining terms  in (\ref{Vopenshifted}), using the coassociator defined in Appendix \ref{CoassociatorAppendix}, brings the scalar potential to a linear combination of only $\ket{(AB)(CD)}_\kappa$-type terms. As above, we find that acting with $(\Fcal^{(4)})^{-1}$ correctly untwists them to the orbifold point scalar potential. It follows that, for all $\SU(4)$ generators $\Rcal^a_{\;b}$, both broken and unbroken, the coproduct (\ref{Coproduct4sites}) annihilates the scalar potential of the deformed theory. We have therefore shown invariance under our deformed  $\SU(4)$ symmetry, as encoded in (\ref{Coproduct4sites}) and (\ref{eq:OrbifoldPointDelta}). 

Of course, in the above, we made a choice to express the shifted quartic terms as linear combinations of the unshifted ones. Equivalently, we could have chosen to convert all the unshifted terms to shifted ones, which would have resulted in an expression related to the orbifold point by an action of $(\Fcal_{\text{shifted}}^{(4)})^{-1}$. We also 
note that if we had misidentified any of the terms in (\ref{Vopenunshifted}) or (\ref{Vopenshifted}), i.e. changed the bracketing from shifted to unshifted without using the coassociator, the construction would not have worked, and we would not have ended up with an overall twist acting on the correct orbifold-point expression. So the construction relies strongly on respecting the quantum plane structure in (\ref{SU2RV}). 

We emphasise that the above computation was expected to work, by the very definition of the coassociator. The reason for explicitly computing the coassociator matrix is mainly to clarify how the construction works in practice. Also, one expects that the understanding of other observables in the full $\SU(4)$ sector of the deformed theory, beyond the scalar potential, will require similar manipulations. Although for the purpose of showing invariance we did not have to actually compute the twisted coproduct, it will be required in general in order to construct other representations, and for that we expect that it would be unavoidable to work with explicit coassociators.

\section{Implications for the spectrum}
\label{sec:spectrum}

In the previous sections we established the invariance of the marginally deformed $\Ncal=2$ Lagrangian under the deformed $\SU(4)$ symmetry, i.e., we showed that the Lagrangian is a deformed $\SU(4)$ singlet. In this section we move on to other representations and explore what, if any, relevance the deformed $\SU(4)$ has for the spectrum of the theory. We will work at the one-loop level, although for the BPS cases that we consider we expect the extension to higher loops to be straightforward. We recall that in the planar context, the question of finding the spectrum of a conformal field theory can be translated to that of diagonalising an associated spin-chain Hamiltonian, see \cite{Beisert:2010jr} for a review.

The one-loop Hamiltonian for spin-chain states made up of the scalar fields of the $\Zset_k$ orbifold theory was derived in \cite{Gadde:2010zi}. For completeness, we reproduce it, in some relevant sectors, in Appendix \ref{HamiltonianAppendix}. It has interesting limits when specialised to closed subsectors. In particular, in the $\SU(2)$ subsector corresponding to the $X,Y$ fields it becomes an alternating Hamiltonian, while in the ``broken'' $\SU(2)$ subsector corresponding to the $X,Z$ or $Y,Z$ fields it is a ``dilute'' Temperley-Lieb-type Hamiltonian. The 1- and 2-magnon problems in these holomorphic sectors were explored in \cite{Pomoni:2021pbj} using a coordinate Bethe ansatz approach. 

In this section, our focus will instead be on what the hidden symmetries tell us about the spectrum of this Hamiltonian. Instead of arbitrary-length chains as in \cite{Pomoni:2021pbj}, we will work with short chains, and we will be interested in going beyond the holomorphic sector to understand states composed of all the scalar fields of the theory. As discussed, working with broken generators requires us to open up the gauge theory traces, and acting once with the broken generators on these states leads to non-closeable states. Due to gauge invariance, it is the closeable states that are related to the physical spectrum of the theory.
However, acting twice with broken generators on a closeable state will always give a state which is closeable. So to see how closeable states are connected using the broken generators, our approach will be to open them up with the same cyclic prescription as for the Lagrangian (which is of course a special case of a closed state), act twice with a broken generator, and then close the states again.

Working with open states introduces ambiguities in the Hamiltonian, as one can add boundary-type terms which vanish when closing the states. We will make use of this ambiguity to modify the ``naive'' open Hamiltonian in order to obtain some desirable features, such as preserving the number of BPS states when deforming away from the orbifold point. 

We clearly don't expect the additional symmetry to tell us all that much about the energy eigenvalues of the theory, since that is not even the case for the $\Ncal=4$ SYM with its unbroken $\SU(4)$ symmetry group. It is only after understanding the integrable structure of $\Ncal=4$ SYM, e.g., by extending to a Yangian symmetry and thus introducing a dependence on the spectral parameter, that one starts to obtain results about the spectrum, for instance through the algebraic or coordinate Bethe ansatz. What the $\SU(4)$ symmetry \emph{does} do is organise the states into multiplets which can each be obtained by acting with lowering operators on a highest-weight state. In the following we will take a few experimental steps towards establishing whether the naively broken $\SU(4)$ generators can be used to transform among states belonging to the multiplets of the $\kappa$-deformed Hamiltonian.

To see the differences between the $\Ncal=4$ SYM case and our current setting, it is perhaps useful to take an algebraic Bethe ansatz perspective, even though of course we don't currently have a spectral-parameter-dependent $R$-matrix. Recall that in this approach, the computation of the conserved charges hinges on the commutation of the transfer matrices for different values of the spectral parameter, $[t(u),t(u')]=0$. Expanding the first transfer matrix around $u=0$ one obtains the Hamiltonian, and expanding the second one around the "quantum plane limit" $u'\ra\infty$ one obtains the Lie algebra of $\SU(4)$ through the $RTT$ relations, with the $R$-matrix of course being the identity in this case. This is an elaborate way of stating that the Hamiltonian commutes with $\SU(4)$, so all the states in an $\SU(4)$ multiplet will have the same energy. In our $\Ncal=2$ setting, apart from the BPS states, we see splittings of the energy eigenvalues as we take $\kappa\neq 1$, both for the open and closed Hamiltonian. Hence the story will clearly be more complicated than that in the $\Ncal=4$ SYM case. However, at least for short multiplets, and similarly to what is referred to as ``dynamical symmetries" in \cite{Davies:1992sva}, our deformed $\SU(4)$ generators do seem to correctly take us between the states in the multiplet (with the correct $\kappa$-dependent coefficients), so it is likely that the deformed $\SU(4)$ still plays a relevant role. In this section we present some of our empirical findings and leave a fuller analysis for future investigations. 

We will start by considering states in holomorphic sectors, and then proceed to the non-holomorphic ones.

\subsection{Holomorphic BPS multiplets} \label{HolomorphicBPS}

The protected spectrum of the marginally deformed $\Zset_2$ orbifold $\Ncal=2$ SYM theory was studied in detail in \cite{Gadde:2010zi}. For holomorphic states, it was shown in that work that the parameter $\kappa$ enters the BPS states in a simple way related to the number of $Z_1$ and $Z_2$ fields. Effectively, up to an overall normalisation, the power of $\kappa$ entering a given monomial which is part of a BPS state is simply $\kappa^{\half(n(Z_1)-n(Z_2))}$ times the coefficient of that monomial at the orbifold point (recall that $\kappa=g_2/g_1$). So the BPS states, at any length, will be symmetrised monomials as usual, but now with these additional $\kappa$-dependent factors. We call this $\kappa$-symmetrisation.

  A simple way to understand these factors of $\kappa$ is to recall that states in the chiral ring are orthogonal to states that include $\p\Wcal/\p\Phi^i$ \cite{Cachazo:2002ry, Mauri:2006uw}. Compared with the orbifold-point theory, the marginally deformed relations (\ref{FDg}) are obtained by rescaling $Z_1$ and $Z_2$ by $g_1$ and $g_2$, respectively. Therefore, to preserve orthogonality, the $Z$ fields in the marginally deformed BPS states should scale by the inverse factors. For instance, for two sites in the $XZ$ sector we have
  \be
  \ket{F_{12}^Y}=g_2 X_{12}Z_2-g_1 Z_{1}X_{12} \;\; \Rightarrow \;\;\ket{\text{BPS}}=\frac{1}{g_2} X_{12}Z_2+\frac{1}{g_1} Z_1 X_{12}
  \ee
Generalising to all states which are symmetric at the orbifold point, we find that BPS states in the marginally deformed theory should scale as   
\be
g_1^{-n(Z_1)}g_2^{-n(Z_2)}=(g_1g_2)^{-\half(n(Z_1)+n(Z_2))}\times \kappa^{\half(n(Z_1)-n(Z_2))}
\ee
The first factor is an overall normalisation which will be the same in each sector with a fixed number of $Z$ fields (and can be dropped), while the second is the $\kappa$-symmetrisation factor.

As an example of how this works, consider a $L=4$ BPS state in the sector with two $Z$ and two $X$ fields. The $\kappa$-symmetrisation prescription tells us that up to overall normalisation, the state for the first index being in gauge group 1 is
{\small
\be
\kappa~ X_{12}X_{21}Z_1Z_1+\kappa^0~X_{12}Z_2X_{21}Z_1+\frac{1}{\kappa}~X_{12}Z_2Z_2X_{21}+\kappa~Z_1X_{12}X_{21}Z_1+\kappa^0~Z_1X_{12}Z_2X_{21}+\kappa~ Z_1Z_1X_{12}X_{21}
\ee
}
which of course can be verified by acting with the $XZ$ sector Hamiltonian and finding that it is indeed an eigenstate with eigenvalue 0.

Another way to express the above is to count, for each monomial, the number of fields with first index in gauge group 1 or 2, which we can call $n(1)$ and $n(2)$. Since any two subsequent $X$ fields (regardless of how many $Z$ fields happen to be between them) will not contribute to the difference $n(1)-n(2)$, it is easy to see that the above formula is equivalent to
\be \label{kappasym1}
\kappa^{\frac{n(1)-n(2)}2}
\ee
To check it for the above state, we see that the first gauge indices for each monomial are $(1,2,1,1),(1,2,2,1),(1,2,2,2),(1,1,2,1),(1,1,2,2)$ and $(1,1,1,2)$, so our formula reproduces the same $\kappa$ factors as above. 

Now consider our two-site dynamical twist (\ref{diagonaltwist}), which was chosen to reproduce the $XZ$ quantum plane. One notices that it also correctly reproduces the $XZ$ sector two-site BPS state if we take $\kappa\ra 1/\kappa$. If we trivially extend it to more sites and write
\be \label{BPSXZ}
\Fcal_{\text{BPS}}^{(L)}= \kappa^{s/2}\otimes \kappa^{s/2}\otimes \cdots \otimes \kappa^{s/2}\;,
\ee
it is straightforward to check that it reproduces the $\kappa$-symmetrisation formula (\ref{kappasym1}). We emphasise that this simple twist applies only to holomorphic BPS states, and the $L$-site extension for other representations would not be expected to take a diagonal form. 

At this stage, we don't have a proof of (\ref{BPSXZ}) by starting from a two-site twist, as that would likely require knowledge of a more universal form of the twist. However, if we assume that $\Delta(\kappa^s)=\1\otimes\1$, i.e. that the coproduct simply washes out the $\Zset_2$ generator $s$, then clearly writing
\be
\Fcal^{(3)}_{\text{BPS}}=(\Fcal_{\text{BPS}}\otimes \1)(\Delta_\op\otimes \id)(\Fcal_{\text{BPS}})=(\1\otimes \Fcal_{\text{BPS}})(\id\otimes \Delta_\op)(\Fcal_{\text{BPS}})
\ee
results in (\ref{BPSXZ}).

Given (\ref{BPSXZ}), we can now define an $L$-site coproduct for the $XZ$-sector BPS states as a twist of the orbifold-point coproduct (\ref{eq:OrbifoldPointDelta}) by
\begin{align} \label{XZcoprodL}
    \Delta^{(L)}_{\text{BPS},\kappa}(\R{a}{b}) &=  \Fcal_{\text{BPS}}^{(L)} \Delta_{\op}^{(L)} (\R{a}{b}) \left(\Fcal^{(L)}_{\text{BPS}}\right)^{-1} \nonumber \\
    &= \sum_{\ell=1}^{L} \left(\kappa^{\frac{s}{2}} \otimes \dots \otimes \kappa^{\frac{s}{2}} \right)  \left(\1\otimes\cdots\otimes \1\otimes\overset{\ell}{\R{a}{b}}\otimes \gamma \otimes\cdots\otimes \gamma \right) \left(\kappa^{-\frac{s}{2}} \otimes \dots \otimes \kappa^{-\frac{s}{2}} \right) \nonumber \\
    &= \sum_{\ell=1}^{L}  \left(\kappa^{\frac{s}{2}-\frac{s}{2}} \otimes \cdots\otimes \kappa^{\frac{s}{2}-\frac{s}{2}} \otimes \kappa^{\frac{s}{2}}\overset{\ell}{\R{a}{b}} \kappa^{-\frac{s}{2}}\otimes \kappa^{\frac{s}{2}}\gamma\kappa^{-\frac{s}{2}} \otimes\cdots\otimes \kappa^{\frac{s}{2}}\gamma\kappa^{-\frac{s}{2}} \right) \nonumber \\
    &= \sum_{\ell=1}^{L}  \left(\1\otimes \cdots\otimes \1 \otimes \overset{\ell}{\R{a}{b}} \otimes \gamma \kappa^{-s} \otimes\cdots\otimes \gamma\kappa^{-s} \right) \nonumber \\
    &= \sum_{\ell=1}^{L}  \left(\1\otimes \cdots\otimes \1 \otimes \overset{\ell}{\R{a}{b}} \otimes K_{\text{BPS}} \otimes\cdots\otimes K_{\text{BPS}} \right) \ .
\end{align}
where $\Rcal^{a}_{\;b}$ is either $\Rcal^{3}_{\;2}=\sigma^+_{XZ}$ or $\Rcal^{2}_{\;3}=\sigma^-_{XZ}$ and we have used that $s\gamma=-\gamma s$, see (\ref{sgamma}). We have defined 
\begin{equation}
    K_{\text{BPS}} = \gamma \ \kappa^{-s} \ .
\end{equation}
where we note that the power of $s$ is opposite to that for the quantum-plane coproduct (\ref{XZcoproduct}). 

\paragraph{XZ sector at three sites}

\mbox{}

As an example of how the coproduct works, let us consider the four $E=0$ eigenstates of the open Hamiltonian in the $XZ$ sector at $L=3$ sites, with the first index in the first gauge group:
\begin{align}
|s_1\rangle &= X_{12}X_{21}X_{12} \ , \nonumber \\
|s_2\rangle &= X_{12}X_{21}Z_{1} + \frac{1}{\kappa} X_{12}Z_{2}X_{21} + Z_{1}X_{12}X_{21} \ , \nonumber \\
|s_3\rangle &= \frac{1}{\kappa^2} X_{12}Z_{2}Z_{2} + \frac{1}{\kappa}Z_{1}X_{12}Z_{2} + Z_{1}Z_{1}X_{12} \ , \nonumber \\
|s_4\rangle &= Z_{1}Z_{1}Z_{1} \ , \label{eq:XZ3sitesBPS}
\end{align}
Acting with the three-site coproduct of the $XZ$-sector raising and lowering operators $\Rcal^3_{\;2}$ and $\Rcal^2_{\;3}$,  
\begin{equation}
    \Delta^{(3)}_{\text{BPS},\kappa}(\Rcal^a_{\;b}) = \1 \otimes \1 \otimes \R{a}{b} + \1 \otimes \R{a}{b} \otimes K_{\text{BPS}} + \R{a}{b} \otimes K_{\text{BPS}} \otimes K_{\text{BPS}} \ , \label{eq:XZ3siteDelta}
\end{equation}
we can confirm that they form an $\mathrm{SU}(2)$ multiplet.

\paragraph{XZ sector at four sites}

\mbox{}

A further example of a BPS multiplet, this time for $L=4$, is illustrated in Fig. \ref{Fig:SU(2)L=4BPS}. We emphasise that the states in these multiplets are related by the action of \emph{broken} raising and lowering operators. So from the usual perspective where only the $\SU(2)_L\times \SU(2)_R\times \Urm(1)_r$ symmetry group is present, the fact that they are in the same multiplet would appear accidental, while from our perspective the $XZ$-sector $\mathfrak{su}(2)$ generators are still present (albeit in a twisted groupoid sense) and can still be used to relate the states. 

\begin{figure}[!ht]
\begin{center}
\resizebox{0.85\textwidth}{!}{ 
\begin{tikzpicture}[node distance={15mm}, thick, main/.style = {draw}] 
\node (1) {$X_{12} X_{21} X_{12} X_{21}$}; 
\node (2) [below of=1] {$ X_{12} Z_2 X_{21} X_{12} + X_{12} X_{21} X_{12} Z_2  +\kappa \left(X_{12} X_{21} Z_1 X_{12} + Z_1 X_{12} X_{21} X_{12}\right) $};
\node (3) [below of=2] {$\frac{1}{\kappa} X_{12} Z_2 Z_2 X_{21} + Z_1 X_{12} Z_2 X_{21} + X_{12} Z_2 X_{21} Z_1 + \kappa Z_1 Z_1 X_{12} X_{21} + \kappa X_{12} X_{21} Z_1 Z_1 + \kappa Z_1 X_{12} X_{21} Z_1 $};
\node(4) [below of=3] {$\frac{1}{\kappa} X_{12} Z_2 Z_2 Z_2 + Z_1 X_{12} Z_2 Z_2 + \kappa Z_1 Z_1 X_{12} Z_2 +  \kappa^2 Z_1 Z_1 Z_1 X_{12} $};
\node(5) [below of=4] {$\kappa^2 Z_1 Z_1 Z_1 Z_1$};
\draw[->] (1) edge [bend left] node[midway, right of = 1] {$\Delta^{(4)}(\R{2}{3})$} (2);
\draw[->] (2) edge [bend left] node[midway, right of = 2] {$\Delta^{(4)}(\R{2}{3})$} (3);
\draw[->] (3) edge [bend left] node[midway, right of = 3] {$\Delta^{(4)}(\R{2}{3})$} (4);
\draw[->] (4) edge [bend left] node[midway, right of = 4] {$\Delta^{(4)}(\R{2}{3})$} (5);
\draw[->] (2) edge [bend left] node[midway, left of = 2] {$\Delta^{(4)}(\R{3}{2})$} (1);
\draw[->] (3) edge [bend left] node[midway, left of = 3] {$\Delta^{(4)}(\R{3}{2})$} (2);
\draw[->] (4) edge [bend left] node[midway, left of = 4] {$\Delta^{(4)}(\R{3}{2})$} (3);
\draw[->] (5) edge [bend left] node[midway, left of = 5] {$\Delta^{(4)}(\R{3}{2})$} (4);
\end{tikzpicture}
}
\caption{The four-site BPS multiplet in the $XZ$ sector. The action of the broken generators defined through the coproduct (\ref{XZcoprodL}) correctly relates all the states in the multiplet.}
\label{Fig:SU(2)L=4BPS}
\end{center}
\end{figure}

For the closeable states in the multiplet, we can reverse the opening-up procedure by adding their $\Zset_2$ conjugates and cyclically identifying the states. For the $L=4$ multiplet in Fig. \ref{Fig:SU(2)L=4BPS}, this leads to the closed states\footnote{To illustrate the closing procedure, here we only considered adding the open states with their $\Zset_2$ conjugates so as to obtain $\Zset_2$-even closed states, which belong to the untwisted sector of the theory. We could of course have combined them in a $\Zset_2$-odd way, resulting in states in the twisted sector.}
\be
\begin{split}
  &\tr_1(X_{12}X_{21}X_{12}X_{21}) \;\;,\;\;\tr_1(Z_1Z_1Z_1Z_1) \;\;,\;\;\;\; \tr_2(Z_2Z_2Z_2Z_2)\;\; \text{and} \\
  &\;\;\tr_1\left(\kappa~ X_{12}X_{21}Z_1Z_1+X_{12}Z_2X_{21}Z_1+\kappa^{-1}~ X_{12}Z_2Z_2 X_{21}\right)\;.
\end{split}
\ee
These are all $E=0$ eigenstates of the closed Hamiltonian. So, as claimed, defining the action of the broken generators through the opening-up procedure and acting an even number of times on a closed state, correctly reproduces the states belonging to the physical spectrum of the theory. 

\paragraph{XYZ sector at three sites}

\mbox{}

The same twist (\ref{BPSXZ}) also acts correctly on BPS states in the full $\SU(3)$ sector, for any length, as can be argued by requiring orthogonality to states including all the holomorphic quantum planes in (\ref{FDg}). Twisting the coproduct of the $YZ$-sector generators $\Rcal^4_{\;2}$ and $\Rcal^2_{\;4}$ leads to the same form of the coproduct as (\ref{XZcoprodL}). Since the orbifold-point coproduct of the unbroken $XY$-sector operators $\Rcal^4_{\;3}$ and $\Rcal^3_{\;4}$ does not contain $\gamma$'s, the twist has no effect. So we can summarise the $L$-site coproduct for the holomorphic $XYZ$ sector as 
\be
\Delta^{(L)}_{\text{BPS},\kappa}(\R{a}{b}) =  
    \sum_{\ell=1}^{L}  \left(\1\otimes \cdots\otimes \1 \otimes \overset{\ell}{\R{a}{b}} \otimes (K_{\text{BPS}})^{a}_{\;b} \otimes\cdots\otimes (K_\text{BPS})^a_{\;b} \right) \ ,
\ee
where we define
\be
(K_{\text{BPS}})^a_{\;b}=\left\{\begin{array}{ll} \gamma \kappa^{-s} &,\; \text{if } \Rcal^a_{\;b} \;\text{is broken}\\  \1 &,\; \text{if } \Rcal^a_{\;b} \;\text{is unbroken}\end{array}\right. \ .
\ee
This coproduct consistently relates any open BPS states in the holomorphic $\SU(3)$ sector. As an example, we can check that
\be
\begin{split}
  &\sigma^-_{XY}\sigma^-_{XZ} (\kappa^\half\ket{X_{12}X_{21}X_{12}})=\sigma^+_{XY}\sigma^+_{YZ} ({\kappa}^\half\ket{Y_{12}Y_{21}Y_{12}})=\sigma^+_{XZ}\sigma^-_{YZ}(\kappa \ket{Z_1Z_1Z_1})\\
  &=\kappa^{\half}~(X_{12}Y_{21}Z_1+Y_{12}X_{21}Z_1+Z_1X_{12}Y_{21}+Z_1Y_{12}X_{21})+\kappa^{-\half}(X_{12}Z_2Y_{21}+Y_{12}Z_2X_{21}) \ ,
\end{split}
\ee
where we used the same convention for the generators as in Appendix \ref{SU4Appendix}.
Here, the normalisations of our initial states are those provided by the twist (\ref{BPSXZ}). Note that one can obtain the final closeable state either by acting on non-closeable states with one broken and one unbroken generator, or on the closeable state $\ket{Z_1Z_1Z_1}$ with two broken generators. One can confirm that the final state is an $E=0$ eigenstate of the open Hamiltonian, and closing by adding the $\Zset_2$ conjugate and identifying cyclically related states one obtains
\be
\tr_1\left(\kappa^{\half}(X_{12}Y_{21}Z_1+Y_{12}X_{21}Z_1)+\kappa^{-\half}(Y_{12}Z_2X_{21}+X_{12}Z_2Y_{21})\right)\;,
\ee
which is indeed an $E=0$ eigenstate of the closed Hamiltonian. 

\subsection{Full $\SU(4)$ multiplets at two sites}

When attempting to  extend the above analysis of BPS states in the holomorphic sector to encompass states in more general representations, as well as the full $\SU(4)$, 
our inability to define the twists in a more universal form is currently a limitation of our approach. However, since we do have a full set of twists at two sites, in this subsection we will use them to define twisted multiplets in the full $\SU(4)$, and compare with the one-loop spectrum of the Hamiltonian.  

\stoptoc
  \subsubsection{The $\mathbf{20'}$ two-site multiplet} \label{20prime}

  We first consider the full BPS multiplet at two sites, which corresponds to the representation $\mathbf{20'}$ in the decomposition (\ref{eq:SO62siteDecomposition}). Here we encounter a slight subtlety, introduced by our need to open up the states and act with the open Hamiltonian. Let us restrict to the subsector of two-site states which are $\mathrm{SU}(2)_L \times \mathrm{SU}(2)_R \times \mathrm{U}(1)_r$  singlets, which we order as 
\begin{equation}
 \{\bar{Z}_1 Z_1, Z_1 \bar{Z}_1, \mathcal{M}_1 , \bar{Z}_2 Z_2, Z_2 \bar{Z}_2, \mathcal{M}_2 \} \ ,
\end{equation}
with 
\begin{align}
\mathcal{M}_1 &= \frac{1}{2} \left( X_{12} \bar{X}_{21} + \bar{X}_{12} X_{21} + Y_{12} \bar{Y}_{21} + \bar{Y}_{12} Y_{21} \right) ,
\end{align}
and $\mathcal{M}_2$ its $\Zset_2$ conjugate.
In this basis, the deformed Hamiltonian given in (\ref{Hneutral}) takes the form 
\begin{equation}
\mathcal{H}_{singlet} = \left(
\begin{array}{cccccc}
 \frac{3}{2\kappa} & -\frac{1}{2\kappa} & \frac{1}{\kappa} & 0 & 0 & 0 \\
 -\frac{1}{2\kappa} & \frac{3}{2\kappa} & \frac{1}{\kappa} & 0 & 0 & 0 \\
 \frac{1}{\kappa} & \frac{1}{\kappa} & 2 \kappa & 0 & 0 & 0 \\
 0 & 0 & 0 & \frac{3\kappa}{2} & -\frac{\kappa}{2} & \kappa \\
 0 & 0 & 0 & -\frac{\kappa}{2} & \frac{3\kappa}{2} & \kappa \\
 0 & 0 & 0 & \kappa & \kappa & \frac{2}{\kappa} \\
\end{array}
\right) \ .
\label{eq:Hsinglet}
\end{equation}
After diagonalising, we find that the state corresponding to the  $\mathrm{SU}(2)_L \times \mathrm{SU}(2)_R \times \mathrm{U}(1)_r$ singlet with $E=0$ at the orbifold point, acquires a \emph{negative eigenvalue} for $0<\kappa<1$:\footnote{We have of course broken the $\Zset_2$ symmetry by considering states with first index in gauge group 1, for the case with first gauge group 2 the corresponding eigenvalue will be negative for $\kappa>1$.}
\begin{equation}
    E(\ket{{\bf (1,1)}_0}) = \frac{1}{2\kappa} + \kappa - \frac{\sqrt{4\kappa^2 - 4 + 9 \kappa^{-2}}}{2} \ .
\end{equation}
This is clearly an artifact of working with the open Hamiltonian, since the corresponding eigenstate of the closed Hamiltonian \emph{does} have $E=0$, in accordance with expectations that the number of BPS states should not change as we deform away from the orbifold point (see \cite{Gadde:2009dj} for a detailed discussion), and in any case we would definitely not expect any states to have negative anomalous dimensions.

Fortunately, it is possible to cure this problem of negative eigenvalue, by adding to the Hamiltonian (\ref{eq:Hsinglet}) a term which ($i$) vanishes at the orbifold point and ($ii$) does not modify the closed chain action of the Hamiltonian. We find
\begin{align}
\delta\mathcal{H}= 2 \ \left(
\begin{array}{cccccc}
 0 & 0 & 0 & 0 & 0 & 0 \\
 0 & 0 & 0 & 0 & 0 & 0 \\
 0 & 0 & \kappa^{-1}-\kappa & 0 & 0 & 0 \\
 0 & 0 & 0 & 0 & 0 & 0 \\
 0 & 0 & 0 & 0 & 0 & 0 \\
 0 & 0 & 0 & 0 & 0 & \kappa - \kappa^{-1} \\
\end{array}
\right) \ ,
\end{align}
and adding this term to the deformed Hamiltonian in (\ref{eq:Hsinglet}) gives an improved open Hamiltonian in the singlet sector 
\begin{equation}
\hat{\mathcal{H}}_{\text{singlet}} = \mathcal{H}_{\text{singlet}} + \delta \mathcal{H} = \left(
\begin{array}{cccccc}
 \frac{3}{2\kappa} & -\frac{1}{2\kappa} & \frac{1}{\kappa} & 0 & 0 & 0 \\
 -\frac{1}{2\kappa} & \frac{3}{2\kappa} & \frac{1}{\kappa} & 0 & 0 & 0 \\
 \frac{1}{\kappa} & \frac{1}{\kappa} & \frac{2}{\kappa} & 0 & 0 & 0 \\
 0 & 0 & 0 & \frac{3\kappa}{2} & -\frac{\kappa}{2} & \kappa \\
 0 & 0 & 0 & -\frac{\kappa}{2} & \frac{3\kappa}{2} & \kappa \\
 0 & 0 & 0 & \kappa & \kappa & 2 \kappa \\
\end{array}
\right) \ .
\label{eq:HSingletmodified}
\end{equation}
We emphasise that this modification in no way affects the physical closed-chain spectrum of the theory. 
Apart from the $\ket{{\bf(1,1)}_0}$ state in the $\mathbf{20'}$, it also affects the full $\mathrm{SO}(6)$ singlet $\mathbf{1}$, which we will look at in Section (\ref{SO6Singlet2sites}).
Having regained our BPS state in the singlet sector, we can combine it with the remaining $E=0$ states of the open deformed $\SU(4)$ Hamiltonian, to form the $\kappa$-deformed version of the $\mathbf{20'}$:
\be\begin{split}
\ket{(\mathbf{1},\mathbf{1})_2} &= \bar{Z}_{i}\bar{Z}_{i}  \\
\ket{(\mathbf{1},\mathbf{1})_{-2}} &= Z_{i}Z_{i}  \\
\ket{(\mathbf{1},\mathbf{1})_0}&=X_{i}\bar{X}_{i+1}+\bar{X}_{i}X_{i-1}+Y_{i}\bar{Y}_{i-1}+\bar{Y}_{i}Y_{i+1}-2\left(Z_{i}\bar{Z}_{i}+\bar{Z}_{i}Z_{i}\right) \\
\ket{(\mathbf{2},\mathbf{2})_1}&= X_{i}\bar{Z}_{i+1}+\kappa^{(-1)^{i+1}}\bar{Z}_{i}X_{i}  \\
\ket{(\mathbf{2},\mathbf{2})_{-1}} &= X_{i}Z_{i+1}+\kappa^{(-1)^{i+1}} Z_{i}X_{i}  \\
\ket{(\mathbf{3},\mathbf{3})_0} &= X_{i}X_{i+1} \ .
\end{split}\ee
Here the states are labelled by their 
$(\mathrm{SU}(2)_L, \mathrm{SU}(2)_R)_{\mathrm{U}(1)_r}$ quantum numbers. For readers familiar with the labelling in  \cite{Dolan:2002zh}, the conversion can be found in table \ref{tab:ConversionDolan20prime}.

\begin{table}[!ht]
    \centering
    \begin{tabular}{|l c c|}
    \hline
       & & \begin{footnotesize}
           primary of
       \end{footnotesize} \\
       \hline
       $(\mathbf{1},\mathbf{1})_2$ & $|0,0,+2\rangle$ & $\mathcal{E}_{2(0,0)}$ \\ 
        $(\mathbf{1},\mathbf{1})_{-2}$ & $|0,0,-2\rangle$ & $\mathcal{\bar{E}}_{-2(0,0)}$ \\ 
        $(\mathbf{1},\mathbf{1})_0$ & $|0,0,0\rangle$ & $\mathcal{\hat{C}}_{0(0,0)}$ \\ 
        $(\mathbf{2},\mathbf{2})_1$ & $|\pm\frac{1}{2},\pm\frac{1}{2},+1\rangle$ & $\mathcal{D}^{(\pm\frac{1}{2})}_{\frac{1}{2}(0,0)}$ \\
        $(\mathbf{2},\mathbf{2})_{-1}$ & $|\pm\frac{1}{2},\pm\frac{1}{2},-1\rangle$ & $\mathcal{\bar{D}}^{(\pm\frac{1}{2})}_{-\frac{1}{2}(0,0)}$ \\ 
        $(\mathbf{3},\mathbf{3})_0$ & $|\pm 1, \pm 1, 0\rangle, |0,0,0\rangle$ & $\mathcal{B}_{1}$ \\
        \hline
    \end{tabular}
    \caption{Conversion table between the notation used in \cite{Dolan:2002zh} and the representations of the unbroken R-symmetry group $(\mathrm{SU}(2)_L, \mathrm{SU}(2)_R)_{\mathrm{U}(1)_r}$ for each multiplet  in the $\mathbf{20'}$.}
    \label{tab:ConversionDolan20prime}
\end{table}

\begin{figure}[h!]
\begin{center}
\begin{tikzpicture}[scale=0.61]

\draw [darkgreen,stealth-stealth](-0.4,2.2) -- (0.4,2.2);
\draw [darkgreen,stealth-stealth](-2.5,0) -- (-1.6,0);
\draw [darkgreen,stealth-stealth](1.6,0) -- (2.5,0);
\draw [darkgreen,stealth-stealth](-0.4,-2.2) -- (0.4,-2.2);
\draw [cyan,stealth-stealth](-0.5,3.5) -- (-1.1,2.9);
\draw [cyan,stealth-stealth](-2.55,1.45) -- (-3.15,0.85);
\draw [cyan,stealth-stealth](-2.55,-1.45) -- (-3.15,-0.85);
\draw [cyan,stealth-stealth](-0.5,-3.5) -- (-1.1,-2.9);
\draw [cyan,stealth-stealth](0.6,-3.5) -- (1.2,-2.9);
\draw [cyan,stealth-stealth](2.65,-1.45) -- (3.25,-0.85);
\draw [cyan,stealth-stealth](2.65,1.45) -- (3.25,0.85);
\draw [cyan,stealth-stealth](0.6,3.5) -- (1.2,2.9);
\draw [cyan,stealth-stealth](-1.3,-1.5) -- (-0.7,-0.9);
\draw [cyan,stealth-stealth](1.05,0.85) -- (1.65,1.45);
\draw [cyan,stealth-stealth](-1.45,1.45) -- (-0.85,0.85);
\draw [cyan,stealth-stealth](1.45,-1.45) -- (0.85,-0.85);

\draw (0,0) node (M02) {{\footnotesize{$
\begin{matrix}
\ket{0,-1,0}\\
2\times\ket{0,0,0}\\
\ket{0,+1,0}
\end{matrix}$}}};
\draw [BurntOrange,stealth-stealth](-1.4,-0.75) -- (-1.4,0.75);

\draw (-4.0,0) node (M+2) {\footnotesize{$
\begin{matrix}
\ket{-1,-1,0}\\
\ket{-1,0,0}\\
\ket{-1,+1,0}
\end{matrix}$}};
\draw [BurntOrange,stealth-stealth](-5.4,-0.75) -- (-5.4,0.75);

\draw (4.0,0) node (M-2) {\footnotesize{$
\begin{matrix}
\ket{+1,-1,0}\\
\ket{+1,0,0}\\
\ket{+1,+1,0}
\end{matrix}$}};
\draw [BurntOrange,stealth-stealth](5.4,-0.75) -- (5.4,0.75);

\draw (0,4) node (E2) {\footnotesize{$\ket{0,0,+2}$}};
\draw (0,-4) node (Ebar2) {\footnotesize{$\ket{0,0,-2}$}};

\draw (-2,2.2) node (Dhat1) {\footnotesize{$
\begin{matrix}
\ket{-\frac12,-\frac12,+1}\\
\ket{-\frac12,+\frac12,+1}
\end{matrix}$}};
\draw [BurntOrange,stealth-stealth](-3.7,1.7) -- (-3.7,2.65);

\draw (2,2.2) node (D1) {\footnotesize{$
\begin{matrix}
\ket{+\frac12,-\frac12,+1}\\
\ket{+\frac12,+\frac12,+1}
\end{matrix}$}};
\draw [BurntOrange,stealth-stealth](3.75,1.7) -- (3.75,2.65);

\draw (-2,-2.2) node (Dbar1) {\footnotesize{$
\begin{matrix}
\ket{-\frac12,-\frac12,-1}\\
\ket{-\frac12,+\frac12,-1}
\end{matrix}$}};
\draw [BurntOrange,stealth-stealth](-3.7,-1.75) -- (-3.7,-2.7);

\draw (2,-2.2) node (Dbarhat1) {\footnotesize{$
\begin{matrix}
\ket{+\frac12,-\frac12,-1}\\
\ket{+\frac12,+\frac12,-1}
\end{matrix}$}};
\draw [BurntOrange,stealth-stealth](3.75,-1.75) -- (3.75,-2.7);

\draw [|-|](-4,5.5) -- node[label= above:$\mathrm{SU}(2)_L$] {} (4,5.5);
\foreach \x in {1,1/2,0,-1/2,-1} \draw (4*\x , 5.6) -- (4*\x , 5.4) node[anchor=north] {$\x$};

\draw (-7.4,0) node (my) {$\mathrm{U}(1)_r$};
\draw [|-|](-5.75,-4) -- (-5.75,4) node[label= above:] {};
\foreach \y in {-2,-1,0,1,2} \draw (-5.65 , 2*\y) -- (-5.85 , 2*\y) node[anchor=east] {$\y$};

\end{tikzpicture}
\caption{Depiction of the open $\mathbf{20'}$ multiplet, with the action of the $\textcolor{cyan}{broken}$ R-symmetry generators as dotted blue arrows and the unbroken \textcolor{darkgreen}{$\mathrm{SU}(2)_L$} as solid green arrows. The states present at each node of this diagram are connected via the action of the unbroken \textcolor{BurntOrange}{$\mathrm{SU}(2)_R$}.
}
\label{fig:multiplet20prime}
\end{center}
\end{figure}
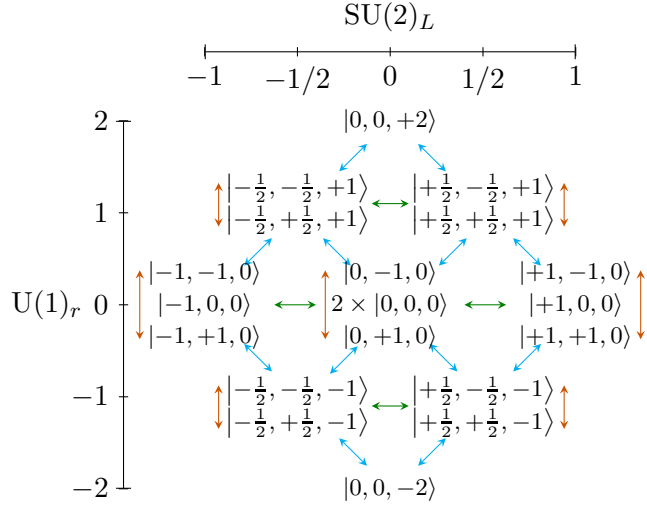

We can now ask whether this deformed $\mathbf{20'}$ multiplet is compatible with the deformed $\SU(4)$ symmetry. Using the two-site twists of Section \ref{TwistsSection}, but with $\kappa\ra 1/\kappa$ as was done to obtain $\Fcal^{(2)}_{\text{BPS}}$ in the $XZ$ sector, we can define two-site twists $\Fcal_{\text{BPS}}^{(2)}$ for the general $\SU(4)$ sector, and define a twisted coproduct in the usual way
\be
\Delta^{(2)}_{\text{BPS},\kappa}(\Rcal^a_{\;b})=\Fcal_{\text{BPS}}^{(2)}~ \Delta_{\op}(\Rcal^a_{\;b})~ (\Fcal_{\text{BPS}}^{(2)})^{-1} \ .
\ee
If $\Rcal^a_{\;b}$ are unbroken generators, this coproduct reduces to the usual Lie algebraic coproduct. A simple computation confirms that this coproduct also works for the broken generators, that is, it correctly relates states in the $\mathbf{20'}$ which would be related by these generators in the unbroken $\SU(4)$ case. This is illustrated in Fig. \ref{fig:multiplet20prime}. 

 \subsubsection{The {\bf 15} two-site multiplet}

Now let us look at the $\mathbf{15}$, the antisymmetric multiplet appearing in the $\mathrm{SO}(6)$ decomposition at two sites. Among other states, this multiplet contains the various unbroken and broken $\SU(2)$ singlet states which formed the basis for our quantum planes (see Section \ref{sec:Extension}). So in a sense our twists were chosen such that one obtains this multiplet by acting on the corresponding multiplet at the orbifold point, and inversely, by construction the inverses of the twists will take the deformed $\mathbf{15}$ to the orbifold point. We can confirm this by finding the corresponding $\kappa$-deformed eigenstates of the two-site $\mathrm{SO}(6)$ Hamiltonian 
\be\begin{split}
\ket{(\mathbf{1},\mathbf{1})_0} &= \kappa^{\frac{(-1)^{i}}{2} }\left(\bar{Z}_{i}Z_{i}-Z_{i}\bar{Z}_{i}\right) \\
\ket{(\mathbf{1},\mathbf{3})_0} &= \kappa^{\frac{(-1)^{i}}{2} }\left(X_{i}Y_{i+1}-Y_{i}X_{i-1}\right)  \\
\ket{(\mathbf{3},\mathbf{1})_0} &= \kappa^{\frac{(-1)^{i}}{2} }\left(X_{i}\bar{Y}_{i+1}-\bar{Y}_{i}X_{i+1}\right)  \\
\ket{(\mathbf{2},\mathbf{2})_1} &= X_{i}\bar{Z}_{i+1}-\kappa^{(-1)^{i}}\bar{Z}_{i}X_{i} \\
\ket{(\mathbf{2},\mathbf{2})_{-1}} &= X_{i}Z_{i+1}-\kappa^{(-1)^{i}} Z_{i}X_{i} \ ,
\end{split}\ee
where we only list the highest-weight state in each representation. Table \ref{tab:Conversion15} indicates the conversion from the $(\mathrm{SU}(2)_L, \mathrm{SU}(2)_R)_{\mathrm{U}(1)_r}$ quantum numbers used here to the notation of \cite{Dolan:2002zh}.

\begin{table}[!ht]
    \centering
    \begin{tabular}{|l c c|}
     \hline
        & & \begin{footnotesize}
           primary of
       \end{footnotesize} \\
       \hline
       $(\mathbf{1},\mathbf{1})_0$ & $|0,0,0\rangle$ & $\mathfrak{E}_{0(0,0)}$ \\ 
        $(\mathbf{1},\mathbf{3})_0$ & $|0, \pm 1, 0\rangle, |0,0,0\rangle$ & $\mathfrak{B}_{1}$ \\ 
        $(\mathbf{3},\mathbf{1})_0$ & $|\pm 1, 0, 0\rangle, |0,0,0\rangle$ & $\mathfrak{\hat{B}}_{1}$ \\ 
        $(\mathbf{2},\mathbf{2})_1$ & $|\pm\frac{1}{2}, \pm\frac{1}{2}, +1\rangle$ & $\mathfrak{D}^{(\pm\frac{1}{2})}_{\frac{1}{2}(0,0)}$ \\ 
        $(\mathbf{2},\mathbf{2})_{-1}$ & $|\pm\frac{1}{2}, \pm\frac{1}{2}, -1\rangle$ & $\mathfrak{\bar{D}}^{(\pm\frac{1}{2})}_{-\frac{1}{2}(0,0)}$ \\ 
        \hline
    \end{tabular}
    \caption{Conversion table to the notation in \cite{Dolan:2002zh} for representations of the unbroken R-symmetry group $(\mathrm{SU}(2)_L, \mathrm{SU}(2)_R)_{\mathrm{U}(1)_r}$ for each multiplet in the $\mathbf{15}$.}
    \label{tab:Conversion15}
\end{table}

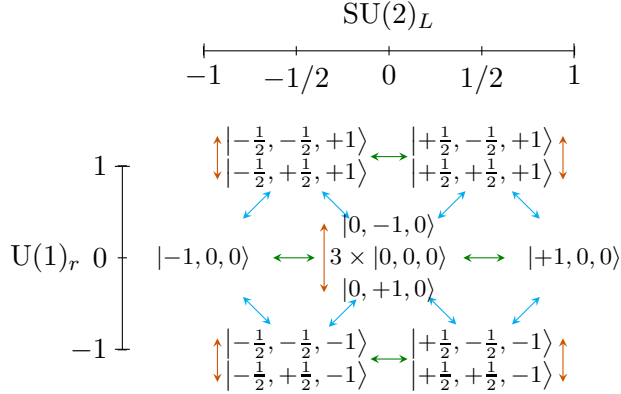
\begin{figure}[h!]
\begin{center}
\begin{tikzpicture}[scale=0.61]

\draw [darkgreen,stealth-stealth](-0.4,2.2) -- (0.4,2.2);
\draw [darkgreen,stealth-stealth](-2.5,0) -- (-1.6,0);
\draw [darkgreen,stealth-stealth](1.6,0) -- (2.5,0);
\draw [darkgreen,stealth-stealth](-0.4,-2.2) -- (0.4,-2.2);
\draw [cyan,stealth-stealth](-2.55,1.45) -- (-3.15,0.85);
\draw [cyan,stealth-stealth](-2.55,-1.45) -- (-3.15,-0.85);
\draw [cyan,stealth-stealth](2.65,-1.45) -- (3.25,-0.85);
\draw [cyan,stealth-stealth](2.65,1.45) -- (3.25,0.85);
\draw [cyan,stealth-stealth](-1.3,-1.5) -- (-0.7,-0.9);
\draw [cyan,stealth-stealth](1.05,0.85) -- (1.65,1.45);
\draw [cyan,stealth-stealth](-1.45,1.45) -- (-0.85,0.85);
\draw [cyan,stealth-stealth](1.45,-1.45) -- (0.85,-0.85);

\draw (0,0) node (M02) {{\footnotesize{$
\begin{matrix}
\ket{0,-1,0}\\
3\times\ket{0,0,0}\\
\ket{0,+1,0}
\end{matrix}$}}};
\draw [BurntOrange,stealth-stealth](-1.4,-0.75) -- (-1.4,0.75);

\draw (-4.0,0) node (M+2) {\footnotesize{$
\begin{matrix}
\ket{-1,0,0}
\end{matrix}$}};

\draw (4.0,0) node (M-2) {\footnotesize{$
\begin{matrix}
\ket{+1,0,0}
\end{matrix}$}};

\draw (-2,2.2) node (Dhat1) {\footnotesize{$
\begin{matrix}
\ket{-\frac12,-\frac12,+1}\\
\ket{-\frac12,+\frac12,+1}
\end{matrix}$}};
\draw [BurntOrange,stealth-stealth](-3.7,1.7) -- (-3.7,2.65);

\draw (2,2.2) node (D1) {\footnotesize{$
\begin{matrix}
\ket{+\frac12,-\frac12,+1}\\
\ket{+\frac12,+\frac12,+1}
\end{matrix}$}};
\draw [BurntOrange,stealth-stealth](3.75,1.7) -- (3.75,2.65);

\draw (-2,-2.2) node (Dbar1) {\footnotesize{$
\begin{matrix}
\ket{-\frac12,-\frac12,-1}\\
\ket{-\frac12,+\frac12,-1}
\end{matrix}$}};
\draw [BurntOrange,stealth-stealth](-3.7,-1.75) -- (-3.7,-2.7);

\draw (2,-2.2) node (Dbarhat1) {\footnotesize{$
\begin{matrix}
\ket{+\frac12,-\frac12,-1}\\
\ket{+\frac12,+\frac12,-1}
\end{matrix}$}};
\draw [BurntOrange,stealth-stealth](3.75,-1.75) -- (3.75,-2.7);

\draw [|-|](-4,4.5) -- node[label= above:$\mathrm{SU}(2)_L$] {} (4,4.5);
\foreach \x in {1,1/2,0,-1/2,-1} \draw (4*\x , 4.6) -- (4*\x , 4.4) node[anchor=north] {$\x$};

\draw (-7.4,0) node (my) {$\mathrm{U}(1)_r$};
\draw [|-|](-5.75,-2) -- (-5.75,2) node[label= above:] {};
\foreach \y in {-1,0,1} \draw (-5.65 , 2*\y) -- (-5.85 , 2*\y) node[anchor=east] {$\y$};

\end{tikzpicture}
\caption{The open $\mathbf{15}$ multiplet, with the action of the $\textcolor{cyan}{broken}$ R-symmetry generators shown as dotted blue arrows and the unbroken \textcolor{darkgreen}{$\mathrm{SU}(2)_L$} as solid green arrows. The states present at each node of this diagram are connected via the action of the unbroken \textcolor{BurntOrange}{$\mathrm{SU}(2)_R$}.
}
\label{multiplet15}
\end{center}
\end{figure}

These states do not all have the same energy, however one can check that indeed the two-site coproduct obtained from the twists in Section \ref{TwistsSection} (of course without taking $\kappa\ra 1/\kappa$), correctly relates all the states in the multiplet. The action of the $\SU(4)$ generators is depicted in Fig. \ref{multiplet15}.  

 \subsubsection{The singlet two-site multiplet}
 \label{SO6Singlet2sites}
 
The last multiplet we need to consider at two sites for the full $\mathrm{SO}(6)$ sector is the singlet $\mathbf{1}$, which in our conventions has $E=3$ at the orbifold point. For the naive open Hamiltonian (\ref{eq:Hsinglet}), this state mixes with the BPS state that is the superconformal primary of the $\ket{{\bf(1,1)}_0}$ multiplet, and is $\kappa$-dependent. The modification of the open Hamiltonian in the singlet sector, which gave us (\ref{eq:HSingletmodified}), resolves this mixing and gives the state
\begin{equation}
|\mathbf{1}\rangle = X_{12}\bar{X}_{21}+\bar{X}_{12} X_{21}+Y_{12} \bar{Y}_{21} +\bar{Y}_{12} Y_{21}+Z_{1} \bar{Z}_1 +\bar{Z}_1 Z_1 \ ,
\end{equation}
with eigenvalue $3/\kappa$, as well as its $\Zset_2$ conjugate with eigenvalue $3\kappa$. Although the eigenvalues do become $\kappa$-dependent, the state itself is the same as at the orbifold point.

Untwisting this state using the two-site twists has no effect, and correspondingly the two-site coproduct will annihilate this state for all the generators $\Rcal^a_{\;b}$. Of course, this is by construction, as (ignoring the $e^V$ factors which do not carry $\SU(4)$ weight) this term is the opened kinetic term in the Lagrangian and our twists were defined such that they leave this term invariant. 

\resumetoc

\section{Conclusions}

In this work we have taken a new perspective on the symmetries of four-dimensional quiver SCFT's with $\Ncal=2$ supersymmetry. Firstly, at the orbifold point of the theory, we recovered the naively broken R-symmetry generators by extending our notion of symmetry from a group to a groupoid. We then used the F- and D-terms of the theory, as well as the unbroken symmetries, to define twist operations which take us away from the orbifold point to obtain the marginally deformed theories. Inverting those twists, we were able to show that the naively broken $\SU(4)$ generators persist in the marginally deformed theories as well, although their action is no longer coassociative, which considerably complicates their study. We concluded with several checks of the new generators by using them to relate states in the physical spectrum of the one-loop Hamiltonian.

Our construction is not free of ambiguities and educated guesses, and further work is needed to fully justify all the steps that we followed. It is possible that alternative twists satisfying a shifted cocycle condition can be found, which would make the non-associativity of dynamical type rather than general quasi-Hopf, and thus greatly facilitate (for instance) the extension of our coproducts from two to more sites and eventually lead to a more rigorous proof of our four-site twists (\ref{eq:F4SB}) and (\ref{eq:F4NSB}). Nevertheless, we hope to have convinced the reader that the quiver SCFT's do have additional useful symmetries which are not visible from a strictly Lie group perspective.

At the orbifold point, our revived $\SU(4)$ symmetry could be an important element in obtaining a fuller understanding of the integrability of this theory. Recall that although a twisted Bethe ansatz for the $\Zset_k$ orbifolds was proposed in \cite{Beisert:2005he}, it has not so far been derived via a Yangian-type symmetry at the level of the Lagrangian. We expect that our new framework will be of help in figuring out such a structure along the lines of \cite{Beisert:2018zxs, Beisert:2017pnr}.

For concreteness, we focused on the $\Zset_2$ orbifold theory, but it is straightforward to extend our construction to the $\Zset_k$ case (and eventually to more general ADE $\mathcal{N}=2$ orbifolds as well as $\mathcal{N}=1$ orbifolds). To obtain the orbifold-point coproduct for the $\Zset_k$ case, one simply replaces the $\gamma$ operator, which in our case satisfies $\gamma^2=1$, by an operator satisfying $\gamma^k=1$ and acts on indices of gauge node $i$ as $\gamma:i\rightarrow i\!+\!1$. The coproduct of the raising operators (\ref{RRaising}) will contain $\gamma$, while that of the lowering ones (\ref{Rlowering}) will contain $\gamma^{-1}$. The twists which marginally deform away from the orbifold point are going to be similar, apart from the presence of more deformation parameters $\kappa_i=g_{i+1}/g_i$. The only real difference for $k>2$ is the absence of the $\SU(2)_L$ symmetry, which however did not play any role in deriving our twists capturing the marginal deformation. We intend to analyse more general orbifolds in detail in future work.

Moreover, an important observation made in Section \ref{HolomorphicBPS} 
is that holomorphic BPS states in the chiral ring are orthogonal to states that include F-terms ($\p\Wcal/\p\Phi^i$) \cite{Cachazo:2002ry, Mauri:2006uw}. To preserve this orthogonality when twisting, the $Z$ fields in BPS states should scale by the inverse factors with respect to states including F-terms. For two sites this translates to the fact that the twists of the singlet and the triplet representations should be related by $\kappa \leftrightarrow 1/ \kappa$. This relation should further generalise to more sites and thus constrain the form of the twist.
 This idea can be simply generalised to any marginal deformation of $\mathcal{N}=1,2$ superconformal orbifolds.

For simplicity, in this paper, we focused our attention on the bosonic subsector. Our treatment can be easily generalised to include fermions and derivatives for the complete Hilbert space following
\cite{Liendo:2011xb}. This is work in progress.

The main future challenge is to understand whether some integrable structures persist when deforming away from the orbifold point. We do not yet have an answer to this question. However, as always, a better understanding of the complete symmetries of the problem is likely to be crucial in this endeavour. 

In this paper, we concentrated our attention on short spin chains and worked in position space, which is the most usual language in which the algebraic structures are naturally presented.
We were able to obtain the twist only for two different limiting values of the rapidity, corresponding to fully symmetrised and antisymmetrised representations, since, as we saw, the twists for the BPS states are obtained from those at the quantum plane limit by $\kappa\ra 1/\kappa$, which we interpret as due to the above two limiting values of the rapidity.

In finding an expression for the twist as a function of the rapidity, we do have a lot of data in momentum space, which, when combined with our current algebraic  understanding, should lead to a unique determination of the twist. Already in \cite{Gadde:2010zi} the one- and two-magnon eigenvalue problem was studied around the infinite length $\phi$-vacuum (obeying the BPS condition $\Delta=r$ and constructed out of $Z$ fields in the language of this paper).
In addition, in
\cite{Pomoni:2021pbj} one- and two-magnon solutions around the  
$Q$-vacuum (obeying the BPS condition $\Delta=2R$ and constructed from the bifundamental fields) were obtained. These solutions have the special feature of pointing to an elliptic structure for the rapidity.
More recently, in \cite{Bozkurt:2024tpz}
 the three-magnon eigenvector around the $\phi$-vacuum was computed, and the four-magnon eigenvector will appear soon \cite{fourmagnon}. These solutions have the novel feature of being long-range, with their coefficients obeying an  infinite tower of Yang-Baxter equations. Combining all these data with the algebraic approach in this paper should allow fixing the twist as a function of the rapidity.

Lastly, it is reasonable to expect that the findings regarding the one-loop Hamiltonian can be generalised to all loops. On the one hand, the exact S-matrix of \cite{Gadde:2010ku} should be derivable from the algebroid symmetry.
However, from the perspective of the strong-coupling regime (i.e. the dual worldsheet description of the gauge theory model we are discussing), the symmetries are only broken due to the boundary conditions of the string. Even the marginal deformation should only arise solely by twisting the boundary conditions of the string (in the orbifold theories we are considering, where the NS-NS flux $H=dB=0$).
It seems possible that our current algebroid approach, which is meaningful from the spin chain point of view (at weak coupling), and the twisting of the boundary conditions of the string, which is natural in the gravity side (at strong coupling),
can be reconciled, possibly via
introducing  a nontrivial connection that spreads the  effect  of the boundary conditions along the spin chain with a  Drinfeld-type twist \cite{Borsato:2021fuy}.

In order to interpolate between the weak coupling and strong coupling, we do not just need to find the correct algebraic description of our model, but furthermore need to compute the redefinition of the gauge coupling (string tension $T_{\text{eff}}=f(g^2)$) obtainable via localisation and  coined as exact effective coupling in \cite{Mitev:2014yba,Mitev:2015oty}, see also \cite{Pomoni:2011jj,Pomoni:2013poa}. On the gravity side, it is the B-field which is responsible for the above redefinition. The recent work \cite{Skrzypek:2022cgg} where a solution of IIB supergravity where the orbifold singularity is resolved is presented, allows to compute subleading corrections in the strong-coupling expansion in agreement with localisation \cite{Billo:2022lrv}. 

Clearly, we have only scratched the surface of the mathematical structures underlying the symmetries of the $\Ncal=2$ superconformal quiver theories. Rigorously understanding how the path groupoid is bound with the R-symmetry algebroid into a larger structure of a 2-category (see Section \ref{sec:Rsymgroupoid}) should impose constraints on the allowed twists and corresponding coassociators and perhaps even completely fix them. Elucidating this, along with its implications for integrability, is the subject of further work.

\mbox{}

{\bf Acknowledgements}

\mbox{}

EP and KZ wish to thank the Niels Bohr Institute, Copenhagen, for a very productive research visit. 
We wish to acknowledge discussions with  Raschid Abedin,  Sibylle Driezen, Charlotte Kristjansen, Jules Lamers, Didina Serban, Christoph Schweigert and Volker Schomerus.
We thank Enrico Andriolo for early collaboration on this project. 
HB, EP and XZ acknowledge support by the Deutsche Forschungsgemeinschaft (DFG, German Research Foundation) under Germany’s Excellence Strategy – EXC 2121 ``Quantum Universe'' - 390833306.
The research of EP is   funded by the  ERC-2021-CoG - BrokenSymmetries 101044226, as well as DFG via SFB 1624 - “Higher structures, moduli spaces and integrability”
- 506632645. 
The research of XZ was funded by National Science  Foundation of China under Grant No. 12347103. 

\newpage

\appendix

\section{The $\SU(4)$ R-symmetry group} \label{SU4Appendix}

In this appendix, we provide a concise overview of the relevant concepts pertaining to  the $\SU(4)$ R-symmetry and its breaking to $\SU(2)_L\times\SU(2)_R\times\Urm(1)_r$.

Let us consider how the $\SU(4)$ R-symmetry of the $\Ncal=4$ theory acts on the fields. It is convenient to combine the six real scalar fields into three complex scalar fields $X$, $Y$, and $Z$, and then organise them into an antisymmetric combination $\varphi^{ab}$, with indices $a,b=\{1,\ldots,4\}$ in the fundamental representation. In our convention,
\be
\varphi^{ab}=-\varphi^{ba}=\frac{1}{2}\epsilon^{abcd}\bar{\varphi}_{cd}=\begin{pmatrix}0 & {Z} & {X} & {Y}\\
-{Z} & 0 & \bar{{Y}} & -\bar{{X}}\\
-{X} & -\bar{{Y}} & 0 & \bar{{Z}}\\
-{Y} & \bar{{X}} & -\bar{{Z}} & 0
\end{pmatrix} .
\label{eq:PhiabMatrix}
\ee
The action of the generators $\Rcal^a_{\;b}$ of $\SU(4)$ on the fields is
\be
R_{\phantom{a}b}^{a}{\varphi}^{cd}=\delta_{\;b}^{c}{\varphi}^{ad}+\delta_{\;b}^{d}{\varphi}^{ca}-\frac{1}{2}\delta_{\;b}^{a}{\varphi}^{cd} ,
\label{eq:SU4GenSingleField}
\ee
which can be written out more explicitly for ease of reference as
\begin{align}
    &\mathcal{R} {Z} = \begin{pmatrix}
        \frac{1}{2}{Z} & 0 & 0 & 0 \\
        0 & \frac{1}{2}{Z} & 0 & 0 \\
        -\bar{{Y}} & {X} & -\frac{1}{2} {Z} & 0 \\
        \bar{{X}} & {Y} & 0 & -\frac{1}{2} {Z}
    \end{pmatrix} \qquad &&\mathcal{R} \bar{{Z}} = \begin{pmatrix}
        -\frac{1}{2}\bar{{Z}} & 0 & {Y} & -{X} \\
        0 & -\frac{1}{2} \bar{{Z}} & -\bar{{X}} & -\bar{{Y}} \\
        0 & 0 & \frac{1}{2} \bar{{Z}} & 0 \\
        0 & 0 & 0 & \frac{1}{2} \bar{{Z}}
    \end{pmatrix} \nonumber \\
    &\mathcal{R} {X} = \begin{pmatrix}
        \frac{1}{2}{X} & 0 & 0 & 0 \\
        \bar{{Y}} & -\frac{1}{2}{X} & {Z} & 0 \\
        0 & 0 & \frac{1}{2}{X} & 0 \\
        -\bar{{Z}} & 0 & {Y} & -\frac{1}{2}{X}
    \end{pmatrix} \qquad &&\mathcal{R} \bar{{X}} = \begin{pmatrix}
        -\frac{1}{2}\bar{{X}} & -{Y} & 0 & -{Z} \\
        0 & \frac{1}{2} \bar{{X}} & 0 & 0 \\
        0 & -\bar{{Z}} & -\frac{1}{2} \bar{{X}} & -\bar{{Y}} \\
        0 & 0 & 0 & \frac{1}{2} \bar{{X}}
    \end{pmatrix} \nonumber \\
    &\mathcal{R} {Y} = \begin{pmatrix}
        \frac{1}{2}{Y} & 0 & 0 & 0 \\
        -\bar{{X}} & -\frac{1}{2}{Y} & 0 & {Z} \\
        \bar{{Z}} & 0 & -\frac{1}{2}{Y} & {X} \\
        0 & 0 & 0 & \frac{1}{2}{Y}
    \end{pmatrix} \qquad &&\mathcal{R} \bar{{Y}} = \begin{pmatrix}
        -\frac{1}{2}\bar{{Y}} & {X} & -{Z} & 0 \\
        0 & \frac{1}{2} \bar{{Y}} & 0 & 0 \\
        0 & 0 & \frac{1}{2} \bar{{Y}} & 0 \\
        0 & -\bar{{Z}} & -\bar{{X}} & -\frac{1}{2} \bar{{Y}}
    \end{pmatrix} \ ,
\end{align}
where the notation is that $\Rcal^3_{\;1} Z=-\bar{Y}$, $\Rcal^4_{\;1} Z=\bar{X}$ etc. 

As shown in (\ref{eq:PhiabMatrix}), the scalar fields belong to the two-index antisymmetric  representation $\mathbf{6}$ of $\SU(4)$, or
equivalently the fundamental (vector) representation of $\SO(6)\simeq \SU(4)$. The tensor products of two and three fields decompose as
\begin{align}
\mathbf{6}\times\mathbf{6} &= \mathbf{20'} + \mathbf{15} + \mathbf{1} \label{eq:SO62siteDecomposition} \\
\mathbf{6}\times\mathbf{6}\times\mathbf{6} &= 2(\mathbf{64}) + \mathbf{50} + \mathbf{10} + \mathbf{\overline{10}} + 3(\mathbf{6}) \;,\label{eq:SO63siteDecomposition}
\end{align}
where the $\mathbf{20'}$ and $\mathbf{50}$ are 1/2 BPS representations, and the singlet is the Konishi operator. The representation $\mathbf{10}$  contains the superpotential, while $\mathbf{\overline{10}}$ contains the conjugate superpotential.

The $\mathrm{SU}(2)_L \times \mathrm{SU}(2)_R \times \mathrm{U}(1)_r$ unbroken subgroup of $\mathrm{SU}(4)_R$ acts as
\be
\begin{split}
    \mathrm{SU}(2)_L \quad &: \quad \sigma^+_L \;=\; \Rcal^3_{\;4}, \ \sigma^-_L \;=\; \Rcal^4_{\;3}, \ \sigma^3_L \;=\; \frac{1}{2}\left( \Rcal^3_{\;3} - \Rcal^4_{\;4} \right) \\
    \mathrm{SU}(2)_R \quad &: \quad \sigma^+_R \;=\; \Rcal^1_{\;2}, \ \sigma^-_R \;=\; \Rcal^2_{\;1}, \ \sigma^3_R \;=\; \frac{1}{2}\left( \Rcal^1_{\;1} - \Rcal^2_{\;2} \right) \\
    \mathrm{U}(1)_r \quad &: \quad \sigma_{r} \;=\; - \left( \Rcal^1_{\;1} + \Rcal^2_{\;2}\right) = \Rcal^3_{\;3} + \Rcal^4_{\;4} \ .
\end{split}
\ee
The resulting quantum numbers for the complex scalars are listed in Table \ref{tab:FieldsQuantumNumbersN=4SYM}.
\begin{table}[ht]
    \centering
    \begin{tabular}{c|c|c|c}
       $\varphi^{cd}$ & $\mathrm{SU}(2)_L$ & $\mathrm{SU}(2)_R$ & $\mathrm{U}(1)_r$ \\ 
       \hline 
       $Z$  & 0 & 0 & $-1$ \\
       $\bar{Z}$ & 0 & 0 & 1 \\
       $X$ & $\frac{1}{2}$ & $\frac{1}{2}$ & 0 \\
       $\bar{X}$ & $-\frac{1}{2}$ & $-\frac{1}{2}$ & 0 \\
       $Y$ & $-\frac{1}{2}$ & $\frac{1}{2}$ & 0 \\
       $\bar{Y}$ & $\frac{1}{2}$ & $-\frac{1}{2}$ & 0 
    \end{tabular}
    \caption{The $\mathrm{SU}(2)_L \times \mathrm{SU}(2)_R \times \mathrm{U}(1)_r$ quantum numbers of the complex scalar fields.}
    \label{tab:FieldsQuantumNumbersN=4SYM}
\end{table}

We note that the $\SU(2)_L$ group is accidental for the $\Zset_2$ case, as it is not present for $\Zset_k$ orbifolds with $k>2$. 

The remaining generators can be classified as raising and lowering operators of broken $\SU(2)$'s,
\be \label{SU2raisinglowering}
\begin{split}
    \mathrm{SU}(2)_{XZ} \quad &: \quad \sigma^+_{XZ} \;=\; \Rcal^3_{\;2}, \ \sigma^-_{XZ} \;=\; \Rcal^2_{\;3}, \ \sigma^3_{XZ} \;=\; \left( \Rcal^3_{\;3} - \Rcal^2_{\;2} \right)
    = \sigma^3_R + \sigma^3_L + \sigma_r ,
    \\
    \mathrm{SU}(2)_{YZ} \quad &: \quad \sigma^+_{YZ} \;=\; \Rcal^2_{\;4}, \ \sigma^-_{YZ} \;=\; \Rcal^4_{\;2}, \ \sigma^3_{YZ} \;=\; \left( \Rcal^2_{\;2} - \Rcal^4_{\;4} \right)
    = -\sigma^3_R + \sigma^3_L - \sigma_r ,\\
    \mathrm{SU}(2)_{\bar{X}Z} \quad &: \quad \sigma^+_{\bar{X}Z} \;=\; \Rcal^1_{\;4}, \ \sigma^-_{\bar{X}Z} \;=\; \Rcal^4_{\;1}, \ \sigma^3_{\bar{X}Z} \;=\; \left(\Rcal^1_{\;1} - \Rcal^4_{\;4} \right) = \sigma^3_R + \sigma^3_L - \sigma_r , \\
    \mathrm{SU}(2)_{\bar{Y}Z} \quad &: \quad \sigma^+_{\bar{Y}Z} \;=\; \Rcal^3_{\;1}, \ \sigma^-_{\bar{Y}Z} \;=\; \Rcal^1_{\;3}, \ \sigma^3_{\bar{Y}Z} \;=\; \left( \Rcal^3_{\;3} - \Rcal^1_{\;1} \right) = - \sigma^3_R + \sigma^3_L + \sigma_r \ ,
\end{split}
\ee
and similarly for their conjugate sectors involving $\bar{Z}$. The $\sigma^\pm$ in this list are the broken generators which in $\Ncal=4$ SYM used to relate the $X,Y$ fields (and their conjugates), which are now bifundamental, to the $(Z,\Zb)$ fields, which are now adjoint in their respective $\SU(N)$ groups. So they are the generators that we wish to resurrect as generators of a groupoid version of $\SU(4)$.

The choice of raising/lowering operators in each $\SU(2)$ sector is motivated by whether the second colour index is raised or lowered under the action of the operator. This is immaterial in the current $\Zset_2$ case, but we choose our convention such that it is compatible with more general $\Zset_k$ orbifolds, where  the $X,\bar{Y}$ fields are paths from node $i$ to $i+1$ while the $Y,\bar{X}$ fields from $i$ to $i-1$, with $i+k$ identified with $i$ \cite{Pomoni:2021pbj}. For instance,
\be
\sigma^+_{XZ} Z_{i}=X_{i,i+1} \;\;\text{and}\;\; \sigma^-_{YZ} Z_i=Y_{i,i-1}\;,
\ee
which agrees with the identifications in (\ref{SU2raisinglowering}).

\section{The path and R-symmetry groupoids} \label{GroupoidAppendix}
\stoptoc
A group is a nonempty set $G\neq \emptyset$ equipped with an operation $\cdot:G\times G\to G$
that composes every ordered pair of elements $(g_1,g_2)$ to form
a unique element $g_3=g_1\cdot g_2$, such that the composition is associative, has an identity element, and has an inverse element for each element in $G$.
A group is a category which has only one object and every arrow has a two-sided inverse under composition.

A groupoid $\mathcal{G}$ can be seen as a group, except that the composition is allowed to  
be a partial function, $\circ: \mathcal{G}\times\mathcal{G}\rightharpoonup\mathcal{G}$. In other words, it is not required that all pairs of elements in $\mathcal{G}$ can be composed.
In categorical language, a groupoid is a small category in which each morphism is an isomorphism. More explicitly, a groupoid $\mathcal{G}$ consists of a set $\mathcal{G}_{0}$
of objects, a set $\mathcal{G}_{1}$ of arrows,
and five structure
maps $\mathcal{S},\mathcal{T}:\mathcal{G}_{1}\rightrightarrows\mathcal{G}_{0}$,
$\circ:\mathcal{G}_{1}\times\mathcal{G}_{1}\rightharpoonup\mathcal{G}_{1}$,
$\mathcal{I}:\mathcal{G}_{0}\to\mathcal{G}_{1}$, $^{-1}:\mathcal{G}_{1}\to\mathcal{G}_{1}$,
obeying the following properties:
\begin{itemize}
\item For each arrow $g\in\mathcal{G}_{1}$, its source and target objects
are respectively $\mathcal{S}(g)$ and $\mathcal{T}(g)$, and we write
$\mathcal{S}\left(g\right)\xrightarrow{g}\mathcal{T}\left(g\right)$.
\item A pair of arrows $\left(g_{2},g_{1}\right)\in\mathcal{G}_{1}\times\mathcal{G}_{1}$
is composable when $\mathcal{T}\left(g_{1}\right)=\mathcal{S}\left(g_{2}\right)$,
and the set of composable arrows is denoted by $\mathcal{G}_{2}\subseteq\mathcal{G}_{1}\times\mathcal{G}_{1}$.
The map $\circ:\mathcal{G}_{2}\to\mathcal{G}_{1}$ is the composition,
such that $\mathcal{S}\left(g_{2}\circ g_{1}\right)=\mathcal{S}\left(g_{1}\right)$,
$\mathcal{T}\left(g_{2}\circ g_{1}\right)=\mathcal{T}\left(g_{2}\right)$.
The composition is associative, $\left(g_{3}\circ g_{2}\right)\circ g_{1}=g_{3}\circ\left(g_{2}\circ g_{1}\right)$.
\item The unit map $\mathcal{I}$ sends every object $x\in\mathcal{G}_{0}$
to the identity arrow $\mathrm{id}_{x}\in\mathcal{G}_{1}$ at $x$,
such that for every $g\in\mathcal{G}_{1}$, $\mathrm{id}_{\mathcal{T}\left(g\right)}\circ g=g\circ\mathrm{id}_{\mathcal{S}\left(g\right)}=g$.
\item The inverse map $^{-1}$ sends every arrow $g\in\mathcal{G}_{1}$
to its inverse $g^{-1}$, such that $g^{-1}\circ g=\mathrm{id}_{\mathcal{S}\left(g\right)}$
and $g\circ g^{-1}=\mathrm{id}_{\mathcal{T}\left(g\right)}$.
\end{itemize}
If $\mathcal{G}_{0}$ contains only a single object, then this definition
reduces to that of a group. On the other hand, given a groupoid $\mathcal{G}$
and one object $x\in\mathcal{G}_{0}$, the subcollection of arrows
$\left\{ \left.g\in\mathcal{G}_{1}\right|x\xrightarrow{g}x\right\} $
forms an automorphism group $\mathrm{Aut}_{\mathcal{G}}\left(x\right)$
of $x$ in $\mathcal{G}$.
In physics, the symmetry group is the set of all symmetry transformations which isomorphically relates one
object to itself, endowed with the group operation of composition. A groupoid is a collection of
symmetry transformations acting between possibly more than one object. 

 Let us now apply the above formal definitions to the two different types of groupoid that we introduced with physics language in Section \ref{OPSymmetries}. The path groupoid, which comprehensively describes the total vector space of all spin-chain states, and the R-symmetry groupoid, which describes the mathematical structure which replaces the $\SU(4)$ R-symmetry Lie group. 

\subsection{Path groupoid}

We begin by defining the path groupoid that is obtained from the quiver in Fig.\ref{fig:z2}.  It consists firstly of a set of objects $\mathcal{G}_0 = \left\{\circled{1} , \circled{2}\right\}$, which correspond to the two colour groups, or equivalently the two nodes of the quiver.  Furthermore, it consists of the set of arrows  $\mathcal{G}_1 = \Big\{\{Z_1, X_{12}, \cdots\}, \{Z_1 Z_1, \cdots \}, \cdots \Big\}$, as in (\ref{eq:vectorspace}).
The set $\mathcal{G}_1$ contains all possible paths, which in the spin-chain picture corresponds to all allowed spin-chain with all possible lengths. Paths made by following the directed arrows correspond to monomials of single site fields with properly contracted gauge  indices. 

The composition $\circ$ is a map from $\mathcal{G}_{2}\subseteq\mathcal{G}_{1}\times\mathcal{G}_{1}$ to $\mathcal{G}_{1}$, defined such that the target ($\mathcal T$) of the first map is the source ($\mathcal{S}$) of the second one. 
Let us now check that $\{\mathcal{G}_0,\mathcal{G}_1$\}, along with the composition $\circ$, satisfies the above  defining properties of a groupoid.
To establish conventions, for the individual fields, which correspond to the shortest possible arrows in $\mathcal{G}_1$, we write
\begin{align}
    \mathcal{S}(X_{12}) &= \circled{1} = \mathcal{T} (X_{21}) \ , \\
    \mathcal{S}(X_{21}) &= \circled{2} = \mathcal{T}(X_{12}) \ ,  \\
    \mathcal{S}(Z_1) &= \circled{1} = \mathcal{T}(Z_1) \ , \\
    \mathcal{S}(Z_2) &= \circled{2} = \mathcal{T}(Z_2) \ ,
\end{align}
and similarly for all other single-site fields. Note that here we use the physicist convention of reading maps from left to right. Longer arrows, or paths, can be defined by longer spin-chain states as described in (\ref{eq:vectorspace}), with for example $\mathcal{S}(X_{12}Z_2)=\circled{1}$ and $\mathcal{T}(X_{12}Z_2)=\circled{2}$. Moreover, for the unit map ($\mathcal{I}$) we have
\begin{align}
    \mathcal{I} \ : \ \circled{$i$} \rightarrow Z_i   \ ,
\end{align}
and the inverse map ($^{-1}$) acting on the single fields gives
\begin{equation}
    ^{-1} \ : \ X_{12} \rightarrow X_{21} \ , Y_{12}\rightarrow Y_{21}\;,\;Z_1\rightarrow Z_1 \ ,
\end{equation}
and their $\Zset_2$ conjugate relations. From the single-field inverse, we can follow the arrows to write the inverses of multi-field states, e.g. ${}^{-1}:X_{12}Z_2\ra Z_2X_{21}$. 

Clearly, following the arrows, all requirements concerning composition and associativity are satisfied in the above sense.

\subsection{The R-symmetry groupoid} \label{sec:Rsymgroupoid}

Let us now define the R-symmetry groupoid, which acts on the above path groupoid and is a generalisation of the $\SU(4)$ Lie group. It is easier for the reader to consider each length $L$ separately. Furthermore, we will restrict our analysis to the
$\SU(2)_{XZ}$ sector, and the extension to the other broken $\SU(2)$ sectors is a relatively straightforward process.  

We start with $L=1$, where the groupoid is defined by the set of objects 
\be
\mathcal{G}^{(1)}_0 = \{ X_{12}, X_{21},Z_1, Z_2\}\;,
\ee
and a set of arrows which are composed as exponential maps of the generators 
\be \label{eq:algebroidXZ}
\mathcal{G}^{(1)}_1 = \{\1^{(1)}, \sigma_+^{(1)}, \sigma_-^{(1)}, \sigma_3^{(1)},\1^{(2)},\sigma_+^{(2)}, \sigma_-^{(2)}, \sigma_3^{(2)},\gamma \}\;,
\ee
where the identity matrix is required if we are working with the universal enveloping algebra. In writing (\ref{eq:algebroidXZ}) we are already referring to the algebroid language. Going from the Lie algebroid to the Lie groupoid works in the same way as going from a Lie algebra to its corresponding Lie group. 
For the source and target map of the algebroid we get 
\begin{align}
    \mathcal{S}(\sigma_+^{(i)}) &= Z_i = \mathcal{T} (\sigma_-^{(i)}) \ , \\
    \mathcal{S}(\sigma_-^{(i)}) &= X_{i,i+1} =  \mathcal{T} (\sigma^{(i)}_+) \ , \\
    \mathcal{S}(\gamma) &= \mathcal{G}^{(1)}_0 = \mathcal{T}(\gamma) \ ,
\end{align}
where the index $i$ labels the colour groups and identified mod(2) i.e. $i\cong i+2$. It is worth noting that the source/target of $\gamma$ is $\{Z_1,X_{12}\}$ or $\{Z_2,X_{21}\}$, but not both at the same time, whereas the source/target of $\{\sigma_3^{(i)},I\}$ is automatically the full $\mathcal{G}^{(1)}_0$.
The unit map $\mathcal{I}$ for $\{Z_i,X_i\}$ is captured by the identity of the universal enveloping algebra, while the inverse map of $\sigma_+^{(i)}$ is $\sigma_-^{(i)}$ and vice versa. The maps $\{\gamma,\sigma_3^{(i)},\1^{(i)}\}$ are their own inverses.
The R-symmetry algebroid respects the $\mathfrak{su}(2)$ algebra as its composition rule, together with the relation
\begin{align}
    \gamma \circ \sigma^{(i)} &= \sigma^{(i+1)} \circ \gamma \ , 
\end{align}
since $\gamma$ will change the gauge indices (i.e. exchange the objects for their $\Zset_2$ conjugates). Moreover, composition of generators in different gauge sectors is not allowed, as that would correspond to invalid paths.

At $L=2$, we have $\mathcal{G}^{(2)}_0 = \{Z_iZ_i,X_{i}Z_{i+1} \pm Z_iX_{i}, X_{i}X_{i+1}\}$. The arrows in $\mathcal{G}^{(2)}_1$ are now generators, whose action has been properly extended to two sites using the coproduct (\ref{eq:CoproductOrbifoldPoint}).

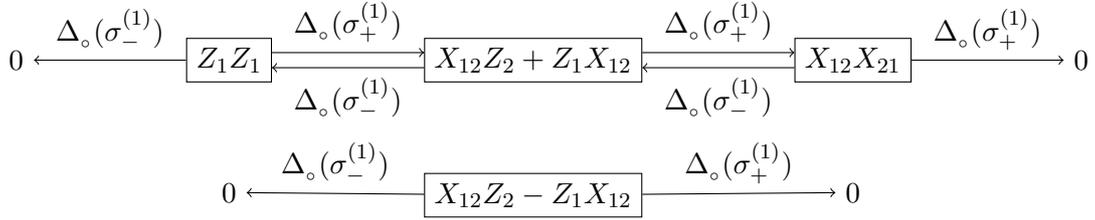
\begin{figure}[h!]
\begin{center}
\begin{tikzpicture}[node distance=1.5cm, auto]
    % Nodes
    \node (Z1Z1) [draw, rectangle] {$Z_1 Z_1$};
    \node (emptya) [left=2cm of Z1Z1] {$0$};
    \node (Z1X12PlusX12Z2) [right=2cm of Z1Z1, draw, rectangle] {$X_{12} Z_2 + Z_1 X_{12}$};
    \node (X12X21) [right=2cm of Z1X12PlusX12Z2, draw, rectangle] {$X_{12} X_{21}$};
    \node (emptyb) [right=2cm of X12X21] {$0$};
    \node (empty1) [below=1.2cm of Z1Z1] {$0$};
    \node (Z1X12MinusX12Z2) [below=1.2cm of Z1X12PlusX12Z2, draw, rectangle] {$X_{12} Z_2 - Z_1 X_{12}$};
    \node (empty2) [below=1.2cm of X12X21] {$0$};

    % Arrows
    \draw[->] ([yshift=0.1cm]Z1Z1.east) -- node[above] {$\Delta_\op(\sigma_+^{(1)})$} ([yshift=0.1cm]Z1X12PlusX12Z2.west);
    \draw[->] ([yshift=0.1cm]Z1X12PlusX12Z2.east) -- node[above] {$\Delta_\op(\sigma_+^{(1)})$} ([yshift=0.1cm]X12X21.west);
    \draw[->] (X12X21) -- node[above] {$\Delta_\op(\sigma_+^{(1)})$} (emptyb);
    \draw[->] (Z1Z1) -- node[above] {$\Delta_\op(\sigma_-^{(1)})$} (emptya);
    \draw[->] ([yshift=-0.1cm]X12X21.west) -- node[below] {$\Delta_\op(\sigma_-^{(1)})$} ([yshift=-0.1cm]Z1X12PlusX12Z2.east);
    \draw[->] ([yshift=-0.1cm]Z1X12PlusX12Z2.west) -- node[below] {$\Delta_\op(\sigma_-^{(1)})$} ([yshift=-0.1cm]Z1Z1.east);

    \draw[<-] (empty1) -- node[above] {$\Delta_\op(\sigma_-^{(1)})$} (Z1X12MinusX12Z2);
    \draw[->] (Z1X12MinusX12Z2) -- node[above] {$\Delta_\op(\sigma_+^{(1)})$} (empty2);

\end{tikzpicture}
\caption{The algebroid structure at two sites.}
\label{fig:RsymAlgebroid2sites}
\end{center}
\end{figure}

For the source and target maps we have
\begin{align}
    \mathcal{S}(\Delta_\op(\sigma_+^{(i)})) &= \{ Z_iZ_i, X_{i}Z_{i+1}\pm Z_i X_i \} \ , \label{Twositemaps1}\\
    \mathcal{S}(\Delta_\op(\sigma_-^{(i)})) &= \{ X_{i}Z_{i+1} \pm Z_i X_i, X_i X_{i+1} \} \ ,\label{Twositemaps2} \\ 
    \mathcal{T}(\Delta_\op(\sigma_+^{(i)})) &= \{ X_i Z_{i+1}+ Z_i X_i,X_i X_{i+1} \} \ ,\label{Twositemaps3} \\
    \mathcal{T}(\Delta_\op(\sigma_-^{(i)})) &= \{Z_iZ_i, X_i Z_{i+1}+ Z_i X_i\} \ ,\label{Twositemaps4}
\end{align}
where the antisymmetric combination $X_i Z_{i+1}- Z_i X_i$ is a singlet under the action of the arrows. Note that the target can also include the zero element, which we do not explicitly write in (\ref{Twositemaps2})  and (\ref{Twositemaps3}), but is depicted in Fig. \ref{fig:RsymAlgebroid2sites}.
The unit and inverse maps are the same as for $L=1$, extended to $L=2$ sites. Furthermore, the composition rule is also captured by the $\mathfrak{su}(2)$ 
algebra relations. Showing this is the purpose of Appendix \ref{Ap:commutators}. 

We wish to remark that the morphisms of the path groupoid 1-category are the objects of the R-symmetry groupoid, viewed as a category. The vertical and horizontal compositions seem to have the structure of a 2-category. Confirming this is the subject of current investigation. 

\stoptoc
\subsection{Compatibility of the path groupoid and R-symmetry algebroid}
\resumetoc

Let us now check that the module (path groupoid) product $m$ is compatible with the R-symmetry algebroid coproduct $\Delta(\R{a}{b})$. We define $m$ as
\be \label{pathgroupoidproduct}
m: \Vscr_{ij}\otimes \Vscr_{kl}\ra \left\{\begin{array}{ll} \Vscr_{il} \;\; & \text{if} \;\; j=k\\
0 & \text{if} \;\; j\neq k\end{array}\right.
\ee
or, more explicitly,
\be
m([\varphi^1_{ij}\otimes \varphi^2_{kl}])=\left\{\begin{array}{ll} \varphi_{ij}\varphi_{jl} & \text{if}\;\; j=k\\
0 & \text{if} \;\; j\neq k\end{array}\right.\;.
\ee
Acting on a product of fields (which by definition have compatible gauge indices) with a broken generator $\Rcal^a_{\;b}$, we have the definition
\be\label{groupoidcompatibility}
\begin{split}
\Rcal^a_{\;b}\triangleright \varphi^1_{ij}\varphi^2_{jk}&=m(\Delta(\Rcal^a_{\;b})\triangleright [\varphi^1_{ij}\otimes \varphi^2_{jk}])
=m((\1\otimes \Rcal^{a}_{\;b}+\Rcal^a_{\;b}\otimes \gamma)\triangleright [\varphi^1_{ij}\otimes \varphi^2_{jk}])\\
&=m( [\varphi^1_{ij}\otimes (\Rcal^a_{\;b}\triangleright\varphi^2)_{j g(k)}]+[(\Rcal^{a}_{\;b}\triangleright \varphi^1)_{i g(j)}\otimes \gamma \triangleright \varphi^2_{jk}])\\
&=m( [\varphi^1_{ij}\otimes (\Rcal^a_{\;b}\triangleright\varphi^2)_{j g(k)}]+[(\Rcal^{a}_{\;b}\triangleright \varphi^1)_{i g(j)}\otimes\varphi^2_{g(j)g(k)}])\\
&=\varphi^1_{ij}(\Rcal^a_{\;b}\triangleright\varphi^2)_{jg(k)}+
(\Rcal^a_{\;b}\triangleright\varphi^1)_{ig(j)}\varphi^2_{g(j)g(k)}\;.
\end{split}
\ee
Here $g$ is the $\Zset_2$ group element which flips each index, i.e. $g(1)=2$, $g(2)=1$, and we used that broken generators flip the second index of the field they act on. 
The algebroid coproduct guarantees that valid paths on the quiver map to valid paths. 
This construction can be straightforwardly extended to encompass more sites.
Here, we simply illustrate the construction using a three-site example. The action $\Rcal^2_{\;3}\triangleright (X_{12}X_{21}Z_1)=Z_1X_{12}Z_2+X_{12}Z_2Z_2$ can be depicted as the following operation on the quiver, where, as in Fig. \ref{fig:z2}, the blue node denotes gauge group 1 and the red node gauge group 2:

\be
\Rcal^{2}_{\;3}\triangleright \left[ \begin{tikzpicture}[scale=0.3,baseline=-0.1cm]
  \draw[fill=blue] (0,0) circle (2ex);
\draw[fill=red] (6,0) circle (2ex);

\draw[->,blue,thick] (-0.5,-0.8) arc (-60:-310:1);

\draw[->,purple,thick] (0.2,0.8) arc (110:70:8);
\draw[->,purple,thick] (5.7,-0.8) arc (-70:-110:8);

\node at (3,2) {$1$};\node at (3,-2) {$2$};\node at (-3,0) {$3$};
  \end{tikzpicture}\;\;\right]\;\;=\;\;
\begin{tikzpicture}[scale=0.3,baseline=-0.1cm]
  \draw[fill=blue] (0,0) circle (2ex);
\draw[fill=red] (6,0) circle (2ex);

\draw[->,blue,thick] (-0.5,-0.8) arc (-60:-310:1);
\draw[->,red,thick] (6.4,0.8) arc (120:-130:1);

\draw[->,purple,thick] (0.2,0.8) arc (110:70:8);

\node at (9,0) {$3$};\node at (3,2) {$2$};\node at (-3,0) {$1$};
\end{tikzpicture}\;\;+\;\;\begin{tikzpicture}[scale=0.3,baseline=-0.1cm]
  \draw[fill=blue] (0,0) circle (2ex);
\draw[fill=red] (6,0) circle (2ex);

\draw[->,red,thick] (6.4,0.8) arc (120:-130:1);
\draw[->,red,thick] (6.4,0.8) arc (120:-110:1);

\draw[->,purple,thick] (0.2,0.8) arc (110:70:8);

\node at (9.2,0) {$2,3$};\node at (3,2) {$1$};;
\end{tikzpicture}
\ee
Here the numbers indicate the order in which the arrows of the path quiver are composed. We see that, after the action of the broken generator, the source has remained node 1, but the target has changed from 1 to 2. 
\resumetoc

\section{Algebroid commutation relations}
\label{Ap:commutators}

In this appendix, we show that the commutators of generic $\SU(4)$ R-symmetry generators obey the $\mathfrak{su}(4)$ commutation relations for any number of sites $L$, both at the orbifold point and in the marginally deformed theory. The computation for the theory at the orbifold point will be presented in detail, after which the differences that emerge when extending the analysis to the marginally deformed case will be discussed. 

\stoptoc
\subsection{Orbifold point}

For concreteness, we will demonstrate that the coproduct $\Delta_{\op}^{(L)}\left(\R{a}{b}\right)$ defined in 
 (\ref{eq:OrbifoldPointDelta}) obeys the $\mathfrak{su}(4)$ commutation relations for the case $L=3$. This example is sufficient to illustrate all the relevant steps,  and the extension to generic $L$ then follows straightforwardly.  Recall that (\ref{eq:OrbifoldPointDelta}) includes the operator $\Omega^a_{\;b}$ which is equal to $\1$ for unbroken and $\gamma$ for broken generators. Clearly, at a given site,
\begin{equation}
    [\Oab, \Omega^c_{\phantom{a}d}] = 0 \; .
\end{equation}
Let us now consider the commutator of two $\SU(4)$ R-symmetry generators for $L=3$. Explicit calculation gives
\begin{align}
    &\left[\Delta_{\op}^{(3)}\left(\R{a}{b}\right),\Delta_{\op}^{(3)}\left(\R{c}{d}\right)\right] = \nonumber \\
    &= \1 \otimes \1 \otimes [\R{a}{b},\R{c}{d}] + \1 \otimes [\R{a}{b},\R{c}{d}] \otimes \Oab \Omega^c_{\phantom{a}d} + [\R{a}{b},\R{c}{d}] \otimes \Oab \Omega^c_{\phantom{a}d} \otimes \Oab \Omega^c_{\phantom{a}d} \nonumber \\
    &+ \1 \otimes \R{a}{b} \otimes [\Oab,\R{c}{d}] + \R{a}{b} \otimes [\Oab,\R{c}{d}] \otimes \Oab \Omega^c_{\phantom{a}d} + \R{a}{b} \otimes \Oab \otimes [\Oab,\R{c}{d}] \nonumber \\
    &+ \1 \otimes \R{c}{d} \otimes [\R{a}{b}, \Omega^c_{\phantom{a}d}] + \R{c}{d} \otimes [\R{a}{b}, \Omega^c_{\phantom{a}d}] \otimes \Oab \Omega^c_{\phantom{a}d} + \R{c}{d} \otimes \Omega^c_{\phantom{a}d} \otimes [\R{a}{b}, \Omega^c_{\phantom{a}d}] \ ,
\end{align}
leaving us with the task of determining the two commutators $[\R{a}{b}, \Omega^c_{\phantom{a}d}]$ and $[\R{a}{b}, \R{c}{d}]$.

If $\R{c}{d}$ is an unbroken generator, $\Omega^c_{\phantom{a}d} = \1$, and the commutator $[\R{a}{b}, \Omega^c_{\phantom{a}d}]=0$. On the other hand, if $\R{c}{d}$ is a broken generator, we have $\Omega^c_{\phantom{a}d} = \gamma$. When the commutator acts on a generic scalar field ${\varphi_i}^{cd}$,   
\be
    [\R{a}{b},\gamma] {\varphi_i}^{cd} = \left(\R{a}{b} \gamma - \gamma \R{a}{b} \right) {\varphi_i}^{cd}  
    =\R{a}{b} \varphi_{i+1}^{cd} - \gamma \left( \delta_{b}^{c}{\varphi_i}^{ad}+\delta_{b}^{d}{\varphi_i}^{ca}-\frac{1}{2}\delta_{b}^{a}{\Phi_i}^{cd} \right) \ ,
\ee
where we applied (\ref{eq:SU4GenSingleField}) for $\R{a}{b} \in \{R_{(i\hspace{1pt}i)},R_{(i\hspace{1pt}i+1)}\}$, as required by  Fig.~\ref{fig:algebroid} since we are acting on ${\varphi_i}^{cd}$. The remaining $\R{a}{b}$  will act on $\varphi_{i+1}^{cd}$ and therefore, it has to be part of $\{R_{(i+1\hspace{1pt}i+1)},R_{(i+1\hspace{1pt}i)}\}$.
We therefore find
\be
    [\R{a}{b},\gamma] {\varphi_i}^{cd} =\left( \delta_{b}^{c}\varphi_{i+1}^{ad}+\delta_{b}^{d}\varphi_{i+1}^{ca}-\frac{1}{2}\delta_{b}^{a}\varphi_{i+1}^{cd} \right) - \left( \delta_{b}^{c}\varphi_{i+1}^{ad}+\delta_{b}^{d}\varphi_{i+1}^{ca}-\frac{1}{2}\delta_{b}^{a}\varphi_{i+1}^{cd} \right) 
    = 0 \ .
\ee
This means that the only non-trivial contribution to the commutator of two $\SU(4)$ R-symmetry generators is
\begin{align}
     \left[\Delta_{\op}^{(3)}\left(\R{a}{b}\right),\Delta_{\op}^{(3)}\left(\R{c}{d}\right)\right] &= \1 \otimes \1 \otimes [\R{a}{b},\R{c}{d}] + \1 \otimes [\R{a}{b},\R{c}{d}] \otimes \Oab \Omega^c_{\phantom{a}d} \nonumber \\
     &+ [\R{a}{b},\R{c}{d}] \otimes \Oab \Omega^c_{\phantom{a}d} \otimes \Oab \Omega^c_{\phantom{a}d} \ .
\end{align}
Applying the single-site $\mathfrak{su}(4)$ commutation relation
\begin{equation}
    [\R{a}{b},\R{c}{d}] = \delta^c_b \R{a}{d} - \delta^a_d \R{c}{b} \ ,
    \label{eq:ComRR}
\end{equation}
we find an interesting behaviour for the interplay of broken and unbroken generators
\be\label{brokenunbroken}\begin{split}
    [(\text{unbroken}),(\text{unbroken})] &= (\text{unbroken}) \ , \\
    [(\text{broken}),(\text{unbroken})] &= (\text{broken}) \ , \\ 
    [(\text{broken}),(\text{broken})] &= (\text{unbroken}) \ ,
\end{split}\ee
which can be checked by plugging in explicit generators into (\ref{eq:ComRR}). \footnote{Of course, to be precise one needs to be cautious not to combine generators of type $\{R_{(i\hspace{1pt}i)},R_{(i\hspace{1pt}i+1)}\}$ with $\{R_{(i+1\hspace{1pt}i+1)},R_{(i+1\hspace{1pt}i)}\}$ in the commutator when acting on fields, as this would be an invalid ``path" in the algebroid structure depicted in Fig.~\ref{fig:algebroid}.}
Furthermore, we have that
\begin{equation}
    \Oab \Omega^c_{\phantom{a}d}=\begin{cases}
\1, & \mathrm{if~}\R{a}{b}~\mathrm{and~}\R{c}{d}~\text{are both broken or both unbroken}\\
\gamma, & \mathrm{if~}\R{a}{b}~\mathrm{is~broken~and~}\R{c}{d}~\mathrm{is~unbroken,~and~vice~versa}
\end{cases} \ .
\end{equation}
Taking all the preceding elements together, we find that for $L=3$ the R-symmetry  generators  respect the $\mathfrak{su}(4)$ commutation relation,
\begin{equation}
   \left[\Delta_{\op}^{(3)}\left(\R{a}{b}\right),\Delta_{\op}^{(3)}\left(\R{c}{d}\right)\right] = \delta^c_b \Delta_{\op}^{(3)}\left(\R{a}{d}\right) - \delta^a_d \Delta_{\op}^{(3)}\left(\R{c}{b}\right) \ .
\end{equation}
As previously stated, the above $L=3$ computation is merely indicative and can be straightforwardly extended to any $L$.

\subsection{Marginally deformed case}
\resumetoc
The twists used in Section  \ref{TwistsSection} to define twisted coproducts are of two types: In some sectors we use matrix-type twists, such as (\ref{XYtwist}) for the XY sector, while in other sectors we use dynamical twists, such as (\ref{diagonaltwist}) for the XZ sector. In both cases the twisted coproduct is defined as 
\be
\Delta^{(L)}_\kappa \left( \R{a}{b} \right) = \Fcal^{(L)} \hspace{1pt} \Delta^{(L)}_\op \left( \R{a}{b} \right) \hspace{1pt} (\Fcal^{(L)})^{-1} \ ,
\ee
and the previous argument also holds if we twist using the full (block-diagonal) twist, since then the commutator of two generators is reduced to the orbifold-point case:
\be
    \left[\Delta^{(L)}_\kappa \left( \R{a}{b} \right),\Delta^{(L)}_\kappa \left( \R{c}{d} \right)\right] = \Fcal^{(L)} \left[ \Delta^{(L)}_\op \left( \R{a}{b} \right),\Delta^{(L)}_\op \left( \R{c}{d} \right)\right] (\Fcal^{(L)})^{-1} \ .
\ee
Similarly, the dynamical coproducts can be verified directly, as they differ from those at the orbifold point merely by the replacement
\begin{equation}
    \Oab \rightarrow K^a_{\phantom{a}b} = \begin{cases}
\1, & \mathrm{if~}\R{a}{b}~\mathrm{is~unbroken}\\
\gamma \kappa^{-s}, & \mathrm{if~}\R{a}{b}~\mathrm{is~broken}
\end{cases} \ .
\end{equation}
Meanwhile, at a given site we still have that 
\begin{equation}
    [K^a_{\phantom{a}b}, K^c_{\phantom{a}d}] = 0 \ ,
\end{equation}
and the argument presented in the preceding subsection remains valid. Furthermore, since $\gamma^2=\1$ and $s\gamma = -\gamma s$, we have that
\begin{align}
    K^a_{\phantom{a}b} K^c_{\phantom{a}d} = \begin{cases}
        \1, & \text{if both generators are broken or both unbroken} \\
        \gamma \kappa^{-s}, & \mathrm{if~one~generator~is~broken~and~the~other~is~unbroken}
    \end{cases} \ .
\end{align}
Therefore, the statement (\ref{brokenunbroken}) still holds in the marginally deformed case.

\section{Opening up procedure}
\label{Ap:OpeningUp}

As explained in Section \ref{OPSymmetries}, the broken R-symmetry generators do not respect the gauge theory trace, since they flip all the gauge indices to the right of the site where they act. So we cannot act  on closed states with broken generators. To bypass this issue, we first work with open states, where a single action of the broken generators \emph{is} well-defined. Our prescription for acting on a closed state will be to \textit{cut open the trace} and average over all possible cutting points, in a cyclic manner: 
\begin{equation}
\tr_i\left(\varphi_{1}\varphi_{2} \dots\varphi_{L} \right) \mapsto \frac{1}{L} \sum_{\tau \in \Zset_L} \ \varphi_{\tau(1)} \varphi_{\tau(2)} \dots \varphi_{\tau(L)} \ ,
\label{eq:CyclicOpeningProcedure}
\end{equation}
where each term in the summand is a cyclic permutation of the fields in the trace and can be viewed as an open state. Given that the first and last gauge indices of each monomial are no longer required to be equal, it is possible to define the action of the broken generators in a consistent manner. 

The necessity for opening up and cyclic symmetrisation of traces has also arisen in other cases where quantum groups have been applied to gauge theory, in particular in the study of marginally deformed gauge theories, see e.g. \cite{Bundzik2007,Mansson:2008xv,Dlamini:2016aaa,Garus:2017bgl,Dlamini:2019zuk}, as well as in the demonstration of the Yangian symmetry of the planar $\Ncal=4$ SYM at the level of the Lagrangian \cite{Beisert:2018zxs}. It arises because relaxing the co-commutativity property of the coproduct is not immediately compatible with the cyclicity of the trace.\footnote{At the level of string theory, we expect it to originate in the symmetrised trace prescription \cite{Tseytlin:1997csa} for the DBI action describing a the stack of D-branes which reduces to Yang-Mills theory in the low energy limit.} This is still true in our case, but is made more acute by the need to work with non-closeable states. 

Since physical states in our theory are  traces in the colour indices, the opening-up prescription means that we will be working with unphysical states. However, acting twice with the broken generators on a closeable state results in a closeable state. Therefore, we will consider the single action of a broken generator to be merely an intermediate step.  After acting twice, we can close the state again by inverting the aforementioned procedure, thus allowing for a comparison of physical states with physical states.

As explained in Section \ref{subsec:nonassociativity}, when opening up the scalar potential (\ref{eq:FullQuarticTerms}), apart from the cyclic order we also need to preserve the bracketing indicating the origin of each monomial as an F- or D-term. Therefore, the open quartic terms with first gauge group 1 are a sum of unshifted and shifted contributions, $\Vcal_1=\Vcal^{(u)}_1+\Vcal^{(s)}_1$ and similarly for their $\Zset_2$ conjugates. For reference, we record these terms below. The unshifted contribution is
{\small \be\label{Vopenunshifted}\begin{split}
&\Vcal^{(u)}_1=\frac{\kappa}{4}\Big(X_{12}Z_2\bar{Z}_2\bar{X}_{21}+X_{12}\bar{Z}_2Z_2\bar{X}_{21}+\bar{X}_{12}Z_2\bar{Z}_2X_{21}+\bar{X}_{12}\bar{Z}_2Z_2X_{21}  \\
    &\quad\qquad\qquad+Y_{12}Z_2\bar{Z}_2\bar{Y}_{21}+Y_{12}\bar{Z}_2Z_2\bar{Y}_{21}+\bar{Y}_{12}Z_2\bar{Z}_2Y_{21}+\bar{Y}_{12}\bar{Z}_2Z_2Y_{21}\Big)\\
&\quad-\frac{1}{4}\Big(X_{12}\bar{Z}_2\bar{X}_{21}Z_1+Z_1X_{12}\bar{Z}_2\bar{X}_{21}+X_{12}Z_2\bar{X}_{21}\bar{Z}_1+\bar{Z}_1X_{12}Z_2\bar{X}_{21}+\bar{X}_{12}\bar{Z}_2X_{21}Z_1  \\
    &\quad\qquad+Z_1\bar{X}_{12}\bar{Z}_2X_{21}+\bar{X}_{12}Z_2X_{21}\bar{Z}_1+\bar{Z}_1\bar{X}_{12}Z_2X_{21}+Y_{12}\bar{Z}_2\bar{Y}_{21}Z_1+Z_1Y_{12}\bar{Z}_2\bar{Y}_{21}  \\
&\quad\qquad+Y_{12}Z_2\bar{Y}_{21}\bar{Z}_1+\bar{Z}_1Y_{12}Z_2\bar{Y}_{21}+\bar{Y}_{12}\bar{Z}_2Y_{21}Z_1+Z_1\bar{Y}_{12}\bar{Z}_2Y_{21}+\bar{Y}_{12}Z_2Y_{21}\bar{Z}_1+\bar{Z}_1\bar{Y}_{12}Z_2Y_{21}\Big)\\
&\quad-\frac{1}{4\kappa}\Big(X_{12}\bar{X}_{21}\bar{X}_{12}X_{21}+\bar{X}_{12}X_{21}X_{12}\bar{X}_{21} +2 X_{12}Y_{21}\bar{X}_{12}\bar{Y}_{21}+2 \bar{X}_{12}\bar{Y}_{21}X_{12}Y_{21}-X_{12}\bar{X}_{21}Y_{12}\bar{Y}_{21}\\
    &\quad\qquad-Y_{12}\bar{Y}_{21}X_{12}\bar{X}_{21}+X_{12}\bar{X}_{21}\bar{Y}_{12}Y_{21}-2 X_{12}Y_{21}\bar{Y}_{12}\bar{X}_{21}-2 \bar{Y}_{12}\bar{X}_{21}X_{12}Y_{21}+\bar{Y}_{12}Y_{21}X_{12}\bar{X}_{21}  \\
    &\quad\qquad-Z_1X_{12}\bar{X}_{21}\bar{Z}_1-\bar{Z}_1X_{12}\bar{X}_{21}Z_1-X_{12}\bar{X}_{21}X_{12}\bar{X}_{21}+\bar{X}_{12}X_{21}Y_{12}\bar{Y}_{21}-2 \bar{X}_{12}\bar{Y}_{21}Y_{12}X_{21} \\
    &\quad\qquad-2 Y_{12}X_{21}\bar{X}_{12}\bar{Y}_{21}+Y_{12}\bar{Y}_{21}\bar{X}_{12}X_{21}-\bar{X}_{12}X_{21}\bar{Y}_{12}Y_{21}-\bar{Y}_{12}Y_{21}\bar{X}_{12}X_{21}-Z_1\bar{X}_{12}X_{21}\bar{Z}_1  \\
    &\quad\qquad-\bar{Z}_1\bar{X}_{12}X_{21}Z_1-\bar{X}_{12}X_{21}\bar{X}_{12}X_{21}+2 Y_{12}X_{21}\bar{Y}_{12}\bar{X}_{21}+2 \bar{Y}_{12}\bar{X}_{21}Y_{12}X_{21}+Y_{12}\bar{Y}_{21}\bar{Y}_{12}Y_{21}  \\
    &\quad\qquad+\bar{Y}_{12}Y_{21}Y_{12}\bar{Y}_{21}-Z_1Y_{12}\bar{Y}_{21}\bar{Z}_1-\bar{Z}_1Y_{12}\bar{Y}_{21}Z_1-Y_{12}\bar{Y}_{21}Y_{12}\bar{Y}_{21}-Z_1\bar{Y}_{12}Y_{21}\bar{Z}_1 \\
    &\quad\qquad-\bar{Z}_1\bar{Y}_{12}Y_{21}Z_1-\bar{Y}_{12}Y_{21}\bar{Y}_{12}Y_{21}-Z_1\bar{Z}_1Z_1\bar{Z}_1+Z_1\bar{Z}_1\bar{Z}_1Z_1+\bar{Z}_1Z_1Z_1\bar{Z}_1-\bar{Z}_1Z_1\bar{Z}_1Z_1 \Big)\;,
\end{split}
\ee}
where for clarity we do not explicitly show the parentheses, which are all of the form $(\varphi^i\varphi^j)(\varphi^k\varphi^l)$. As for the shifted terms, they are 
{\small\be \label{Vopenshifted}
\begin{split}
\Vcal^{(s)}_1&=\frac{\kappa}4\Big(-\bracketing{X_{12}}{X_{21}}{\bar{X}_{12}}{\bar{X}_{21}}-\bracketing{\bar{X}_{12}}{\bar{X}_{21}}{X_{12}}{X_{21}}-2 \bracketing{X_{12}}{\bar{Y}_{21}}{\bar{X}_{12}}{Y_{21}}-2 \bracketing{\bar{X}_{12}}{Y_{21}}{X_{12}}{\bar{Y}_{21}}  \\
    &\qquad+\bracketing{X_{12}}{\bar{Y}_{21}}{Y_{12}}{\bar{X}_{21}}+\bracketing{Y_{12}}{\bar{X}_{21}}{X_{12}}{\bar{Y}_{21}}+2 \bracketing{X_{12}}{\bar{X}_{21}}{\bar{Y}_{12}}{Y_{21}}-\bracketing{X_{12}}{Y_{21}}{\bar{Y}_{12}}{\bar{X}_{21}}  \\
    &\qquad-\bracketing{\bar{Y}_{12}}{\bar{X}_{21}}{X_{12}}{Y_{21}}+2 \bracketing{\bar{Y}_{12}}{Y_{21}}{X_{12}}{\bar{X}_{21}}+\bracketing{X_{12}}{\bar{X}_{21}}{X_{12}}{\bar{X}_{21}}+2 \bracketing{\bar{X}_{12}}{X_{21}}{Y_{12}}{\bar{Y}_{21}}  \\
    &\qquad-\bracketing{\bar{X}_{12}}{\bar{Y}_{21}}{Y_{12}}{X_{21}}-\bracketing{Y_{12}}{X_{21}}{\bar{X}_{12}}{\bar{Y}_{21}}+2 \bracketing{Y_{12}}{\bar{Y}_{21}}{\bar{X}_{12}}{X_{21}}+\bracketing{\bar{X}_{12}}{Y_{21}}{\bar{Y}_{12}}{X_{21}} \\
&\qquad+\bracketing{\bar{Y}_{12}}{X_{21}}{\bar{X}_{12}}{Y_{21}}+\bracketing{\bar{X}_{12}}{X_{21}}{\bar{X}_{12}}{X_{21}}-2 \bracketing{Y_{12}}{\bar{X}_{21}}{\bar{Y}_{12}}{X_{21}}-2 \bracketing{\bar{Y}_{12}}{X_{21}}{Y_{12}}{\bar{X}_{21}}\\
&\qquad-\bracketing{Y_{12}}{Y_{21}}{\bar{Y}_{12}}{\bar{Y}_{21}}-\bracketing{\bar{Y}_{12}}{\bar{Y}_{21}}{Y_{12}}{Y_{21}}+\bracketing{Y_{12}}{\bar{Y}_{21}}{Y_{12}}{\bar{Y}_{21}}+\bracketing{\bar{Y}_{12}}{Y_{21}}{\bar{Y}_{12}}{Y_{21}}\Big)\\
&-\frac{1}{4}\Big(\bracketing{X_{12}}{\bar{Z}_2}{\bar{X}_{21}}{Z_1}+\bracketing{Z_1}{X_{12}}{\bar{Z}_2}{\bar{X}_{21}}+\bracketing{X_{12}}{Z_2}{\bar{X}_{21}}{\bar{Z}_1}+\bracketing{\bar{Z}_1}{X_{12}}{Z_2}{\bar{X}_{21}}  \\
    &\qquad+\bracketing{\bar{X}_{12}}{\bar{Z}_2}{X_{21}}{Z_1}+\bracketing{Z_1}{\bar{X}_{12}}{\bar{Z}_2}{X_{21}}+\bracketing{\bar{X}_{12}}{Z_2}{X_{21}}{\bar{Z}_1}+\bracketing{\bar{Z}_1}{\bar{X}_{12}}{Z_2}{X_{21}}  \\
    &\qquad+\bracketing{Y_{12}}{\bar{Z}_2}{\bar{Y}_{21}}{Z_1}+\bracketing{Z_1}{Y_{12}}{\bar{Z}_2}{\bar{Y}_{21}}+\bracketing{Y_{12}}{Z_2}{\bar{Y}_{21}}{\bar{Z}_1}+\bracketing{\bar{Z}_1}{Y_{12}}{Z_2}{\bar{Y}_{21}}  \\
    &\qquad+\bracketing{\bar{Y}_{12}}{\bar{Z}_2}{Y_{21}}{Z_1}+\bracketing{Z_1}{\bar{Y}_{12}}{\bar{Z}_2}{Y_{21}}+\bracketing{\bar{Y}_{12}}{Z_2}{Y_{21}}{\bar{Z}_1}+\bracketing{\bar{Z}_1}{\bar{Y}_{12}}{Z_2}{Y_{21}}\Big)\\
&+\frac{1}{4\kappa}\Big(\bracketing{X_{12}}{\bar{X}_{21}}{Z_1}{\bar{Z}_1}+\bracketing{X_{12}}{\bar{X}_{21}}{\bar{Z}_1}{Z_1}+\bracketing{Z_1}{\bar{Z}_1}{X_{12}}{\bar{X}_{21}}+\bracketing{\bar{Z}_1}{Z_1}{X_{12}}{\bar{X}_{21}} \\
    &\qquad+\bracketing{\bar{X}_{12}}{X_{21}}{Z_1}{\bar{Z}_1} +\bracketing{\bar{X}_{12}}{X_{21}}{\bar{Z}_1}{Z_1}+\bracketing{Z_1}{\bar{Z}_1}{\bar{X}_{12}}{X_{21}}+\bracketing{\bar{Z}_1}{Z_1}{\bar{X}_{12}}{X_{21}} \\
    &\qquad+\bracketing{Y_{12}}{\bar{Y}_{21}}{Z_1}{\bar{Z}_1}+\bracketing{Y_{12}}{\bar{Y}_{21}}{\bar{Z}_1}{Z_1} +\bracketing{Z_1}{\bar{Z}_1}{Y_{12}}{\bar{Y}_{21}}+\bracketing{\bar{Z}_1}{Z_1}{Y_{12}}{\bar{Y}_{21}} \\
&\qquad+\bracketing{\bar{Y}_{12}}{Y_{21}}{Z_1}{\bar{Z}_1}+\bracketing{\bar{Y}_{12}}{Y_{21}}{\bar{Z}_1}{Z_1}+\bracketing{Z_1}{\bar{Z}_1}{\bar{Y}_{12}}{Y_{21}}+\bracketing{\bar{Z}_1}{Z_1}{\bar{Y}_{12}}{Y_{21}}\\
&\qquad-\bracketing{Z_1}{Z_1}{\bar{Z}_1}{\bar{Z}_1}+\bracketing{Z_1}{\bar{Z}_1}{Z_1}{\bar{Z}_1}+\bracketing{\bar{Z}_1}{Z_1}{\bar{Z}_1}{Z_1}-\bracketing{\bar{Z}_1}{\bar{Z}_1}{Z_1}{Z_1}\Big)\;.
\end{split}
\ee}
We note that $\Vcal_1^{(u)}$ and $\Vcal_1^{(s)}$ contain the same number of terms, but the coefficients of the same terms (if we were to ignore the parentheses) are in general different. For example, the terms $(X_{12}\bar{X}_{21})(\bar{Y}_{12}Y_{21})$ and $\bracketing{X_{12}}{\bar{X}_{21}}{\bar{Y}_{12}}{Y_{21}}$ have coefficients $-1/(4\kappa)$ and $\kappa/2$, respectively.

Our goal in Section \ref{scalarinvariance} was to re-express, using the coassociator defined there, all the terms in (\ref{Vopenshifted}) as linear combinations of unshifted terms, which can then be added to (\ref{Vopenunshifted}) in order to be untwisted with a single $(\Fcal^{(4)})^{-1}$. In (\ref{PhiZZZbZbaction}) this was shown explicitly for the last four terms in (\ref{Vopenshifted}), which were added to the last four terms of (\ref{Vopenunshifted}) in (\ref{eq:4sitesZZbStandardBracketing}). Repeating this procedure for the rest of (\ref{Vopenshifted}) leads to an expression which is an overall $\Fcal^{(4)}$ twist of the orbifold-point scalar potential, and is thus annihilated by the coproduct (\ref{Coproduct4sites}).

\section{Quantum planes and twists} \label{QuantumPlanesAppendix}

This appendix presents some of the background underlying our definitions of quantum planes via twists and the twisted action of the algebra generators on these quantum planes. For introductions to these topics, we refer to \cite{Majid,kassel2012quantum}. 

Recall that an algebra $\Acal$ is defined as a vector space together with an associative product $\cdot: \Acal\otimes\Acal\ra \Acal$, a coalgebra is defined by a coassociative coproduct $\Delta:\Acal\ra \Acal\otimes \Acal$, while a bialgebra contains both operations in a compatible way, i.e. $\Delta(X\cdot 
Y)=\Delta(X)\cdot\Delta(Y)$. If the product is the Lie bracket, we then require
\be \label{Deltacommutator}
   [\Delta(X),\Delta(Y)]=\Delta([X,Y])\;\;, \;\;\text{for}\;\; X,Y\in \Acal \ .
   \ee
   The definition of a bialgebra also includes the unit and counit maps, inherited from the algebra and coalgebra definitions respectively, which also need to be compatible. A \emph{Hopf algebra} is a bialgebra with an additional operation, the antipode, which is similar to an inverse. 

   Lie algebras (or rather their universal enveloping algebras) are Hopf algebras where the product is the matrix commutator and the coproducts are simply
   \be
   \Delta(\1)=\1\otimes \1 \;\;\text{and}\;\; \Delta(X)=X\otimes \1+\1\otimes X 
   \ee
which clearly satisfies (\ref{Deltacommutator}). This coproduct is cocommutative, i.e. defining an operation $\tau$ which exchanges the two copies of the algebra, we have $\tau(\Delta(X))=\Delta(X)$. More general Hopf algebras possess noncommutative coproducts. A notable special case is when the two coproducts are related by a similarity transformation with a matrix $R:\Acal\otimes \Acal\ra\Acal\otimes\Acal$, i.e.  $\tau(\Delta(X))=R\Delta(X)R^{-1}$. This is known as a quasitriangular structure, and such quasitriangular Hopf algebras are typically called quantum groups. 

Given a Hopf algebra, one can obtain a new Hopf algebra via the process of twisting. A Drinfeld twist is an invertible map $\Fcal:\Acal\otimes \Acal\ra \Acal\otimes\Acal$ under which the coproduct becomes
   \be \label{twistedcoproduct}
   \Delta_F(X)=\Fcal\Delta(X)\Fcal^{-1} \ .
   \ee
This can be seen to preserve the quasitriangular structure. The unit, counit and antipode are also twisted accordingly, but we will not need to consider them here.  If the twist satisfies a cocycle condition, $(\Fcal\otimes \1)(\Delta\otimes \id)(\Fcal)=(\1\otimes \Fcal)(\id\otimes \Delta)(\Fcal)$, the resulting algebra is coassociative, i.e. one has mapped a Hopf algebra to a new Hopf algebra. 
However, one can also consider more general twists that do not satisfy the cocycle condition. These lead to quasi-Hopf algebras \cite{Drinfeld90}, which are not coassociative. See \cite{Mack:1991tg} for a discussion of the applicability of quasi-Hopf symmetry as an internal symmetry in physics, \cite{Roche:1990hs} for its relevance to the classification of orbifolds of 2d RCFT, and \cite{Dlamini:2019zuk} for previous work on quasi-Hopf symmetry in 4d superconformal theories.
   
A special case of a quasi-Hopf algebra arises when the twist satisfies a \emph{shifted} cocycle condition, where $\Fcal$ depends on an additional, dynamical parameter. Twists with this property lead to the dynamical Yang-Baxter equation \cite{Felder:1994be} which was argued in \cite{Pomoni:2021pbj} to be relevant for the spectral problem of the $\Ncal=2$ orbifold theories.
   
In this work, we argue that the R-symmetry at the orbifold point is related to that in the marginally deformed theories by twists that we can read off from the F- and D-term relations. To understand how to work with twisted coproducts, we start by reviewing how algebra generators act on their module (representation space), which we call $\Vscr$. Calling $m: \Vscr\otimes \Vscr\ra \Vscr$ the product operation on $\Vscr$, and
for $v_1,v_2 \in \Vscr$, the action of a generator on a product state is defined as
   \be \label{Xaction}
   X\triangleright (v_1v_2)=X\triangleright m(v_1\otimes v_2)=m(\Delta(X)\triangleright [v_1\otimes v_2]) \ .
   \ee
When twisting the coproduct as in (\ref{twistedcoproduct}), to obtain a covariant action on the module one often introduces a twisted module product $m_{\Fcal^{-1}}(v_1\otimes v_2)\coloneq m(\Fcal^{-1}\triangleright (v_1\otimes v_2))$, since then
   \be \label{twistedmodule}\begin{split}
   X\triangleright m_{\Fcal^{-1}}(v_1\otimes v_2)&= X\triangleright m\left(\Fcal^{-1}\triangleright [v_1\otimes v_2]\right)=m\left(\Delta(X)\triangleright\Fcal^{-1}\triangleright[v_1\otimes v_2]\right)\\&=m_{\Fcal^{-1}}\left(\Delta_\Fcal(X)\triangleright[v_1\otimes v_2]\right) \ .\end{split}
   \ee
This twisted module product is often called a star product, and this is the approach followed e.g. in \cite{Dlamini:2016aaa,Dlamini:2019zuk} to understand states in the marginally deformed $\Ncal=4$ SYM theory. In this work we will not define star products but work instead with a dual quantum plane formalism \cite{Manin88}, where the coordinates themselves are noncommutative. As in \cite{Mansson:2008xv,Pomoni:2021pbj}, the coordinates of the quantum planes are identified with the scalar fields of our theory.  So our twists will be defined to act on states of the undeformed (orbifold point) theory and produce states of the marginally deformed theory, schematically:
   \be \label{twistedstate}
   \ket{\text{state}}_\Fcal=\Fcal \triangleright\ket{\text{state}}_\op \ .
   \ee
Here we are abusing notation, as twists can only act on $\Vscr\otimes\Vscr$, while a state lives in a single copy of $\Vscr$. Writing an orbifold-point quadratic state as $\ket{\text{state}}_\op=c_{ij}\varphi^i \varphi^j=c_{ij} m(\varphi^i\otimes \varphi^j)$, then what we actually mean by (\ref{twistedstate}) is
   \be
    \ket{\text{state}}_\Fcal= c_{ij}m(\Fcal \triangleright [\varphi^i\otimes \varphi^j])=c_{ij}m((\Fcal^T)^{ij}_{\;kl} (\varphi^k\otimes \varphi^l))=(\Fcal^T)^{ij}_{\;kl} c_{ij}\varphi^k\varphi^l\;,
    \ee
where the last expressions use the explicit tensor components of the twist. The transposition arises because $\triangleright$ means matrix multiplication of $\Fcal=(\Fcal^{ij}_{\;kl})$ with the vectors
    \be
    \varphi^1\otimes \varphi^1=\left(\begin{array}{c}1\\ 0\\ 0\\ 0\end{array}\right)\;\;,\;\;
    \varphi^1\otimes \varphi^2=\left(\begin{array}{c}0\\ 1\\ 0\\ 0\end{array}\right)\;\;,\;\;
    \varphi^2\otimes \varphi^1=\left(\begin{array}{c}0\\ 0\\ 1\\ 0\end{array}\right)\;\;,\;\;
    \varphi^2\otimes \varphi^2=\left(\begin{array}{c}0\\ 0\\ 0\\ 1\end{array}\right)\;,
    \ee
(where for simplicity we take $\Vscr$ to be 2-dimensional) which gives $(F\triangleright \varphi^i\otimes \varphi^j)=\Fcal^{kl}_{\;ij}\varphi^k\otimes \varphi^l=(\Fcal^T)^{ij}_{\;kl}\varphi^k\otimes\varphi^l$. We note that after the twist the fields $\varphi^i\varphi^j$ are no longer commutative (even if we disregard the fact that they are $N\times N$ matrices) but satisfy quantum plane relations.\footnote{The string-theory picture is that of the coordinates of the transverse space to the stack(s) of $D$-branes defining the gauge theory becoming non-commutative (in the sense of the open-string metric \cite{Seiberg:1999vs}) as one deforms away from the orbifold point, and the scalar fields inherit this non-commutativity.} 
    
Now we need to consider the action of the twisted algebra generators on these twisted states. Let us start by expanding the standard Lie-algebraic action (\ref{Xaction}) of a generator $X$ on a product state,
\be\begin{split}
X\triangleright\ket{\text{state}}_\op&=
X\triangleright c_{ij} \varphi^i\varphi^j=c_{ij} m(\Delta_\op(X)\triangleright [\varphi^i\otimes \varphi^j])
=c_{ij} m(X\varphi^i\otimes\varphi^j+\varphi^i\otimes X\varphi^j)\\ &=c_{ij}(((X^T)^i_k\varphi^k)\varphi^j+\varphi^i((X^T)^j_k\varphi^k))=\tilde{c}_{kl}\varphi^k\varphi^l\;,
\end{split}
\ee
where $\tilde{c}_{kl}=c_{ij}((X^T)^i_k\delta^j_l+\delta^k_i(X^T)^j_l$ denote the coefficients of the new state produced by the action of $X$. Here the transpositions arise for a similar reason as above, i.e. that to express for instance $\sigma_-\doublet{1}{0}=\doublet{0}{1}$ in the basis $\varphi^1=\doublet{1}{0}$, $\varphi^2=\doublet{0}{1}$ one needs to write $\sigma_- \varphi^1=(\sigma_-^T)^1_{\;k}\varphi^k$. Note that in the above, the module product $m$ is the path groupoid product (\ref{pathgroupoidproduct}), which is nonzero only for products of fields allowed by the gauge structure, and correspondingly the coproduct should contain additional $\gamma$ operators, as shown in (\ref{groupoidcompatibility}). However, to avoid overburdening the notation, we are not explicitly indicating the groupoid structure.

We can now define the twisted action of $X$ on the corresponding twisted state as
   \be\label{twistedXaction}\begin{split}
   X \triangleright_{{}_\Fcal} \ket{\text{state}}_\Fcal&=c_{ij} m(\Delta_\Fcal(X) \triangleright\Fcal\triangleright [\varphi^i\otimes \varphi^j])=c_{ij}m(\Fcal\triangleright \Delta_0(X)\triangleright [\varphi^i\otimes \varphi^j])\\&=c_{ij}m(\Fcal\triangleright [(X^T)^i_m\varphi^m\otimes \varphi^j+\varphi^i\otimes (X^T)^j_m\varphi^m])\\&=(\Fcal^T)^{mj}_{\;kl} c_{ij}((X^T)^i_m\varphi^k)\varphi^l+(\Fcal^T)^{im}_{kl}c_{ij}\varphi^k((X^T)^j_m\varphi^l))\\
   &=(\Fcal^T)^{mn}_{kl}\tilde{c}_{mn}\varphi^k\varphi^l=\Fcal\triangleright X\triangleright\ket{\text{state}}_\op \;.
\end{split}
   \ee   
   What this formula tells us is that 
   the action of first twisting the state and then acting with the twisted action of $X$ is equivalent to first acting with the undeformed $X$ on the untwisted state and then twisting. This is the dual statement to that in (\ref{twistedmodule}), and shows that, under twisting, all the multiplets of the undeformed theory map to multiplets of the deformed theory.
   
   To help clarify the above formulas, let us consider the example of the $XZ$ sector, where the two-site twist (\ref{diagonaltwist}) in matrix form is
   \be
   \Fcal=\left(\begin{array}{cccc}1&0&0&0\\0&1&0&0\\ 0&0&\kappa^{-1}&0\\ 0&0&0&\kappa^{-1}\end{array}\right)\;\;\quad\text{in the basis} \;\;\left(\begin{array}{c}X_{12}X_{21}\\ X_{12}Z_2\\Z_1X_{12}\\Z_1Z_1\end{array}\right)\;.
   \ee
   Acting on the highest weight state $X_{12}X_{21}=c_{11}\varphi^1\varphi^1$ (with $c_{11}=1$) with the lowering operator $\sigma_-$, (\ref{twistedXaction}) evaluates to
   \be
   \sigma_- \triangleright_{{}_\Fcal} (X_{12}X_{21})=c_{11}(\Fcal^T)^{21}_{\;2l}((\sigma^T_-)^1_2\varphi^2)\varphi^l+c_{11}(\Fcal^T)^{12}_{\;k2}\varphi^k((\sigma_-^T)^1_2\varphi^2)\;,
   \ee
   which correctly produces the state $\kappa^{-1}Z_1X_{12}+X_{12}Z_2$ that we would have obtained by acting on $X_{12}X_{21}$ with the undeformed coproduct of $\sigma_-$ and then twisting.

   The transposition of $\Fcal$ when working with indices is not very consequential for us, as all our twists are symmetric, $\Fcal^{ij}_{\;kl}=\Fcal^{kl}_{\;ij}$. However, it becomes important when extending to more sites, as in Section \ref{Superpotentialinvariance} where e.g. the matrix $(\Fcal\otimes\1)(\Delta\otimes \id)(\Fcal)$ appears in transposed form when acting on states. 

 Having explained how our generators act on states by use of the twisted coproduct, when acting with broken generators in the main text we will not use the more precise notation above, but just show the twisted coproducts acting on states.

\section{The coassociator for the scalar potential} \label{CoassociatorAppendix}

As discussed in Section \ref{sec:Coassociator}, in order to rewrite the shifted states in the scalar potential to unshifted ones, we need to use the coassociator $\Phi=\Fcal^{(4)}(\Fcal^{(4)}_{\text{shifted}})^{-1}$. In this Appendix we will write down the coassociator explicitly in matrix form, by acting on all the shifted monomials, which gives linear combinations of unshifted monomials. An example of such a computation was illustrated in (\ref{PhiZZZZ}). The matrices we write down contain the components of $\Phi^T$, which are relevant for unshifting each shifted monomial, as in (\ref{monomialrotation}). 

Let us note that since all our twists satisfy $\Fcal(\kappa)^{-1}=\Fcal(\kappa^{-1})$, we have that
\be
\Phi^{-1}(\kappa)=\Fcal^{(4)}_{\text{shifted}}(\kappa)(\Fcal^{(4)}(\kappa))^{-1}=(\Fcal^{(4)}_{\text{shifted}}(\kappa^{-1}))^{-1} \Fcal^{(4)}(\kappa^{-1})=\Phi^T(\kappa^{-1})
\ee
i.e. combining a $\Zset_2$ transformation with transposition gives the inverse of $\Phi$. This property can be shown to hold for all the matrices below. 

Since our twists also preserve the number of fields of each type, the coassociator factorises into blocks with fixed numbers of $Z$ and $\Zb$ fields, so we will present the components in each block separately. We will also focus on states with first index in gauge group 1, those with first index in gauge group 2 follow by $\Zset_2$ conjugation. Finally, in this appendix we do not indicate the $\kappa$ subscript on the states, as no orbifold-point states appear.

\subsection*{The $ZZ\bar{Z} \bar{Z}$-sector}

For the pure $ZZ\bar{Z} \bar{Z}$-sector, the coassociator allowing us to express states of the form $\bracing{\bracketing{A}{B}{C}{D}}$ in terms of linear combinations of $\bracing{(AB)(CD)}$ states, 
\begin{equation}
     \begin{pmatrix}
        \bracing{\bracketing{Z_1}{Z_1}{\bar{Z}_1}{\bar{Z}_1}} \\
        \bracing{\bracketing{Z_1}{\bar{Z}_1}{Z_1}{\bar{Z}_1}} \\
        \bracing{\bracketing{Z_1}{\bar{Z}_1}{\bar{Z}_1}{Z_1}} \\
        \bracing{\bracketing{\bar{Z}_1}{Z_1}{Z_1}{\bar{Z}_1}} \\
        \bracing{\bracketing{\bar{Z}_1}{Z_1}{\bar{Z}_1}{Z_1}} \\
        \bracing{\bracketing{\bar{Z}_1}{\bar{Z}_1}{Z_1}{Z_1}}
    \end{pmatrix}
    = [\Phi^T_{ZZ\Zb\Zb}]_{{}_{6\times 6}}\begin{pmatrix}
    \bracing{(Z_1 Z_1)(\bar{Z}_1 \bar{Z}_1)} \\
    \bracing{(Z_1 \bar{Z}_1)(Z_1 \bar{Z}_1)} \\
    \bracing{(Z_1 \bar{Z}_1)(\bar{Z}_1 Z_1)} \\
    \bracing{(\bar{Z}_1 Z_1)(Z_1 \bar{Z}_1)} \\
    \bracing{(\bar{Z}_1 Z_1)(\bar{Z}_1 Z_1)} \\
    \bracing{(\bar{Z}_1 \bar{Z}_1)(Z_1 Z_1)} 
\end{pmatrix} \ ,
\end{equation}
takes the explicit matrix form
\be
\Phi^T_{ZZ\Zb\Zb}=
\left(
\begin{array}{cccccc}
 \frac{\left(\sqrt{\kappa }+1\right)^2}{4 \kappa ^2} & -\frac{\kappa ^2-1}{8 \kappa } & -\frac{(\kappa -1)^2}{8 \kappa } & -\frac{(\kappa -1)^2}{8 \kappa } & -\frac{\kappa ^2-1}{8 \kappa } & \frac{\left(\sqrt{\kappa }-1\right)^2}{4 \kappa ^2} \\
 -\frac{\kappa -1}{4 \kappa ^2} & \frac{\kappa ^2+6 \kappa +1}{8 \kappa } & \frac{\kappa ^2-1}{8 \kappa } & \frac{\kappa ^2-1}{8 \kappa } & \frac{(\kappa -1)^2}{8 \kappa } & -\frac{\kappa -1}{4 \kappa ^2} \\
 0 & \frac{1}{4} (\kappa -1) \kappa  & \frac{1}{4} \left(\sqrt{\kappa }+1\right)^2 \kappa  & \frac{1}{4} \left(\sqrt{\kappa }-1\right)^2 \kappa  & \frac{1}{4} (\kappa -1) \kappa  & 0 \\
 0 & \frac{1}{4} (\kappa -1) \kappa  & \frac{1}{4} \left(\sqrt{\kappa }-1\right)^2 \kappa  & \frac{1}{4} \left(\sqrt{\kappa }+1\right)^2 \kappa  & \frac{1}{4} (\kappa -1) \kappa  & 0 \\
 -\frac{\kappa -1}{4 \kappa ^2} & \frac{(\kappa -1)^2}{8 \kappa } & \frac{\kappa ^2-1}{8 \kappa } & \frac{\kappa ^2-1}{8 \kappa } & \frac{\kappa ^2+6 \kappa +1}{8 \kappa } & -\frac{\kappa -1}{4 \kappa ^2} \\
 \frac{\left(\sqrt{\kappa }-1\right)^2}{4 \kappa ^2} & -\frac{\kappa ^2-1}{8 \kappa } & -\frac{(\kappa -1)^2}{8 \kappa } & -\frac{(\kappa -1)^2}{8 \kappa } & -\frac{\kappa ^2-1}{8 \kappa } & \frac{\left(\sqrt{\kappa }+1\right)^2}{4 \kappa ^2} \\
\end{array}
\right)\;.
\ee
It has a determinant of $1$ and, as mentioned, its inverse corresponds to its $\Zset_2$-conjugate transposed. We note that not all the six shifted monomials appear in the scalar potential, but we need to consider all of them in order for $\Phi^T$ to be a square matrix. It is useful to show the action of $\Phi_{ZZ\Zb\Zb}$ on the actual linear combination appearing in (\ref{Vopenshifted}). It is
\begin{equation}
     \left(1,-1,0,0,-1,1\right) \xrightarrow{\Phi_{ZZ\bar{Z}\bar{Z}}} \left(\frac{1}{\kappa },\frac{1}{2} (-\kappa -1),\frac{1-\kappa }{2},\frac{1-\kappa }{2},\frac{1}{2} (-\kappa -1),\frac{1}{\kappa }\right) \ ,
\end{equation}
which is the same as (\ref{PhiZZZbZbaction}). 
The coassociator in this sector admits an even simpler form if we perform a basis transformation. Choosing the following basis for shifted and unshifted states:
\begin{equation}
    \begin{pmatrix}
        \bracing{\bracketing{Z_1}{Z_1}{\bar{Z}_1}{\bar{Z}_1}}-\bracing{\bracketing{\bar{Z}_1}{\bar{Z}_1}{Z_1}{Z_1}}\\
        \bracing{\bracketing{Z_1}{Z_1}{\bar{Z}_1}{\bar{Z}_1}}+\bracing{\bracketing{\bar{Z}_1}{\bar{Z}_1}{Z_1}{Z_1}}\\
        \bracing{\bracketing{Z_1}{\bar{Z}_1}{Z_1}{\bar{Z}_1}}-\bracing{\bracketing{\bar{Z}_1}{Z_1}{\bar{Z}_1}{Z_1}}\\
        \bracing{\bracketing{Z_1}{\bar{Z}_1}{Z_1}{\bar{Z}_1}}+\bracing{\bracketing{\bar{Z}_1}{Z_1}{\bar{Z}_1}{Z_1}}\\
        \bracing{\bracketing{Z_1}{\bar{Z}_1}{\bar{Z}_1}{Z_1}}-\bracing{\bracketing{\bar{Z}_1}{Z_1}{Z_1}{\bar{Z}_1}}\\
        \bracing{\bracketing{Z_1}{\bar{Z}_1}{\bar{Z}_1}{Z_1}}+\bracing{\bracketing{\bar{Z}_1}{Z_1}{Z_1}{\bar{Z}_1}}
    \end{pmatrix} \;\;\text{and}\;\;\begin{pmatrix}
        \bracing{(Z_1Z_1)(\bar{Z}_1\bar{Z}_1)}-\bracing{(\bar{Z}_1\bar{Z}_1)(Z_1Z_1)}\\
        \bracing{(Z_1Z_1)(\bar{Z}_1\bar{Z}_1)}+\bracing{(\bar{Z}_1\bar{Z}_1)(Z_1Z_1)}\\
        \bracing{(Z_1\bar{Z}_1)(Z_1\bar{Z}_1)}-\bracing{(\bar{Z}_1Z_1)(\bar{Z}_1Z_1)}\\
        \bracing{(Z_1\bar{Z}_1)(Z_1\bar{Z}_1)}+\bracing{(\bar{Z}_1Z_1)(\bar{Z}_1Z_1)}\\
        \bracing{(Z_1\bar{Z}_1)(\bar{Z}_1Z_1)}-\bracing{(\bar{Z}_1Z_1)(Z_1\bar{Z}_1)}\\
        \bracing{(Z_1\bar{Z}_1)(\bar{Z}_1Z_1)}+\bracing{(\bar{Z}_1Z_1)(Z_1\bar{Z}_1)}
    \end{pmatrix} \ ,
\end{equation}
we find that $\Phi_{ZZ\bar{Z}\bar{Z}}$ takes the simplified form
\begin{equation}
   \left(
\begin{array}{cccccc}
 \frac{1}{\kappa ^{3/2}} & 0 & 0 & 0 & 0 & 0 \\
 0 & \frac{\kappa +1}{2 \kappa ^2} & 0 & -\frac{\kappa ^2-1}{4 \kappa } & 0 & -\frac{(\kappa -1)^2}{4 \kappa } \\
 0 & 0 & 1 & 0 & 0 & 0 \\
 0 & -\frac{\kappa -1}{2 \kappa ^2} & 0 & \frac{(\kappa +1)^2}{4 \kappa } & 0 & \frac{\kappa ^2-1}{4 \kappa } \\
 0 & 0 & 0 & 0 & \kappa ^{3/2} & 0 \\
 0 & 0 & 0 & \frac{1}{2} (\kappa -1) \kappa  & 0 & \frac{1}{2} \kappa  (\kappa +1) \\
\end{array}
\right) \ .
\end{equation}

\subsection*{The $XZ\bar{X}\bar{Z}$-sector}

For the $XZ\bar{X}\bar{Z}$-sector the coassociator factorises into three blocks of dimensions $8 \times 8, \hspace{1pt} 16 \times 16$ and $8 \times 8$ with determinants $\{\kappa^5,1,\kappa^5\}$, respectively. For the first $8 \times 8$ block, choosing the bases
\begin{equation}
    \begin{pmatrix}
        \bracing{\bracketing{X_{12}}{\bar{X}_{21}}{Z_1}{\bar{Z}_1}} \\
        \bracing{\bracketing{X_{12}}{\bar{X}_{21}}{\bar{Z}_1}{Z_1}} \\
        \bracing{\bracketing{\bar{X}_{12}}{X_{21}}{Z_1}{\bar{Z}_1}} \\
        \bracing{\bracketing{\bar{X}_{12}}{X_{21}}{\bar{Z}_1}{Z_1}} \\
        \bracing{\bracketing{Y_{12}}{\bar{Y}_{21}}{Z_1}{\bar{Z}_1}} \\
        \bracing{\bracketing{Y_{12}}{\bar{Y}_{21}}{\bar{Z}_1}{Z_1}} \\
        \bracing{\bracketing{\bar{Y}_{12}}{Y_{21}}{\bar{Z}_1}{Z_1}} \\
        \bracing{\bracketing{\bar{Y}_{12}}{Y_{21}}{Z_1}{\bar{Z}_1}} \end{pmatrix}\;\;\text{and}\;\;\begin{pmatrix}
    \bracing{(X_{12} \bar{X}_{21})(Z_1 \bar{Z}_1)} \\
    \bracing{(X_{12}\bar{X}_{21})(\bar{Z}_1Z_1)} \\
    \bracing{(\bar{X}_{12}X_{21})(Z_1\bar{Z}_1)} \\
    \bracing{(\bar{X}_{12}X_{21})(\bar{Z}_1Z_1)} \\
    \bracing{(Y_{12}\bar{Y}_{21})(Z_1\bar{Z}_1)} \\
    \bracing{(Y_{12}\bar{Y}_{21})(\bar{Z}_1Z_1)} \\
    \bracing{(\bar{Y}_{12}Y_{21})(\bar{Z}_1Z_1)} \\
    \bracing{(\bar{Y}_{12}Y_{21})(Z_1\bar{Z}_1)}
    \end{pmatrix}\;,
\end{equation}
we calculate $\Phi^T_{X\bar{X}Z\bar{Z}}$ to be
\begin{equation}
\resizebox{14cm}{!}{%
    $\frac{1}{8} \left( \begin{array}{cccccccc}
    3 \kappa +4 \sqrt{\kappa }+1 & 3 \kappa -2 \sqrt{\kappa }-1 & \kappa -1 & \left(\sqrt{\kappa }-1\right)^2 & 1-\kappa  & -\left(\sqrt{\kappa }-1\right)^2 & \left(\sqrt{\kappa }-1\right)^2 & \kappa -1 \\
    3 \kappa -2 \sqrt{\kappa }-1 & 3 \kappa +4 \sqrt{\kappa }+1 & \left(\sqrt{\kappa }-1\right)^2 & \kappa -1 & -\left(\sqrt{\kappa }-1\right)^2 & 1-\kappa  & \kappa -1 & \left(\sqrt{\kappa }-1\right)^2 \\
    \kappa -1 & \left(\sqrt{\kappa }-1\right)^2 & 3 \kappa +4 \sqrt{\kappa }+1 & 3 \kappa -2 \sqrt{\kappa }-1 & \kappa -1 & \left(\sqrt{\kappa }-1\right)^2 & -\left(\sqrt{\kappa }-1\right)^2 & 1-\kappa  \\
    \left(\sqrt{\kappa }-1\right)^2 & \kappa -1 & 3 \kappa -2 \sqrt{\kappa }-1 & 3 \kappa +4 \sqrt{\kappa }+1 & \left(\sqrt{\kappa }-1\right)^2 & \kappa -1 & 1-\kappa  & -\left(\sqrt{\kappa }-1\right)^2 \\
    1-\kappa  & -\left(\sqrt{\kappa }-1\right)^2 & \kappa -1 & \left(\sqrt{\kappa }-1\right)^2 & 3 \kappa +4 \sqrt{\kappa }+1 & 3 \kappa -2 \sqrt{\kappa }-1 & \left(\sqrt{\kappa }-1\right)^2 & \kappa -1 \\
    -\left(\sqrt{\kappa }-1\right)^2 & 1-\kappa  & \left(\sqrt{\kappa }-1\right)^2 & \kappa -1 & 3 \kappa -2 \sqrt{\kappa }-1 & 3 \kappa +4 \sqrt{\kappa }+1 & \kappa -1 & \left(\sqrt{\kappa }-1\right)^2 \\
    \left(\sqrt{\kappa }-1\right)^2 & \kappa -1 & -\left(\sqrt{\kappa }-1\right)^2 & 1-\kappa  & \left(\sqrt{\kappa }-1\right)^2 & \kappa -1 & 3 \kappa +4 \sqrt{\kappa }+1 & 3 \kappa -2 \sqrt{\kappa }-1 \\
    \kappa -1 & \left(\sqrt{\kappa }-1\right)^2 & 1-\kappa  & -\left(\sqrt{\kappa }-1\right)^2 & \kappa -1 & \left(\sqrt{\kappa }-1\right)^2 & 3 \kappa -2 \sqrt{\kappa }-1 & 3 \kappa +4 \sqrt{\kappa }+1 \\
    \end{array} \right) $%
    } \ .
\end{equation}

As above, it is more insightful to consider the action of $\Phi_{X\bar{X}Z\bar{Z}}$ on the actual shifted linear combination appearing in (\ref{Vopenshifted}). We find
\begin{equation}  \label{PhiXXbZZbaction}
     \frac{1}{\kappa}\left(1,1,1,1,1,1,1,1\right) \xrightarrow{\Phi_{X\bar{X}Z\bar{Z}}} (1,1,1,1,1,1,1,1) \ ,
\end{equation}
i.e. the coassociator simply strips away an overall $\kappa$-dependence.

As for the $ZZ\bar{Z}\bar{Z}$-sector, also for $\Phi_{X\bar{X}Z\bar{Z}}$ we can perform a change of basis for shifted states:
\begin{equation}
    \resizebox{14cm}{!}{%
    $\begin{pmatrix}
        -\bracing{\bracketing{X_{12}}{\bar{X}_{21}}{Z_1}{\bar{Z}_1}}+\bracing{\bracketing{X_{12}}{\bar{X}_{21}}{\bar{Z}_1}{Z_1}}+\bracing{\bracketing{\bar{X}_{12}}{X_{21}}{Z_1}{\bar{Z}_1}}-\bracing{\bracketing{\bar{X}_{12}}{X_{21}}{\bar{Z}_1}{Z_1}}-\bracing{\bracketing{Y_{12}}{\bar{Y}_{21}}{Z_1}{\bar{Z}_1}}+\bracing{\bracketing{Y_{12}}{\bar{Y}_{21}}{\bar{Z}_1}{Z_1}}+\bracing{\bracketing{\bar{Y}_{12}}{Y_{21}}{Z_1}{\bar{Z}_1}}-\bracing{\bracketing{\bar{Y}_{12}}{Y_{21}}{\bar{Z}_1}{Z_1}}\\
        \bracing{\bracketing{X_{12}}{\bar{X}_{21}}{Z_1}{\bar{Z}_1}}-2 \bracing{\bracketing{X_{12}}{\bar{X}_{21}}{\bar{Z}_1}{Z_1}}+\bracing{\bracketing{\bar{X}_{12}}{X_{21}}{Z_1}{\bar{Z}_1}}-\bracing{\bracketing{Y_{12}}{\bar{Y}_{21}}{Z_1}{\bar{Z}_1}}+\bracing{\bracketing{\bar{Y}_{12}}{Y_{21}}{Z_1}{\bar{Z}_1}}\\
        -\bracing{\bracketing{X_{12}}{\bar{X}_{21}}{\bar{Z}_1}{Z_1}}+\bracing{\bracketing{\bar{X}_{12}}{X_{21}}{Z_1}{\bar{Z}_1}}-\bracing{\bracketing{Y_{12}}{\bar{Y}_{21}}{Z_1}{\bar{Z}_1}}+\bracing{\bracketing{\bar{Y}_{12}}{Y_{21}}{\bar{Z}_1}{Z_1}}\\
        \bracing{\bracketing{X_{12}}{\bar{X}_{21}}{Z_1}{\bar{Z}_1}}-\bracing{\bracketing{X_{12}}{\bar{X}_{21}}{\bar{Z}_1}{Z_1}}-\bracing{\bracketing{Y_{12}}{\bar{Y}_{21}}{Z_1}{\bar{Z}_1}}+\bracing{\bracketing{Y_{12}}{\bar{Y}_{21}}{\bar{Z}_1}{Z_1}}\\
        -\bracing{\bracketing{X_{12}}{\bar{X}_{21}}{Z_1}{\bar{Z}_1}}+\bracing{\bracketing{X_{12}}{\bar{X}_{21}}{\bar{Z}_1}{Z_1}}-\bracing{\bracketing{\bar{X}_{12}}{X_{21}}{Z_1}{\bar{Z}_1}}+\bracing{\bracketing{\bar{X}_{12}}{X_{21}}{\bar{Z}_1}{Z_1}}\\
        \bracing{\bracketing{X_{12}}{\bar{X}_{21}}{Z_1}{\bar{Z}_1}}+\bracing{\bracketing{X_{12}}{\bar{X}_{21}}{\bar{Z}_1}{Z_1}}+\bracing{\bracketing{\bar{Y}_{12}}{Y_{21}}{Z_1}{\bar{Z}_1}}+\bracing{\bracketing{\bar{Y}_{12}}{Y_{21}}{\bar{Z}_1}{Z_1}}\\
        -\bracing{\bracketing{X_{12}}{\bar{X}_{21}}{Z_1}{\bar{Z}_1}}-\bracing{\bracketing{X_{12}}{\bar{X}_{21}}{\bar{Z}_1}{Z_1}}+\bracing{\bracketing{Y_{12}}{\bar{Y}_{21}}{Z_1}{\bar{Z}_1}}+\bracing{\bracketing{Y_{12}}{\bar{Y}_{21}}{\bar{Z}_1}{Z_1}}\\
        \bracing{\bracketing{X_{12}}{\bar{X}_{21}}{Z_1}{\bar{Z}_1}}+\bracing{\bracketing{X_{12}}{\bar{X}_{21}}{\bar{Z}_1}{Z_1}}+\bracing{\bracketing{\bar{X}_{12}}{X_{21}}{Z_1}{\bar{Z}_1}}+\bracing{\bracketing{\bar{X}_{12}}{X_{21}}{\bar{Z}_1}{Z_1}}
    \end{pmatrix}$%
    }\;,
\end{equation}
and similarly for the unshifted ones:
\begin{equation}
    \resizebox{14cm}{!}{%
    $\begin{pmatrix}
        -\bracing{(X_{12}\bar{X}_{21})(Z_1\bar{Z}_1)}+\bracing{(X_{12}\bar{X}_{21})(\bar{Z}_1Z_1)}+\bracing{(\bar{X}_{12}X_{21})(Z_1\bar{Z}_1)}-\bracing{(\bar{X}_{12}X_{21})(\bar{Z}_1Z_1)}-\bracing{(Y_{12}\bar{Y}_{21})(Z_1\bar{Z}_1)}+\bracing{(Y_{12}\bar{Y}_{21})(\bar{Z}_1Z_1)}+\bracing{(\bar{Y}_{12}Y_{21})(Z_1\bar{Z}_1)}-\bracing{(\bar{Y}_{12}Y_{21})(\bar{Z}_1Z_1)}\\
        \bracing{(X_{12}\bar{X}_{21})(Z_1\bar{Z}_1)}-2 \bracing{(X_{12}\bar{X}_{21})(\bar{Z}_1Z_1)}+\bracing{(\bar{X}_{12}X_{21})(Z_1\bar{Z}_1)}-\bracing{(Y_{12}\bar{Y}_{21})(Z_1\bar{Z}_1)}+\bracing{(\bar{Y}_{12}Y_{21})(Z_1\bar{Z}_1)}\\
        -\bracing{(X_{12}\bar{X}_{21})(\bar{Z}_1Z_1)}+\bracing{(\bar{X}_{12}X_{21})(Z_1\bar{Z}_1)}-\bracing{(Y_{12}\bar{Y}_{21})(Z_1\bar{Z}_1)}+\bracing{(\bar{Y}_{12}Y_{21})(\bar{Z}_1Z_1)}\\
        \bracing{(X_{12}\bar{X}_{21})(Z_1\bar{Z}_1)}-\bracing{(X_{12}\bar{X}_{21})(\bar{Z}_1Z_1)}-\bracing{(Y_{12}\bar{Y}_{21})(Z_1\bar{Z}_1)}+\bracing{(Y_{12}\bar{Y}_{21})(\bar{Z}_1Z_1)}\\
        -\bracing{(X_{12}\bar{X}_{21})(Z_1\bar{Z}_1)}+\bracing{(X_{12}\bar{X}_{21})(\bar{Z}_1Z_1)}-\bracing{(\bar{X}_{12}X_{21})(Z_1\bar{Z}_1)}+\bracing{(\bar{X}_{12}X_{21})(\bar{Z}_1Z_1)}\\
        \bracing{(X_{12}\bar{X}_{21})(Z_1\bar{Z}_1)}+\bracing{(X_{12}\bar{X}_{21})(\bar{Z}_1Z_1)}+\bracing{(\bar{Y}_{12}Y_{21})(Z_1\bar{Z}_1)}+\bracing{(\bar{Y}_{12}Y_{21})(\bar{Z}_1Z_1)}\\
        -\bracing{(X_{12}\bar{X}_{21})(Z_1\bar{Z}_1)}-\bracing{(X_{12}\bar{X}_{21})(\bar{Z}_1Z_1)}+\bracing{(Y_{12}\bar{Y}_{21})(Z_1\bar{Z}_1)}+\bracing{(Y_{12}\bar{Y}_{21})(\bar{Z}_1Z_1)}\\
        \bracing{(X_{12}\bar{X}_{21})(Z_1\bar{Z}_1)}+\bracing{(X_{12}\bar{X}_{21})(\bar{Z}_1Z_1)}+\bracing{(\bar{X}_{12}X_{21})(Z_1\bar{Z}_1)}+\bracing{(\bar{X}_{12}X_{21})(\bar{Z}_1Z_1)}
    \end{pmatrix}$%
    }
\end{equation}
In this basis, $\Phi_{X\bar{X}Z\bar{Z}}$ is now diagonal:\footnote{The top component of the shifted bracketing basis is a collection of terms like $F\times \bar{F}$, whereas the corresponding term of the unshifted basis can be thought of as $G_1^0 \cdot E_1$. This indicates mixing of the F- and D-term contributions due to the coassociator when changing the bracketing.}
\begin{equation} \label{PhiXXbZZbdiagonal}
   \left(
\begin{array}{cccccccc}
 1 & 0 & 0 & 0 & 0 & 0 & 0 & 0 \\
 0 & \sqrt{\kappa } & 0 & 0 & 0 & 0 & 0 & 0 \\
 0 & 0 & \sqrt{\kappa } & 0 & 0 & 0 & 0 & 0 \\
 0 & 0 & 0 & \sqrt{\kappa } & 0 & 0 & 0 & 0 \\
 0 & 0 & 0 & 0 & \sqrt{\kappa } & 0 & 0 & 0 \\
 0 & 0 & 0 & 0 & 0 & \kappa  & 0 & 0 \\
 0 & 0 & 0 & 0 & 0 & 0 & \kappa  & 0 \\
 0 & 0 & 0 & 0 & 0 & 0 & 0 & \kappa  \\
\end{array}
\right) \ .
\end{equation}
Adding the actions on the last three basis elements (which give the actual combination appearing in (\ref{Vopenshifted})) explains the simple form of (\ref{PhiXXbZZbaction}). 

For the $16\times 16$ sector, the coassociator is simply the identity matrix:
\begin{equation}
    \begin{pmatrix}
        \bracing{\bracketing{X_{12}}{Z_2}{\bar{X}_{21}}{\bar{Z}_1}} \\
        \bracing{\bracketing{X_{12}}{\bar{Z}_2}{\bar{X}_{21}}{Z_1}} \\
        \bracing{\bracketing{\bar{X}_{12}}{Z_2}{X_{21}}{\bar{Z}_1}} \\
        \bracing{\bracketing{\bar{X}_{12}}{\bar{Z}_2}{X_{21}}{Z_1}} \\
        \bracing{\bracketing{Y_{12}}{Z_2}{\bar{Y}_{21}}{\bar{Z}_1}} \\
        \bracing{\bracketing{Y_{12}}{\bar{Z}_2}{\bar{Y}_{21}}{Z_1}} \\
        \bracing{\bracketing{\bar{Y}_{12}}{Z_2}{Y_{21}}{\bar{Z}_1}} \\
        \bracing{\bracketing{\bar{Y}_{12}}{\bar{Z}_2}{Y_{21}}{Z_1}} \\
        \bracing{\bracketing{Z_1}{X_{12}}{\bar{Z}_2}{\bar{X}_{21}}} \\
        \bracing{\bracketing{Z_1}{\bar{X}_{12}}{\bar{Z}_2}{X_{21}}} \\
        \bracing{\bracketing{Z_1}{Y_{12}}{\bar{Z}_2}{\bar{Y}_{21}}} \\
        \bracing{\bracketing{Z_1}{\bar{Y}_{12}}{\bar{Z}_2}{Y_{21}}} \\
        \bracing{\bracketing{\bar{Z}_1}{X_{12}}{Z_2}{\bar{X}_{21}}} \\
        \bracing{\bracketing{\bar{Z}_1}{\bar{X}_{21}}{Z_2}{X_{21}}} \\
        \bracing{\bracketing{\bar{Z}_1}{Y_{12}}{Z_2}{\bar{Y}_{21}}} \\
        \bracing{\bracketing{\bar{Z}_1}{\bar{Y}_{12}}{Z_2}{Y_{21}}} 
    \end{pmatrix}
    = \1_{16\times 16}\begin{pmatrix}
        \bracing{(X_{12}Z_2)(\bar{X}_{21}\bar{Z}_1)} \\
        \bracing{(X_{12}\bar{Z}_2)(\bar{X}_{21}Z_1)} \\
        \bracing{(\bar{X}_{12}Z_2)(X_{21}\bar{Z}_1)} \\
        \bracing{(\bar{X}_{12}\bar{Z}_2)(X_{21}Z_1)} \\
        \bracing{(Y_{12}Z_2)(\bar{Y}_{21}\bar{Z}_1)} \\
        \bracing{(Y_{12}\bar{Z}_2)(\bar{Y}_{21}Z_1)} \\
        \bracing{(\bar{Y}_{12}Z_2)(Y_{21}\bar{Z}_1)} \\
        \bracing{(\bar{Y}_{12}\bar{Z}_2)(Y_{21}Z_1)} \\
        \bracing{(Z_1X_{12})(\bar{Z}_2\bar{X}_{21})} \\
        \bracing{(Z_1\bar{X}_{12})(\bar{Z}_2X_{21})} \\
        \bracing{(Z_1Y_{12})(\bar{Z}_2\bar{Y}_{21})} \\
        \bracing{(Z_1\bar{Y}_{12})(\bar{Z}_2Y_{21})} \\
        \bracing{(\bar{Z}_1X_{12})(Z_2\bar{X}_{21})} \\
        \bracing{(\bar{Z}_1\bar{X}_{21})(Z_2X_{21})} \\
        \bracing{(\bar{Z}_1Y_{12})(Z_2\bar{Y}_{21})} \\
        \bracing{(\bar{Z}_1\bar{Y}_{12})(Z_2Y_{21})} 
\end{pmatrix} 
\end{equation}
So in this sector we can freely change bracketings also way from the orbifold point. 
These terms appear with equal coefficients in (\ref{Vopenshifted}) and of course we have
\begin{equation}
     \left(1,1,1,1,1,1,1,1,1,1,1,1,1,1,1,1\right) \xrightarrow{\Phi_{XZ\bar{X}\bar{Z}}} (1,1,1,1,1,1,1,1,1,1,1,1,1,1,1,1) \ .
\end{equation}

The final $8\times 8$ block appearing in the coassociator for the $XZ\bar{X}\bar{Z}$-sector acts between
\begin{equation}
    \begin{pmatrix}
        \bracing{\bracketing{Z_1}{\bar{Z}_1}{X_{12}}{\bar{X}_{21}}} \\
        \bracing{\bracketing{Z_1}{\bar{Z}_1}{\bar{X}_{12}}{X_{21}}} \\
        \bracing{\bracketing{Z_1}{\bar{Z}_1}{Y_{12}}{\bar{Y}_{21}}} \\
        \bracing{\bracketing{Z_1}{\bar{Z}_1}{\bar{Y}_{12}}{Y_{21}}} \\
        \bracing{\bracketing{\bar{Z}_1}{Z_1}{X_{12}}{\bar{X}_{21}}} \\
        \bracing{\bracketing{\bar{Z}_1}{Z_1}{\bar{X}_{12}}{X_{21}}} \\
        \bracing{\bracketing{\bar{Z}_1}{Z_1}{Y_{12}}{\bar{Y}_{21}}} \\
        \bracing{\bracketing{\bar{Z}_1}{Z_1}{\bar{Y}_{12}}{Y_{21}}}
    \end{pmatrix} \;\;\text{and}\;\;\begin{pmatrix}
    \bracing{(Z_1 \bar{Z}_1)(X_{12} \bar{X}_{21})} \\
    \bracing{(Z_1\bar{Z}_1)(\bar{X}_{12}X_{21})} \\
    \bracing{(Z_1\bar{Z}_1)(Y_{12}\bar{Y}_{21})} \\
    \bracing{(Z_1\bar{Z}_1)(\bar{Y}_{12}Y_{21})} \\
    \bracing{(\bar{Z}_1Z_1)(X_{12}\bar{X}_{21})} \\
    \bracing{(\bar{Z}_1Z_1)(\bar{X}_{12}X_{21})} \\
    \bracing{(\bar{Z}_1Z_1)(Y_{12}\bar{Y}_{21})} \\
    \bracing{(\bar{Z}_1Z_1)(\bar{Y}_{12}Y_{21})}
\end{pmatrix} \ ,
\end{equation}
with $\Phi_{Z\bar{Z}X\bar{X}}^T$ being
\begin{equation}
\resizebox{14cm}{!}{%
    $\frac{1}{8} \left(
\begin{array}{cccccccc}
 3 \kappa +4 \sqrt{\kappa }+1 & \kappa -1 & 1-\kappa  & \kappa -1 & 3 \kappa -2 \sqrt{\kappa }-1 & \left(\sqrt{\kappa }-1\right)^2 & -\left(\sqrt{\kappa }-1\right)^2 & \left(\sqrt{\kappa }-1\right)^2 \\
 \kappa -1 & 3 \kappa +4 \sqrt{\kappa }+1 & \kappa -1 & 1-\kappa  & \left(\sqrt{\kappa }-1\right)^2 & 3 \kappa -2 \sqrt{\kappa }-1 & \left(\sqrt{\kappa }-1\right)^2 & -\left(\sqrt{\kappa }-1\right)^2 \\
 1-\kappa  & \kappa -1 & 3 \kappa +4 \sqrt{\kappa }+1 & \kappa -1 & -\left(\sqrt{\kappa }-1\right)^2 & \left(\sqrt{\kappa }-1\right)^2 & 3 \kappa -2 \sqrt{\kappa }-1 & \left(\sqrt{\kappa }-1\right)^2 \\
 \kappa -1 & 1-\kappa  & \kappa -1 & 3 \kappa +4 \sqrt{\kappa }+1 & \left(\sqrt{\kappa }-1\right)^2 & -\left(\sqrt{\kappa }-1\right)^2 & \left(\sqrt{\kappa }-1\right)^2 & 3 \kappa -2 \sqrt{\kappa }-1 \\
 3 \kappa -2 \sqrt{\kappa }-1 & \left(\sqrt{\kappa }-1\right)^2 & -\left(\sqrt{\kappa }-1\right)^2 & \left(\sqrt{\kappa }-1\right)^2 & 3 \kappa +4 \sqrt{\kappa }+1 & \kappa -1 & 1-\kappa  & \kappa -1 \\
 \left(\sqrt{\kappa }-1\right)^2 & 3 \kappa -2 \sqrt{\kappa }-1 & \left(\sqrt{\kappa }-1\right)^2 & -\left(\sqrt{\kappa }-1\right)^2 & \kappa -1 & 3 \kappa +4 \sqrt{\kappa }+1 & \kappa -1 & 1-\kappa  \\
 -\left(\sqrt{\kappa }-1\right)^2 & \left(\sqrt{\kappa }-1\right)^2 & 3 \kappa -2 \sqrt{\kappa }-1 & \left(\sqrt{\kappa }-1\right)^2 & 1-\kappa  & \kappa -1 & 3 \kappa +4 \sqrt{\kappa }+1 & \kappa -1 \\
 \left(\sqrt{\kappa }-1\right)^2 & -\left(\sqrt{\kappa }-1\right)^2 & \left(\sqrt{\kappa }-1\right)^2 & 3 \kappa -2 \sqrt{\kappa }-1 & \kappa -1 & 1-\kappa  & \kappa -1 & 3 \kappa +4 \sqrt{\kappa }+1 \\
\end{array}
\right)$%
    } \ .
\end{equation}
Despite the complexity of this matrix, we again find that the actual contribution appearing in the scalar potential (\ref{Vopenshifted}) is mapped as
\begin{equation}
     \frac{1}{\kappa}\left(1,1,1,1,1,1,1,1\right) \xrightarrow{\Phi_{Z\bar{Z}X\bar{X}}} (1,1,1,1,1,1,1,1) \ ,
\end{equation}
i.e. the coassociator again just strips away a relative $\kappa$-dependence.  Also $\Phi_{Z\bar{Z}X\bar{X}}$  becomes diagonal, and equal to (\ref{PhiXXbZZbdiagonal}), when changing to the basis
\begin{equation}
    \resizebox{14cm}{!}{%
    $\begin{pmatrix}
    \bracing{\bracketing{Z_1}{\bar{Z}_1}{X_{12}}{\bar{X}_{21}}}-\bracing{\bracketing{\bar{Z}_1}{Z_1}{X_{12}}{\bar{X}_{21}}}-\bracing{\bracketing{Z_1}{\bar{Z}_1}{\bar{X}_{12}}{X_{21}}}+\bracing{\bracketing{\bar{Z}_1}{Z_1}{\bar{X}_{12}}{X_{21}}}+\bracing{\bracketing{Z_1}{\bar{Z}_1}{Y_{12}}{\bar{Y}_{21}}}-\bracing{\bracketing{\bar{Z}_1}{Z_1}{Y_{12}}{\bar{Y}_{21}}}-\bracing{\bracketing{Z_1}{\bar{Z}_1}{\bar{Y}_{12}}{Y_{21}}}+\bracing{\bracketing{\bar{Z}_1}{Z_1}{\bar{Y}_{12}}{Y_{21}}}\\
    -\frac{1}{2} \bracing{\bracketing{Z_1}{\bar{Z}_1}{X_{12}}{\bar{X}_{21}}}+\frac{1}{2} \bracing{\bracketing{Z_1}{\bar{Z}_1}{\bar{X}_{12}}{X_{21}}}-\frac{1}{2} \bracing{\bracketing{Z_1}{\bar{Z}_1}{Y_{12}}{\bar{Y}_{21}}}-\frac{1}{2} \bracing{\bracketing{Z_1}{\bar{Z}_1}{\bar{Y}_{12}}{Y_{21}}}+\bracing{\bracketing{\bar{Z}_1}{Z_1}{\bar{Y}_{12}}{Y_{21}}}\\
   \frac{1}{2} \bracing{\bracketing{Z_1}{\bar{Z}_1}{X_{12}}{\bar{X}_{21}}}-\frac{1}{2} \bracing{\bracketing{Z_1}{\bar{Z}_1}{\bar{X}_{12}}{X_{21}}}-\frac{1}{2} \bracing{\bracketing{Z_1}{\bar{Z}_1}{Y_{12}}{\bar{Y}_{21}}}+\bracing{\bracketing{\bar{Z}_1}{Z_1}{Y_{12}}{\bar{Y}_{21}}}-\frac{1}{2} \bracing{\bracketing{Z_1}{\bar{Z}_1}{\bar{Y}_{12}}{Y_{21}}}\\
    -\frac{1}{2} \bracing{\bracketing{Z_1}{\bar{Z}_1}{X_{12}}{\bar{X}_{21}}}-\frac{1}{2} \bracing{\bracketing{Z_1}{\bar{Z}_1}{\bar{X}_{12}}{X_{21}}}+\bracing{\bracketing{\bar{Z}_1}{Z_1}{\bar{X}_{12}}{X_{21}}}-\frac{1}{2} \bracing{\bracketing{Z_1}{\bar{Z}_1}{Y_{12}}{\bar{Y}_{21}}}+\frac{1}{2} \bracing{\bracketing{Z_1}{\bar{Z}_1}{\bar{Y}_{12}}{Y_{21}}}\\
   -\frac{1}{2} \bracing{\bracketing{Z_1}{\bar{Z}_1}{X_{12}}{\bar{X}_{21}}}+\bracing{\bracketing{\bar{Z}_1}{Z_1}{X_{12}}{\bar{X}_{21}}}-\frac{1}{2} \bracing{\bracketing{Z_1}{\bar{Z}_1}{\bar{X}_{12}}{X_{21}}}+\frac{1}{2} \bracing{\bracketing{Z_1}{\bar{Z}_1}{Y_{12}}{\bar{Y}_{21}}}-\frac{1}{2} \bracing{\bracketing{Z_1}{\bar{Z}_1}{\bar{Y}_{12}}{Y_{21}}}\\
    \bracing{\bracketing{Z_1}{\bar{Z}_1}{X_{12}}{\bar{X}_{21}}}+\bracing{\bracketing{\bar{Z}_1}{Z_1}{X_{12}}{\bar{X}_{21}}}+\bracing{\bracketing{Z_1}{\bar{Z}_1}{\bar{Y}_{12}}{Y_{21}}}+\bracing{\bracketing{\bar{Z}_1}{Z_1}{\bar{Y}_{12}}{Y_{21}}}\\
    -\bracing{\bracketing{Z_1}{\bar{Z}_1}{X_{12}}{\bar{X}_{21}}}-\bracing{\bracketing{\bar{Z}_1}{Z_1}{X_{12}}{\bar{X}_{21}}}+\bracing{\bracketing{Z_1}{\bar{Z}_1}{Y_{12}}{\bar{Y}_{21}}}+\bracing{\bracketing{\bar{Z}_1}{Z_1}{Y_{12}}{\bar{Y}_{21}}}\\
    \bracing{\bracketing{Z_1}{\bar{Z}_1}{X_{12}}{\bar{X}_{21}}}+\bracing{\bracketing{\bar{Z}_1}{Z_1}{X_{12}}{\bar{X}_{21}}}+\bracing{\bracketing{Z_1}{\bar{Z}_1}{\bar{X}_{12}}{X_{21}}}+\bracing{\bracketing{\bar{Z}_1}{Z_1}{\bar{X}_{12}}{X_{21}}}
    \end{pmatrix}$%
    }
\end{equation}
and
\begin{equation}
    \resizebox{14cm}{!}{%
    $\begin{pmatrix}
    \bracing{(Z_1\bar{Z}_1)(X_{12}\bar{X}_{21})}-\bracing{(\bar{Z}_1Z_1)(X_{12}\bar{X}_{21})}-\bracing{(Z_1\bar{Z}_1)(\bar{X}_{12}X_{21})}+\bracing{(\bar{Z}_1Z_1)(\bar{X}_{12}X_{21})}+\bracing{(Z_1\bar{Z}_1)(Y_{12}\bar{Y}_{21})}-\bracing{(\bar{Z}_1Z_1)(Y_{12}\bar{Y}_{21})}-\bracing{(Z_1\bar{Z}_1)(\bar{Y}_{12}Y_{21})}+\bracing{(\bar{Z}_1Z_1)(\bar{Y}_{12}Y_{21})}\\
    -\frac{1}{2} \bracing{(Z_1\bar{Z}_1)(X_{12}\bar{X}_{21})}+\frac{1}{2} \bracing{(Z_1\bar{Z}_1)(\bar{X}_{12}X_{21})}-\frac{1}{2} \bracing{(Z_1\bar{Z}_1)(Y_{12}\bar{Y}_{21})}-\frac{1}{2} \bracing{(Z_1\bar{Z}_1)(\bar{Y}_{12}Y_{21})}+\bracing{(\bar{Z}_1Z_1)(\bar{Y}_{12}Y_{21})}\\
    \frac{1}{2} \bracing{(Z_1\bar{Z}_1)(X_{12}\bar{X}_{21})}-\frac{1}{2} \bracing{(Z_1\bar{Z}_1)(\bar{X}_{12}X_{21})}-\frac{1}{2} \bracing{(Z_1\bar{Z}_1)(Y_{12}\bar{Y}_{21})}+\bracing{(\bar{Z}_1Z_1)(Y_{12}\bar{Y}_{21})}-\frac{1}{2} \bracing{(Z_1\bar{Z}_1)(\bar{Y}_{12}Y_{21})}\\
    -\frac{1}{2} \bracing{(Z_1\bar{Z}_1)(X_{12}\bar{X}_{21})}-\frac{1}{2} \bracing{(Z_1\bar{Z}_1)(\bar{X}_{12}X_{21})}+\bracing{(\bar{Z}_1Z_1)(\bar{X}_{12}X_{21})}-\frac{1}{2} \bracing{(Z_1\bar{Z}_1)(Y_{12}\bar{Y}_{21})}+\frac{1}{2} \bracing{(Z_1\bar{Z}_1)(\bar{Y}_{12}Y_{21})}\\
    -\frac{1}{2} \bracing{(Z_1\bar{Z}_1)(X_{12}\bar{X}_{21})}+\bracing{(\bar{Z}_1Z_1)(X_{12}\bar{X}_{21})}-\frac{1}{2} \bracing{(Z_1\bar{Z}_1)(\bar{X}_{12}X_{21})}+\frac{1}{2} \bracing{(Z_1\bar{Z}_1)(Y_{12}\bar{Y}_{21})}-\frac{1}{2} \bracing{(Z_1\bar{Z}_1)(\bar{Y}_{12}Y_{21})}\\
    \bracing{(Z_1\bar{Z}_1)(X_{12}\bar{X}_{21})}+\bracing{(\bar{Z}_1Z_1)(X_{12}\bar{X}_{21})}+\bracing{(Z_1\bar{Z}_1)(\bar{Y}_{12}Y_{21})}+\bracing{(\bar{Z}_1Z_1)(\bar{Y}_{12}Y_{21})}\\
    -\bracing{(Z_1\bar{Z}_1)(X_{12}\bar{X}_{21})}-\bracing{(\bar{Z}_1Z_1)(X_{12}\bar{X}_{21})}+\bracing{(Z_1\bar{Z}_1)(Y_{12}\bar{Y}_{21})}+\bracing{(\bar{Z}_1Z_1)(Y_{12}\bar{Y}_{21})}\\
    \bracing{(Z_1\bar{Z}_1)(X_{12}\bar{X}_{21})}+\bracing{(\bar{Z}_1Z_1)(X_{12}\bar{X}_{21})}+\bracing{(Z_1\bar{Z}_1)(\bar{X}_{12}X_{21})}+\bracing{(\bar{Z}_1Z_1)(\bar{X}_{12}X_{21})}
    \end{pmatrix}$%
    } \ ,
\end{equation}
for the shifted and unshifted states, respectively.

\subsection*{$XY\bar{X}\bar{Y}$-sector}

This is a 36-dimensional sector, where the  coassociator maps between the following basis elements
\begin{footnotesize}
\begin{equation}
    \begin{pmatrix}
    \bracing{\bracketing{X_{12}}{X_{21}}{\bar{X}_{12}}{\bar{X}_{21}}}\\
    \bracing{\bracketing{X_{12}}{\bar{X}_{21}}{X_{12}}{\bar{X}_{21}}}\\
    \bracing{\bracketing{X_{12}}{\bar{X}_{21}}{\bar{X}_{12}}{X_{21}}}\\
    \bracing{\bracketing{X_{12}}{\bar{X}_{21}}{Y_{12}}{\bar{Y}_{21}}}\\
    \bracing{\bracketing{X_{12}}{\bar{X}_{21}}{\bar{Y}_{12}}{Y_{21}}}\\
    \bracing{\bracketing{X_{12}}{Y_{21}}{\bar{X}_{12}}{\bar{Y}_{21}}}\\
    \bracing{\bracketing{X_{12}}{Y_{21}}{\bar{Y}_{12}}{\bar{X}_{21}}}\\
    \bracing{\bracketing{X_{12}}{\bar{Y}_{21}}{\bar{X}_{12}}{Y_{21}}}\\
    \bracing{\bracketing{X_{12}}{\bar{Y}_{21}}{Y_{12}}{\bar{X}_{21}}}\\
    \bracing{\bracketing{\bar{X}_{12}}{X_{21}}{X_{12}}{\bar{X}_{21}}}\\
    \bracing{\bracketing{\bar{X}_{12}}{X_{21}}{\bar{X}_{12}}{X_{21}}}\\
    \bracing{\bracketing{\bar{X}_{12}}{X_{21}}{Y_{12}}{\bar{Y}_{21}}}\\
    \bracing{\bracketing{\bar{X}_{12}}{X_{21}}{\bar{Y}_{12}}{Y_{21}}}\\
    \bracing{\bracketing{\bar{X}_{12}}{\bar{X}_{21}}{X_{12}}{X_{21}}}\\
    \bracing{\bracketing{\bar{X}_{12}}{Y_{21}}{X_{12}}{\bar{Y}_{21}}}\\
    \bracing{\bracketing{\bar{X}_{12}}{Y_{21}}{\bar{Y}_{12}}{X_{21}}}\\
    \bracing{\bracketing{\bar{X}_{12}}{\bar{Y}_{21}}{X_{12}}{Y_{21}}}\\
    \bracing{\bracketing{\bar{X}_{12}}{\bar{Y}_{21}}{Y_{12}}{X_{21}}}\\
    \bracing{\bracketing{Y_{12}}{X_{21}}{\bar{X}_{12}}{\bar{Y}_{21}}}\\
    \bracing{\bracketing{Y_{12}}{X_{21}}{\bar{Y}_{12}}{\bar{X}_{21}}}\\
    \bracing{\bracketing{Y_{12}}{\bar{X}_{21}}{X_{12}}{\bar{Y}_{21}}}\\
    \bracing{\bracketing{Y_{12}}{\bar{X}_{21}}{\bar{Y}_{12}}{X_{21}}}\\
    \bracing{\bracketing{Y_{12}}{Y_{21}}{\bar{Y}_{12}}{\bar{Y}_{21}}}\\
    \bracing{\bracketing{Y_{12}}{\bar{Y}_{21}}{X_{12}}{\bar{X}_{21}}}\\
    \bracing{\bracketing{Y_{12}}{\bar{Y}_{21}}{\bar{X}_{12}}{X_{21}}}\\
    \bracing{\bracketing{Y_{12}}{\bar{Y}_{21}}{Y_{12}}{\bar{Y}_{21}}}\\
    \bracing{\bracketing{Y_{12}}{\bar{Y}_{21}}{\bar{Y}_{12}}{Y_{21}}}\\
    \bracing{\bracketing{\bar{Y}_{12}}{X_{21}}{\bar{X}_{12}}{Y_{21}}}\\
    \bracing{\bracketing{\bar{Y}_{12}}{X_{21}}{Y_{12}}{\bar{X}_{21}}}\\
    \bracing{\bracketing{\bar{Y}_{12}}{\bar{X}_{21}}{X_{12}}{Y_{21}}}\\
    \bracing{\bracketing{\bar{Y}_{12}}{\bar{X}_{21}}{Y_{12}}{X_{21}}}\\
    \bracing{\bracketing{\bar{Y}_{12}}{Y_{21}}{X_{12}}{\bar{X}_{21}}}\\
    \bracing{\bracketing{\bar{Y}_{12}}{Y_{21}}{\bar{X}_{12}}{X_{21}}}\\
    \bracing{\bracketing{\bar{Y}_{12}}{Y_{21}}{Y_{12}}{\bar{Y}_{21}}}\\
    \bracing{\bracketing{\bar{Y}_{12}}{Y_{21}}{\bar{Y}_{12}}{Y_{21}}}\\
    \bracing{\bracketing{\bar{Y}_{12}}{\bar{Y}_{21}}{Y_{12}}{Y_{21}}}
    \end{pmatrix} \xrightarrow{\Phi_{XY\bar{X}\bar{Y}}}
         \begin{pmatrix}
    \bracing{(X_{12}X_{21})(\bar{X}_{12}\bar{X}_{21})}\\
    \bracing{(X_{12}\bar{X}_{21})(X_{12}\bar{X}_{21})}\\
    \bracing{(X_{12}\bar{X}_{21})(\bar{X}_{12}X_{21})}\\
    \bracing{(X_{12}\bar{X}_{21})(Y_{12}\bar{Y}_{21})}\\
    \bracing{(X_{12}\bar{X}_{21})(\bar{Y}_{12}Y_{21})}\\
    \bracing{(X_{12}Y_{21})(\bar{X}_{12}\bar{Y}_{21})}\\
    \bracing{(X_{12}Y_{21})(\bar{Y}_{12}\bar{X}_{21})}\\
    \bracing{(X_{12}\bar{Y}_{21})(\bar{X}_{12}Y_{21})}\\
    \bracing{(X_{12}\bar{Y}_{21})(Y_{12}\bar{X}_{21})}\\
    \bracing{(\bar{X}_{12}X_{21})(X_{12}\bar{X}_{21})}\\
    \bracing{(\bar{X}_{12}X_{21})(\bar{X}_{12}X_{21})}\\
    \bracing{(\bar{X}_{12}X_{21})(Y_{12}\bar{Y}_{21})}\\
    \bracing{(\bar{X}_{12}X_{21})(\bar{Y}_{12}Y_{21})}\\
    \bracing{(\bar{X}_{12}\bar{X}_{21})(X_{12}X_{21})}\\
    \bracing{(\bar{X}_{12}Y_{21})(X_{12}\bar{Y}_{21})}\\
    \bracing{(\bar{X}_{12}Y_{21})(\bar{Y}_{12}X_{21})}\\
    \bracing{(\bar{X}_{12}\bar{Y}_{21})(X_{12}Y_{21})}\\
    \bracing{(\bar{X}_{12}\bar{Y}_{21})(Y_{12}X_{21})}\\
    \bracing{(Y_{12}X_{21})(\bar{X}_{12}\bar{Y}_{21})}\\
    \bracing{(Y_{12}X_{21})(\bar{Y}_{12}\bar{X}_{21})}\\
    \bracing{(Y_{12}\bar{X}_{21})(X_{12}\bar{Y}_{21})}\\
    \bracing{(Y_{12}\bar{X}_{21})(\bar{Y}_{12}X_{21})}\\
    \bracing{(Y_{12}Y_{21})(\bar{Y}_{12}\bar{Y}_{21})}\\
    \bracing{(Y_{12}\bar{Y}_{21})(X_{12}\bar{X}_{21})}\\
    \bracing{(Y_{12}\bar{Y}_{21})(\bar{X}_{12}X_{21})}\\
    \bracing{(Y_{12}\bar{Y}_{21})(Y_{12}\bar{Y}_{21})}\\
    \bracing{(Y_{12}\bar{Y}_{21})(\bar{Y}_{12}Y_{21})}\\
    \bracing{(\bar{Y}_{12}X_{21})(\bar{X}_{12}Y_{21})}\\
    \bracing{(\bar{Y}_{12}X_{21})(Y_{12}\bar{X}_{21})}\\
    \bracing{(\bar{Y}_{12}\bar{X}_{21})(X_{12}Y_{21})}\\
    \bracing{(\bar{Y}_{12}\bar{X}_{21})(Y_{12}X_{21})}\\
    \bracing{(\bar{Y}_{12}Y_{21})(X_{12}\bar{X}_{21})}\\
    \bracing{(\bar{Y}_{12}Y_{21})(\bar{X}_{12}X_{21})}\\
    \bracing{(\bar{Y}_{12}Y_{21})(Y_{12}\bar{Y}_{21})}\\
    \bracing{(\bar{Y}_{12}Y_{21})(\bar{Y}_{12}Y_{21})}\\
    \bracing{(\bar{Y}_{12}\bar{Y}_{21})(Y_{12}Y_{21})}
\end{pmatrix} \ .
\end{equation}
\end{footnotesize}

As we have not been able to find a simple block-diagonal form for the $36\times 36$-dimensional matrix $\Phi^T_{XY\Xb\Yb}$, we will revert to using tensor language to express its elements.  Denoting (just in this section) $\{X_i=1,\bar{X}_i=\bar{1}, Y_i=2, \bar{Y}_i=\bar{2}\}$, we can write (\ref{monomialrotation}) as 
\begin{equation}
    \bracing{\bracketing{\varphi^i}{\varphi^j}{\varphi^k}{\varphi^l}} =  (\Phi^T)_{mnrs}^{ijkl} \bracing{(\varphi^m\varphi^n)(\varphi^r\varphi^s)}= \Phi^{mnrs}_{\;ijkl} \bracing{(\varphi^m\varphi^n)(\varphi^r\varphi^s)} \;,
\end{equation}
and will present the components of $\Phi^{mnrs}_{\;ijkl}$ below. We note that the coassociator maps to itself under:
\begin{itemize}
    \item Complex conjugation of the elements, such that $\PhiM_{a\bar{a}b\bar{b}}^{c\bar{c}d\bar{d}} = \PhiM_{\bar{a}a\bar{b}b}^{\bar{c}c\bar{d}d}$ .
    \item Exchanging $1\leftrightarrow 2$ and $\bar{1}\leftrightarrow \bar{2}$ at the same time, e.g. $\PhiM_{2\bar{2}1\bar{1}}^{1\bar{1}2\bar{2}} = \PhiM_{1\bar{1}2\bar{2}}^{2\bar{2}1\bar{1}}$ .
\end{itemize}
In the list below, we will only write out elements up to these relations. 

After taking out an overall factor of $\frac{1}{16\kappa}$, the tensor elements are
\begin{tiny}
\begin{align}
    \small \frac{1}{4} \left(21 \kappa +24 \sqrt{\kappa }+\frac{1}{\kappa }+18\right) &= \PhiM_{1\bar{1}1\bar{1}}^{1\bar{1}1\bar{1}} \\
    \frac{\left(\sqrt{\kappa }-1\right)^4}{\kappa } &= \PhiM_{12\bar{1}\bar{2}}^{\bar{1}\bar{2}12} = \PhiM_{1\bar{2}\bar{1}2}^{\bar{1}21\bar{2}} \\
    \frac{\left(\sqrt{\kappa }+1\right)^4}{\kappa } &= \PhiM_{12\bar{1}\bar{2}}^{12\bar{1}\bar{2}}=\PhiM_{1\bar{2}\bar{1}2}^{1\bar{2}\bar{1}2} \\
    \left(3 \sqrt{\kappa }+1\right)^2 &= \PhiM_{\bar{1}11\bar{1}}^{\bar{1}11\bar{1}} = \PhiM_{11\bar{1}\bar{1}}^{11\bar{1}\bar{1}} \\
    \frac{5 \kappa }{2}+5 \sqrt{\kappa }+\frac{1}{2 \kappa }+\frac{1}{\sqrt{\kappa }}+7 &= \PhiM_{\bar{1}1\bar{2}2}^{\bar{1}1\bar{2}2} = \PhiM_{\bar{1}12\bar{2}}^{\bar{1}12\bar{2}} = \PhiM_{1\bar{2}2\bar{1}}^{1\bar{2}2\bar{1}} = \PhiM_{12\bar{2}\bar{1}}^{12\bar{2}\bar{1}} \\
    \frac{\left(\sqrt{\kappa }-1\right)^2 \left(5 \kappa +4 \sqrt{\kappa }+1\right)}{2 \kappa } &= \PhiM_{\bar{1}1\bar{2}2}^{2\bar{2}1\bar{1}} = \PhiM_{\bar{1}12\bar{2}}^{\bar{2}21\bar{1}} = \PhiM_{1\bar{2}2\bar{1}}^{\bar{2}1\bar{1}2} = \PhiM_{12\bar{2}\bar{1}}^{21\bar{1}\bar{2}} \\
    \frac{1}{4} \left(-3 \kappa +8 \sqrt{\kappa }+\frac{1}{\kappa }-6\right) &= \PhiM_{12\bar{2}\bar{1}}^{\bar{1}12\bar{2}}=\PhiM_{12\bar{2}\bar{1}}^{2\bar{2}\bar{1}1}=\PhiM _{1\bar{2}2\bar{1}}^{\bar{1}1\bar{2}2}=\PhiM_{1\bar{2}2\bar{1}}^{\bar{2}2\bar{1}1}=\PhiM_{11\bar{1}\bar{1}}^{2\bar{2}\bar{2}2}=\PhiM_{11\bar{1}\bar{1}}^{\bar{2}22\bar{2}} \\
    -\frac{3 \kappa }{2}+\sqrt{\kappa }+\frac{1}{2 \kappa }+\frac{1}{\sqrt{\kappa }}-1 &= -\PhiM_{\bar{1}1\bar{2}2}^{\bar{1}11\bar{1}} = -\PhiM_{\bar{1}1\bar{2}2}^{2\bar{2}\bar{2}2} = \PhiM_{\bar{1}12\bar{2}}^{\bar{1}11\bar{1}} = \PhiM_{\bar{1}12\bar{2}}^{\bar{2}22\bar{2}} = - \PhiM_{11\bar{1}\bar{1}}^{1\bar{2}2\bar{1}} = \PhiM_{11\bar{1}\bar{1}}^{12\bar{2}\bar{1}}= \PhiM_{11\bar{1}\bar{1}}^{21\bar{1}\bar{2}} = - \PhiM_{11\bar{1}\bar{1}}^{\bar{2}1\bar{1}2} \\
    \frac{\left(\sqrt{\kappa }-1\right) \left(\sqrt{\kappa }+1\right)^3}{\kappa } &= \PhiM_{\bar{1}1\bar{2}2}^{\bar{1}\bar{2}12}=\PhiM_{\bar{1}1\bar{2}2}^{21\bar{2}\bar{1}}=\PhiM _{\bar{1}12\bar{2}}^{\bar{1}21\bar{2}}=\PhiM_{\bar{1}12\bar{2}}^{\bar{2}12\bar{1}}=\PhiM_{12\bar{1}\bar{2}}^{12\bar{2}\bar{1}}=\PhiM_{12\bar{1}\bar{2}}^{21\bar{1}\bar{2}}=\PhiM _{1\bar{2}\bar{1}2}^{1\bar{2}2\bar{1}}=\PhiM_{1\bar{2}\bar{1}2}^{\bar{2}1\bar{1}2} \\
    \frac{\left(\sqrt{\kappa }-1\right)^2 \left(3 \kappa +4 \sqrt{\kappa }+1\right)}{2 \kappa } &= \PhiM_{12\bar{1}\bar{2}}^{1\bar{1}\bar{1}1}=\PhiM_{12\bar{1}\bar{2}}^{\bar{2}22\bar{2}}=-\PhiM_{1\bar{2}\bar{1}2}^{1\bar{1}\bar{1}1}=-\PhiM_{1\bar{2}\bar{1}2}^{2\bar{2}\bar{2}2}=-\PhiM_{11\bar{1}\bar{1}}^{12\bar{1}\bar{2}}=-\PhiM_{11\bar{1}\bar{1}}^{21\bar{2}\bar{1}}=\PhiM_{11\bar{1}\bar{1}}^{1\bar{2}\bar{1}2}=\PhiM_{11\bar{1}\bar{1}}^{\bar{2}12\bar{1}} \\
    \frac{5 \kappa }{2}+\sqrt{\kappa }-\frac{1}{2 \kappa }-\frac{1}{\sqrt{\kappa }}-2 &= \PhiM_{12\bar{1}\bar{2}}^{1\bar{1}2\bar{2}}=\PhiM_{12\bar{1}\bar{2}}^{\bar{2}2\bar{1}1}=\PhiM_{12\bar{2}\bar{1}}^{12\bar{1}\bar{2}}=\PhiM_{12\bar{2}\bar{1}}^{21\bar{2}\bar{1}}=\PhiM_{1\bar{2}\bar{1}2}^{1\bar{1}\bar{2}2}=\PhiM_{1\bar{2}\bar{1}2}^{2\bar{2}\bar{1}1}=\PhiM_{1\bar{2}2\bar{1}}^{1\bar{2}\bar{1}2} =\PhiM_{1\bar{2}2\bar{1}}^{\bar{2}12\bar{1}} \\
    \frac{1}{4} \left(5 \kappa -8 \sqrt{\kappa }+\frac{1}{\kappa }+2\right) &= \PhiM_{2\bar{2}2\bar{2}}^{\bar{2}2\bar{2}2}=\PhiM_{2\bar{2}2\bar{2}}^{1\bar{1}1\bar{1}}=\PhiM_{2\bar{2}2\bar{2}}^{\bar{1}1\bar{1}1}=\PhiM_{12\bar{2}\bar{1}}^{\bar{2}21\bar{1}}=\PhiM_{12_{2}\bar{1}}^{1\bar{1}\bar{2}2}=\PhiM_{1\bar{2}2\bar{1}}^{2\bar{2}1\bar{1}}=\PhiM_{1\bar{2}2\bar{1}}^{1\bar{1}2\bar{2}}=\PhiM_{11\bar{1}\bar{1}}^{1\bar{1}\bar{1}1}=\PhiM_{11\bar{1}\bar{1}}^{\bar{1}11\bar{1}} \\
    \kappa +\frac{1}{\kappa }-2 &= \PhiM_{\bar{1}1\bar{2}2}^{12\bar{2}\bar{1}}=\PhiM_{\bar{1}1\bar{2}2}^{\bar{1}\bar{2}21}=\PhiM_{\bar{1}1\bar{2}2}^{21\bar{1}\bar{2}}=\PhiM_{\bar{1}1\bar{2}2}^{\bar{2}\bar{1}12}=\PhiM_{\bar{1}12\bar{2}}^{1\bar{2}2\bar{1}}=\PhiM_{\bar{1}12\bar{2}}^{\bar{1}2\bar{2}1}=\PhiM_{\bar{1}12\bar{2}}^{2\bar{1}1\bar{2}} \nonumber \\
    &=\PhiM_{\bar{1}12\bar{2}}^{\bar{2}1\bar{1}2}=\PhiM_{12\bar{1}\bar{2}}^{21\bar{2}\bar{1}}=\PhiM_{12\bar{1}\bar{2}}^{\bar{2}\bar{1}21}=\PhiM_{1\bar{2}\bar{1}2}^{2\bar{1}\bar{2}1}=\PhiM_{1\bar{2}\bar{1}2}^{\bar{2}12\bar{1}} \\
    \frac{1}{4} \left(-7 \kappa +\frac{1}{\kappa }+6\right) &= -\PhiM_{2\bar{2}2\bar{2}}^{2\bar{2}\bar{2}2}=-\PhiM_{2\bar{2}2\bar{2}}^{\bar{2}22\bar{2}}=-\PhiM_{2\bar{2}2\bar{2}}^{\bar{1}12\bar{2}}=-\PhiM_{2\bar{2}2\bar{2}}^{2\bar{2}\bar{1}1}=\PhiM_{2\bar{2}2\bar{2}}^{1\bar{1}2\bar{2}}=\PhiM_{2\bar{2}2\bar{2}}^{2\bar{2}1\bar{1}}=-\PhiM_{12\bar{2}\bar{1}}^{\bar{2}2\bar{2}2}=-\PhiM_{12\bar{2}\bar{1}}^{1\bar{1}1\bar{1}}=\PhiM_{1\bar{2}2\bar{1}}^{2\bar{2}2\bar{2}} \nonumber \\
    &=\PhiM_{1\bar{2}2\bar{1}}^{1\bar{1}1\bar{1}}=-\PhiM_{11\bar{1}\bar{1}}^{1\bar{1}1\bar{1}}=-\PhiM_{11\bar{1}\bar{1}}^{\bar{1}1\bar{1}1} \\
    -3 \kappa +2 \sqrt{\kappa }+1 &= -\PhiM_{1\bar{1}\bar{1}1}^{1\bar{1}2\bar{2}}=-\PhiM_{1\bar{1}\bar{1}1}^{\bar{2}2\bar{1}1}=\PhiM_{1\bar{1}\bar{1}1}^{1\bar{1}\bar{2}2}=\PhiM_{1\bar{1}\bar{1}1}^{2\bar{2}\bar{1}1}=-\PhiM_{1\bar{1}\bar{1}1}^{1\bar{1}1\bar{1}}=-\PhiM_{1\bar{1}\bar{1}1}^{\bar{1}1\bar{1}1}=\PhiM_{2\bar{2}\bar{2}2}^{1\bar{1}\bar{2}2}=\PhiM_{2\bar{2}\bar{2}2}^{2\bar{2}\bar{1}1}=-\PhiM_{2\bar{2}2\bar{2}}^{22\bar{2}\bar{2}} \nonumber \\
    &=-\PhiM_{2\bar{2}2\bar{2}}^{\bar{2}\bar{2}22}=\PhiM_{12\bar{2}\bar{1}}^{11\bar{1}\bar{1}}=\PhiM_{12\bar{2}\bar{1}}^{22\bar{2}\bar{2}}=-\PhiM_{1\bar{2}2\bar{1}}^{1,1\bar{1}\bar{1}}=-\PhiM_{1\bar{2}2\bar{1}}^{\bar{2}\bar{2}22} \\
    \frac{\left(\sqrt{\kappa }-1\right)^2 (\kappa +1)}{2 \kappa } &= -\PhiM_{\bar{1}1\bar{2}2}^{1\bar{1}\bar{1}1}=-\PhiM_{\bar{1}1\bar{2}2}^{\bar{2}2,2\bar{2}}=-\PhiM_{11\bar{1}\bar{1}}^{\bar{1}2\bar{2}1}=-\PhiM_{11\bar{1}\bar{1}}^{2\bar{1}1\bar{2}}=\PhiM_{2\bar{2}1\bar{1}}^{1\bar{1}2\bar{2}}=\PhiM_{2\bar{2}1\bar{1}}^{\bar{2}2\bar{1}1} =\PhiM_{\bar{1}12\bar{2}}^{1\bar{1}\bar{1}1}=\PhiM_{\bar{1}12\bar{2}}^{2\bar{2}\bar{2}2}=\PhiM_{\bar{1}12\bar{2}}^{1\bar{1}\bar{2}2}=\PhiM_{\bar{1}12\bar{2}}^{2\bar{2}\bar{1}1} \nonumber \\
    &=\PhiM_{12\bar{2}\bar{1}}^{\bar{1}\bar{2}21}=\PhiM_{12\bar{2}\bar{1}}^{\bar{2}\bar{1}12}=\PhiM_{1\bar{2}2\bar{1}}^{\bar{1}2\bar{2}1}=\PhiM_{1\bar{2}2\bar{1}}^{2\bar{1}1\bar{2}}=\PhiM_{11\bar{1}\bar{1}}^{\bar{1}\bar{2}21}=\PhiM_{11\bar{1}\bar{1}}^{\bar{2}\bar{1}12} \\
    \frac{1}{2} \left(-\kappa -4 \sqrt{\kappa }+\frac{1}{\kappa }+4\right) &= \PhiM_{\bar{1}1\bar{2}2}^{\bar{2}2\bar{2}2}=\PhiM_{\bar{1}1\bar{2}2}^{\bar{1}1\bar{1}1}=\PhiM_{2\bar{2}2\bar{2}}^{1\bar{2}2\bar{1}}=\PhiM_{2\bar{2}2\bar{2}}^{2\bar{1}1\bar{2}}=-\PhiM_{\bar{1}1\bar{2}2}^{\bar{1}12\bar{2}}=-\PhiM_{\bar{1}1\bar{2}2}^{1\bar{1}\bar{2}2}=-\PhiM_{\bar{1}12\bar{2}}^{\bar{1}1\bar{2}2}=-\PhiM_{\bar{1}12\bar{2}}^{2\bar{2}2\bar{2}}=-\PhiM_{\bar{1}12\bar{2}}^{1\bar{1}2\bar{2}}=-\PhiM_{\bar{1}12\bar{2}}^{\bar{1}1\bar{1}1} \nonumber \\
    &=-\PhiM_{2\bar{2}2\bar{2}}^{\bar{1}\bar{2}21}=-\PhiM_{2\bar{2}2\bar{2}}^{21\bar{1}\bar{2}}=-\PhiM_{12\bar{2}\bar{1}}^{1\bar{2}2\bar{1}}=-\PhiM_{12\bar{2}\bar{1}}^{\bar{1}2\bar{2}1}=-\PhiM_{1\bar{2}2\bar{1}}^{12\bar{2}\bar{1}}=-\PhiM_{1\bar{2}2\bar{1}}^{\bar{1}\bar{2}21} \\
    \frac{1}{2} \left(-\kappa +4 \sqrt{\kappa }+\frac{1}{\kappa }-4\right) &= \PhiM_{\bar{1}1\bar{2}2}^{2\bar{2}2\bar{2}}=\PhiM_{\bar{1}1\bar{2}2}^{1\bar{1}1\bar{1}}=\PhiM_{2\bar{2}2\bar{2}}^{\bar{1}2\bar{2}1}=\PhiM_{2\bar{2}2\bar{2}}^{\bar{2}1\bar{1}2}=-\PhiM_{\bar{1}1\bar{2}2}^{\bar{2}21\bar{1}}=-\PhiM_{\bar{1}1\bar{2}2}^{2\bar{2}\bar{1}1}=-\PhiM_{\bar{1}12\bar{2}}^{2\bar{2}1\bar{1}}=-\PhiM_{\bar{1}12\bar{2}}^{\bar{2}2\bar{2}2}=-\PhiM_{\bar{1}12\bar{2}}^{\bar{2}2\bar{1}1}=-\PhiM_{\bar{1}12\bar{2}}^{1\bar{1}1\bar{1}} \nonumber \\
    &=-\PhiM_{2\bar{2}2\bar{2}}^{12\bar{2}\bar{1}}=-\PhiM_{2\bar{2}2\bar{2}}^{\bar{2}\bar{1}12}=-\PhiM_{12\bar{2}\bar{1}}^{2\bar{1}1\bar{2}}=-\PhiM_{12\bar{2}\bar{1}}^{\bar{2}1\bar{1}2}=-\PhiM_{1\bar{2}2\bar{1}}^{21\bar{1}\bar{2}}=-\PhiM_{1\bar{2}2\bar{1}}^{\bar{2}\bar{1}12}\\
    \left(\sqrt{\kappa }-1\right)^2 &= -\PhiM_{1\bar{1}\bar{1}1}^{\bar{1}1\bar{2}2}=-\PhiM_{1\bar{1}\bar{1}1}^{2\bar{2}1\bar{1}}=-\PhiM_{1\bar{1}\bar{1}1}^{2\bar{2}2\bar{2}}=-\PhiM_{1\bar{1}\bar{1}1}^{\bar{2}2\bar{2}2}=-\PhiM_{2\bar{2}2\bar{2}}^{11\bar{1}\bar{1}}=-\PhiM_{2\bar{2}2\bar{2}}^{\bar{1}\bar{1}11}=-\PhiM_{1\bar{2}2\bar{1}}^{\bar{1}\bar{1}11}=-\PhiM_{1\bar{2}2\bar{1}}^{22\bar{2}\bar{2}}=\PhiM_{1\bar{1}\bar{1}1}^{\bar{1}11\bar{1}} \nonumber \\
    &=\PhiM_{1\bar{1}\bar{1}1}^{2\bar{2}\bar{2}2}=\PhiM_{1\bar{1}\bar{1}1}^{\bar{2}22\bar{2}}=\PhiM_{1\bar{1}\bar{1}1}^{\bar{1}12\bar{2}}=\PhiM_{1\bar{1}\bar{1}1}^{\bar{2}21\bar{1}}=\PhiM_{12\bar{2}\bar{1}}^{\bar{1}\bar{1}11}=\PhiM_{12\bar{2}\bar{1}}^{\bar{2}\bar{2}22}=\PhiM_{11\bar{1}\bar{1}}^{\bar{1}\bar{1}11}=\PhiM_{11\bar{1}\bar{1}}^{22\bar{2}\bar{2}}=\PhiM_{11\bar{1}\bar{1}}^{\bar{2}\bar{2}22} \\
    \frac{\left(\sqrt{\kappa }-1\right)^3 \left(\sqrt{\kappa }+1\right)}{\kappa } &= -\frac{1}{2} \PhiM_{12\bar{1}\bar{2}}^{\bar{1}11\bar{1}}=-\frac{1}{2} \PhiM_{12\bar{1}\bar{2}}^{2\bar{2}\bar{2}2}=-\frac{1}{2} \PhiM_{11\bar{1}\bar{1}}^{\bar{1}21\bar{2}}=-\frac{1}{2} \PhiM_{11\bar{1}\bar{1}}^{2\bar{1}\bar{2}1}=\PhiM_{\bar{1}1\bar{2}2}^{12\bar{1}\bar{2}}=\PhiM_{\bar{1}1\bar{2}2}^{\bar{2}\bar{1}21}=\PhiM_{\bar{1}12\bar{2}}^{1\bar{2}\bar{1}2}=\PhiM_{\bar{1}12\bar{2}}^{2\bar{1}\bar{2}1}=\PhiM_{12\bar{1}\bar{2}}^{\bar{1}\bar{2}21} \nonumber \\
    &=\PhiM_{12\bar{1}\bar{2}}^{\bar{2}\bar{1}12}=\PhiM_{1\bar{2}\bar{1}2}^{\bar{1}2\bar{2}1}=\PhiM_{1\bar{2}\bar{1}2}^{2\bar{1}1\bar{2}}=\frac{1}{2} \PhiM_{12\bar{1}\bar{2}}^{\bar{1}1\bar{2}2}=\frac{1}{2} \PhiM_{12\bar{1}\bar{2}}^{2\bar{2}1\bar{1}}=\frac{1}{2} \PhiM_{12\bar{2}\bar{1}}^{\bar{1}\bar{2}12}=\frac{1}{2} \PhiM_{12\bar{2}\bar{1}}^{\bar{2}\bar{1}21}=\frac{1}{2} \PhiM_{1\bar{2}\bar{1}2}^{\bar{1}11\bar{1}}=\frac{1}{2} \PhiM_{1\bar{2}\bar{1}2}^{\bar{2}22\bar{2}} \nonumber \\
    &=\frac{1}{2} \PhiM_{1\bar{2}\bar{1}2}^{\bar{1}12\bar{2}}=\frac{1}{2} \PhiM_{1\bar{2}\bar{1}2}^{\bar{2}21\bar{1}}=\frac{1}{2} \PhiM_{1\bar{2}2\bar{1}}^{\bar{1}21\bar{2}}=\frac{1}{2} \PhiM_{1\bar{2}2\bar{1}}^{2\bar{1}\bar{2}1}=\frac{1}{2} \PhiM_{11\bar{1}\bar{1}}^{\bar{1}\bar{2}12}=\frac{1}{2} \PhiM_{11\bar{1}\bar{1}}^{\bar{2}\bar{1}21} \\
    \frac{(\kappa -1)^2}{2 \kappa } &= -\PhiM_{2\bar{2}2\bar{2}}^{1\bar{2}\bar{1}2}=-\PhiM_{2\bar{2}2\bar{2}}^{\bar{1}21\bar{2}}=-\PhiM_{2\bar{2}2\bar{2}}^{2\bar{1}\bar{2}1}=-\PhiM_{2\bar{2}2\bar{2}}^{\bar{2}12\bar{1}}=-\PhiM_{12\bar{1}\bar{2}}^{2\bar{2}2\bar{2}}=-\PhiM_{12\bar{1}\bar{2}}^{\bar{2}2\bar{2}2}=-\PhiM_{12\bar{1}\bar{2}}^{1\bar{1}1\bar{1}}=-\PhiM_{12\bar{1}\bar{2}}^{\bar{1}1\bar{1}1}=\PhiM_{2\bar{2}2\bar{2}}^{12\bar{1}\bar{2}} \nonumber \\
    &=\PhiM_{2\bar{2}2\bar{2}}^{\bar{1}\bar{2}12}=\PhiM_{2\bar{2}2\bar{2}}^{21\bar{2}\bar{1}}=\PhiM_{2\bar{2}2\bar{2}}^{\bar{2}\bar{1}21}=\PhiM_{12\bar{1}\bar{2}}^{\bar{1}12\bar{2}}=\PhiM_{12\bar{1}\bar{2}}^{\bar{2}21\bar{1}}=\PhiM_{12\bar{1}\bar{2}}^{1\bar{1}\bar{2}2}=\PhiM_{12\bar{1}\bar{2}}^{2\bar{2}\bar{1}1}=\PhiM_{12\bar{2}\bar{1}}^{1\bar{2}\bar{1}2}=\PhiM_{12\bar{2}\bar{1}}^{\bar{1}21\bar{2}}=\PhiM_{12\bar{2}\bar{1}}^{2\bar{1}\bar{2}1}=\PhiM_{12\bar{2}\bar{1}}^{\bar{2}12\bar{1}} \nonumber \\
    &=\PhiM_{1\bar{2}\bar{1}2}^{\bar{1}1\bar{2}2}=\PhiM_{1\bar{2}\bar{1}2}^{2\bar{2}1\bar{1}}=\PhiM_{1\bar{2}\bar{1}2}^{2\bar{2}2\bar{2}}=\PhiM_{1\bar{2}\bar{1}2}^{\bar{2}2\bar{2}2}=\PhiM_{1\bar{2}\bar{1}2}^{1\bar{1}2\bar{2}}=\PhiM_{1\bar{2}\bar{1}2}^{\bar{2}2\bar{1}1}=\PhiM_{1\bar{2}\bar{1}2}^{1\bar{1}1\bar{1}}=\PhiM_{1\bar{2}\bar{1}2}^{\bar{1}1\bar{1}1}=\PhiM_{1\bar{2}2\bar{1}}^{12\bar{1}\bar{2}}=\PhiM_{1\bar{2}2\bar{1}}^{\bar{1}\bar{2}12}=\PhiM_{1\bar{2}2\bar{1}}^{21\bar{2}\bar{1}} \nonumber \\
    &=\PhiM_{1\bar{2}2\bar{1}}^{\bar{2}\bar{1}21} =\frac12\PhiM_{2\bar{2}2\bar{2}}^{\bar{1}1\bar{2}2}=\frac12\PhiM_{2\bar{2}2\bar{2}}^{\bar{2}2\bar{1}1}=\frac12\PhiM_{12\bar{2}\bar{1}}^{1\bar{1}\bar{1}1}=\frac12\PhiM_{12\bar{2}\bar{1}}^{\bar{1}11\bar{1}}=\frac12\PhiM_{12\bar{2}\bar{1}}^{2\bar{2}\bar{2}2}=\frac12\PhiM_{12\bar{2}\bar{1}}^{\bar{2}22\bar{2}}=\frac12\PhiM_{1\bar{2}2\bar{1}}^{\bar{2}2\bar{2}2}=\frac12\PhiM_{1\bar{2}2\bar{1}}^{\bar{1}1\bar{1}1} \nonumber \\
    &=\frac12\PhiM_{11\bar{1}\bar{1}}^{\bar{1}12\bar{2}}=\frac12\PhiM_{11\bar{1}\bar{1}}^{\bar{2}21\bar{1}}=\frac12\PhiM_{11\bar{1}\bar{1}}^{1\bar{1}\bar{2}2}=\frac12\PhiM_{11\bar{1}\bar{1}}^{2\bar{2}\bar{1}1}=-\frac12\PhiM_{2\bar{2}2\bar{2}}^{1\bar{1}\bar{1}1}=-\frac12\PhiM_{2\bar{2}2\bar{2}}^{\bar{1}11\bar{1}}=-\frac12\PhiM_{2\bar{2}2\bar{2}}^{\bar{2}21\bar{1}}=-\frac12\PhiM_{2\bar{2}2\bar{2}}^{1\bar{1}\bar{2}2} \nonumber \\
    &=-\frac12\PhiM_{12\bar{2}\bar{1}}^{\bar{1}1\bar{2}2}=-\frac12\PhiM_{12\bar{2}\bar{1}}^{2\bar{2}1\bar{1}}=-\frac12\PhiM_{12\bar{2}\bar{1}}^{2\bar{2}2\bar{2}}=-\frac12\PhiM_{12\bar{2}\bar{1}}^{1\bar{1}2\bar{2}}=-\frac12\PhiM_{12\bar{2}\bar{1}}^{\bar{2}2\bar{1}1}=-\frac12\PhiM_{12\bar{2}\bar{1}}^{\bar{1}1\bar{1}1}=-\frac12\PhiM_{2\bar{1}1\bar{2}}^{1\bar{1}\bar{1}1}=-\frac12\PhiM_{2\bar{1}1\bar{2}}^{\bar{1}11\bar{1}} \nonumber \\
    &=-\frac12\PhiM_{2\bar{1}1\bar{2}}^{2\bar{2}\bar{2}2}=-\frac12\PhiM_{2\bar{1}1\bar{2}}^{\bar{2}22\bar{2}}=-\frac12\PhiM_{2\bar{1}1\bar{2}}^{\bar{1}12\bar{2}}=-\frac12\PhiM_{2\bar{1}1\bar{2}}^{\bar{2}21\bar{1}}=-\frac12\PhiM_{2\bar{1}1\bar{2}}^{1\bar{1}\bar{2}2}=-\frac12\PhiM_{2\bar{1}1\bar{2}}^{2\bar{2}\bar{1}1}=-\frac12\PhiM_{11\bar{1}\bar{1}}^{\bar{1}1\bar{2}2}=-\frac12\PhiM_{11\bar{1}\bar{1}}^{2\bar{2}1\bar{1}} \nonumber \\
    &=-\frac12\PhiM_{11\bar{1}\bar{1}}^{2\bar{2}2\bar{2}}=-\frac12\PhiM_{11\bar{1}\bar{1}}^{\bar{2}2\bar{2}2}=-\frac12\PhiM_{11\bar{1}\bar{1}}^{1\bar{1}2\bar{2}}=-\frac12\PhiM_{11\bar{1}\bar{1}}^{\bar{2}2\bar{1}1}
\end{align} 
\end{tiny}
As a matrix, $\Phi_{XY\bar{X}\bar{Y}}$ has a determinant of $\kappa^{-24}$. Considering the actual contribution to the scalar potential in this sector in (\ref{Vopenshifted}), the map to unshifted states is:
\begin{small}
    \[
    \begin{split}
         \frac{\kappa}{4} \left( 1, -1 , 0 , 0 , -2 , 0 , 1 , 2 , -1 , 0 , -1 , -2 , 0 , 1 , 2 , -1 , 0 , 1 , 1 , 0 , -1 ,  2 , 1 , 0 , -2 , -1 , 0 , -1 , 2 , 1 , 0 , -2 , 0 , 0 , -1 , 1 \right) \\
         \xrightarrow{\Phi_{XY\bar{X}\bar{Y}}} \hspace{7cm} \\
         \Big(\frac{1}{4},\frac{1-5 \kappa }{16 \kappa },\frac{\kappa -1}{16 \kappa },-\frac{\kappa -1}{16 \kappa },-\frac{7 \kappa +1}{16 \kappa },\frac{\kappa -1}{8 \kappa },\frac{\kappa +1}{8 \kappa },\frac{\kappa +3}{8 \kappa },\frac{\kappa -3}{8 \kappa },\frac{\kappa -1}{16 \kappa },\frac{1-5 \kappa }{16 \kappa },-\frac{7 \kappa +1}{16 \kappa },-\frac{\kappa -1}{16 \kappa },\frac{1}{4}, \\
         \qquad \frac{\kappa +3}{8 \kappa },\frac{\kappa -3}{8 \kappa },\frac{\kappa -1}{8 \kappa },\frac{\kappa +1}{8 \kappa },\frac{\kappa +1}{8 \kappa },\frac{\kappa -1}{8 \kappa },\frac{\kappa -3}{8 \kappa },\frac{\kappa +3}{8 \kappa },\frac{1}{4},-\frac{\kappa -1}{16 \kappa },-\frac{7 \kappa +1}{16 \kappa },\frac{1-5 \kappa }{16 \kappa },\frac{\kappa -1}{16 \kappa },\frac{\kappa -3}{8 \kappa }, \\
         \qquad \qquad\frac{\kappa +3}{8 \kappa },\frac{\kappa +1}{8 \kappa },\frac{\kappa -1}{8 \kappa },-\frac{7 \kappa +1}{16 \kappa },-\frac{\kappa -1}{16 \kappa },\frac{\kappa -1}{16 \kappa },\frac{1-5 \kappa }{16 \kappa },\frac{1}{4}\Big)
    \end{split}
    \]
\end{small}
Of course, this becomes the identity action when $\kappa=1$. We again see that some monomials do not appear in the actual quartic terms, but it is important to know how the coassociator acts on them, as they would be expected to appear in other representations beyond the singlet.

\section{The One-Loop Hamiltonian} \label{HamiltonianAppendix}

The one-loop Hamiltonian for spin chains made up of the scalar fields of the $\Ncal=2$ $\Zset_2$ orbifold theory was derived in \cite{Gadde:2010zi}, to which we refer for the full details. It is a nearest-neighbour Hamiltonian, which at the orbifold point $\kappa=1$ essentially reduces to two (equal) copies of the $\Ncal=4$ SYM $\SO(6)$ Hamiltonian \cite{Minahan:2002ve}.  
Here we just record the Hamiltonian in two subsectors, the holomorphic sector made up of the $X,Y,Z$ fields and the $\SO(6)$ neutral sector.

In the holomorphic $\mathrm{SU}(3)_{XYZ}$ sector, in the basis 
\begin{equation}
    \{Z_1 Z_1,X_{12} Z_2,Z_1 X_{12},Y_{12} Z_2,Z_1 Y_{12},X_{12} X_{21},X_{12} Y_{21},Y_{12} X_{21},Y_{12} Y_{21}
    \}
\end{equation}
the Hamiltonian is
\begin{equation}
    \mathcal{H}^{XYZ}_1 = \left(
\begin{array}{ccccccccc}
 0 & 0 & 0 & 0 & 0 & 0 & 0 & 0 & 0 \\
 0 & \kappa  & -1 & 0 & 0 & 0 & 0 & 0 & 0 \\
 0 & -1 & \frac{1}{\kappa } & 0 & 0 & 0 & 0 & 0 & 0 \\
 0 & 0 & 0 & \kappa  & -1 & 0 & 0 & 0 & 0 \\
 0 & 0 & 0 & -1 & \frac{1}{\kappa } & 0 & 0 & 0 & 0 \\
 0 & 0 & 0 & 0 & 0 & 0 & 0 & 0 & 0 \\
 0 & 0 & 0 & 0 & 0 & 0 & \frac{1}{\kappa } & -\frac{1}{\kappa } & 0 \\
 0 & 0 & 0 & 0 & 0 & 0 & -\frac{1}{\kappa } & \frac{1}{\kappa } & 0 \\
 0 & 0 & 0 & 0 & 0 & 0 & 0 & 0 & 0 \\
\end{array}
\right) \ .
\end{equation}
and its $\kappa\ra 1/\kappa$ conjugate when acting on the $\Zset_2$-conjugate basis. This Hamiltonian is relevant for the discussion of holomorphic BPS states in Section \ref{HolomorphicBPS}. 

In the $\mathrm{SO}(6)$ neutral sector, in the basis
\begin{equation} 
    \{X_{12} \bar{X}_{21},\bar{X}_{12} X_{21},Y_{12} \bar{Y}_{21},\bar{Y}_{12} Y_{21},Z_1 \bar{Z}_1,\bar{Z}_1 Z_1
    \}    
\end{equation}
the Hamiltonian takes the form
\begin{equation} \label{Hneutral}
\mathcal{H}^{\text{neutral}}_1 = \frac{1}{2\kappa} \left(
\begin{array}{cccccc}
 2 \kappa ^2+1 & -1 & 1 & 2 \kappa ^2-1 & 1 & 1 \\
 -1 & 2 \kappa ^2+1 & 2 \kappa ^2-1 & 1 & 1 & 1 \\
 1 & 2 \kappa ^2-1 & 2 \kappa ^2+1 & -1 & 1 & 1 \\
 2 \kappa ^2-1 & 1 & -1 & 2 \kappa ^2+1 & 1 & 1 \\
 1 & 1 & 1 & 1 & 3 & -1 \\
 1 & 1 & 1 & 1 & -1 & 3 \\
\end{array}
\right) \ .
\end{equation}
However, as discussed in Section \ref{20prime}, diagonalising this Hamiltonian leads to a negative eigenvalue, which is an artifact of working with open (non-physical) states. This can be fixed by adding a modification which vanishes at $\kappa=1$ and does not affect the spectrum of the closed Hamiltonian. In the monomial basis, the modified Hamiltonian is 
\begin{equation} \label{Hneutralmod}
\hat{\mathcal{H}}^{\text{neutral}}_1 = \frac{1}{2\kappa} \left(
\begin{array}{cccccc}
 2 \left(\frac{\kappa }{2}+\frac{1}{\kappa }\right) \kappa  & -\kappa ^2 & 1 & 1 & 1 & 1 \\
 -\kappa ^2 & 2 \left(\frac{\kappa }{2}+\frac{1}{\kappa }\right) \kappa  & 1 & 1 & 1 & 1 \\
 1 & 1 & 2 \left(\frac{\kappa }{2}+\frac{1}{\kappa }\right) \kappa  & -\kappa ^2 & 1 & 1 \\
 1 & 1 & -\kappa ^2 & 2 \left(\frac{\kappa }{2}+\frac{1}{\kappa }\right) \kappa  & 1 & 1 \\
 1 & 1 & 1 & 1 & 2 \left(\frac{\kappa }{2}+\frac{1}{\kappa }\right) \kappa  & -\kappa ^2 \\
 1 & 1 & 1 & 1 & -\kappa ^2 & 2 \left(\frac{\kappa }{2}+\frac{1}{\kappa }\right) \kappa  \\
\end{array}
\right) \ .
\end{equation}
and $\kappa\ra1/\kappa$ when acting on the $\Zset_2$-conjugate basis. 
This modified Hamiltonian is what was used in the analysis of the two-site ${\bf 20'}, {\bf 15}$ and singlet representations. 

\bibliography{hiddensym}
\bibliographystyle{utphys}

\end{document}